\documentclass[a4paper]{Frontiers1}

\usepackage{url,hyperref,lineno,microtype,subcaption}
\usepackage[onehalfspacing]{setspace}
\usepackage{amsmath,amssymb,mathrsfs,mathptmx,bm}
\usepackage{caption,graphicx,multirow}

\def\keyFont{\fontsize{8}{11}\helveticabold}
\def\firstAuthorLast{Loskot {et~al.}}

\def\Authors{Pavel Loskot\,$^{1,*}$, Komlan Atitey\,$^{1}$, and Lyudmila
  Mihaylova\,$^{2}$}
\def\Address{$^{1}$College of Engineering, Swansea University, Swansea, United
  Kingdom\\
  $^{2}$Dept. of Automatic Control and Systems Engineering, University of
  Sheffield, Sheffield, United Kingdom}
\def\corrAuthor{Pavel Loskot}
\def\corrEmail{p.loskot@swan.ac.uk}

\newcommand{\toc}[2]{\ref{#1}\ {\uppercase{#2}}\dotfill\pageref{#1}\\}

\newcommand{\xrot}[1]{\footnotesize\rotatebox[origin=l]{90}{#1}}
\newcommand{\xcell}[2][t]{\renewcommand{\arraystretch}{1.}%
  \begin{tabular}[#1]{@{}l@{}}#2\end{tabular}}

\newcommand{\T}{\rule{0pt}{2.6ex}}
\newcommand{\TT}{\rule{0pt}{3.2ex}}

\newcommand{\BB}{\rule[-1.8ex]{0pt}{0pt}}

\newcommand{\iitem}{\textbullet\quad}
\newcommand{\scite}[1]{{\small\cite{#1}}}
\newcommand{\scref}[1]{Section \ref{#1}} 
\newcommand{\sucref}[1]{sub-section \ref{#1}} 
\newcommand{\eref}[1]{(\ref{#1})} 
\newcommand{\tref}[1]{Table \ref{#1}} 
\newcommand{\fref}[1]{Figure \ref{#1}} 

\newcommand{\bibtex}{{\normalfont B\kern-.05em{\scshape i\kern-.025em b}%
    \kern-.08em \TeX}}

\newenvironment{instab}[4][2.5]{\begin{table}[!t]\begin{center}
      \caption{#4\medskip}\label{#3}\renewcommand{\arraystretch}{#1}%
      \begin{tabular}{#2}}{\end{tabular}\end{center}\end{table}}
\newenvironment{instabcont}[4][2.5]{\begin{table}[!t]\ContinuedFloat
    \begin{center}\caption{#4\medskip}\label{#3}
      \renewcommand{\arraystretch}{#1}%
      \begin{tabular}{#2}}{\end{tabular}\end{center}\end{table}}

\newcommand{\E}[1]{\operatorname{E}\!\left[#1\right]}
\newcommand{\var}[1]{\operatorname{var}\!\left[#1\right]}
\newcommand{\EE}[2]{\operatorname{E}_{#1}\!\left[#2\right]}
\newcommand{\Prob}[1]{\operatorname{Pr}\!\left(#1\right)}
\newcommand{\argmin}{\operatorname{argmin}}
\newcommand{\argmax}{\operatorname{argmax}}
\newcommand{\df}{{\,\operatorname{d}}}

\newcommand{\Zs}{\bm{0}}

\newcommand{\vK}{\bm{K}}
\newcommand{\vKh}{\hat{\vK}}

\newcommand{\vQ}{\bm{Q}}
\newcommand{\vKt}{\tilde{\bm{K}}}
\newcommand{\vC}{\bm{C}}
\newcommand{\vCt}{\tilde{\bm{C}}}
\newcommand{\vw}{\bm{w}}

\begin{document}
\onecolumn\firstpage{1}
\newcommand{\headerA}[1]{ \hline \T & \multicolumn{3}{c}{\xcell{Physical\\
      laws}} & \multicolumn{4}{c}{\xcell{Random\\ processes}} &
  \multicolumn{8}{c}{\xcell{Mathematical\\ models}} &
  \multicolumn{4}{c}{\xcell{Interaction\\ models}} &
  \multicolumn{6}{c}{\xcell{CME based\\ models}} \\
  {#1} & \xrot{kinetic rate laws} & \xrot{mass action kinetics} &
  \xrot{mechanistic models} & \xrot{Markov process} & \xrot{Poisson process} &
  \xrot{birth-death process} & \xrot{telegraph process} & \xrot{state space
    representation} & \xrot{ODEs, PDEs, SDEs, DDEs\quad} & \xrot{rational
    model} & \xrot{differential algebraic eqns.} & \xrot{tensor representation}
  & \xrot{S-system model} & \xrot{polynomial model} & \xrot{manifold map} &
  \xrot{Petri nets} & \xrot{Boolean networks} & \xrot{neural networks} &
  \xrot{agent based models} & \xrot{Langevin equation} & \xrot{Fokker-Planck
    equation} & \xrot{reaction rate equation} & \xrot{moment closure} &
  \xrot{linear noise approximation} & \xrot{system size expansion} \\ \hline }

\newcommand{\headerB}[1]{ \hline \T & \multicolumn{4}{c}{Tasks} &
  \multicolumn{5}{c}{Measures} & \multicolumn{4}{c}{\xcell{Bayesian\\ methods}}
  & \multicolumn{3}{c}{\xcell{Monte\\ Carlo}} &
  \multicolumn{3}{c}{\xcell{Kalman\\ filter}} &
  \multicolumn{8}{c}{Model fitting} & {XLR} \\
  {#1} & \xrot{identifi., observab., reachability} & \xrot{optimum experiment
    design} & \xrot{bifurcation analysis} & \xrot{inference, identification} &
  \xrot{sensitivity analysis} & \xrot{confidence/credible intervals} &
  \xrot{Akaike/Fisher/mutual info.} & \xrot{entropy} & \xrot{sum of squared
    errors} & \xrot{MAP, ML, likelihood} & \xrot{approximate Bayesian comput.}
  & \xrot{expectation-maximization} & \xrot{variational Bayesian inference} &
  \xrot{MCMC} & \xrot{Metropol./import. sampling} & \xrot{sequential MC,
    particle filters} & \xrot{Kalman filter} & \xrot{extended Kalman filter} &
  \xrot{unscented Kalman filter} & \xrot{LS and regression} & \xrot{genetic
    algorithms} & \xrot{optimization programming} & \xrot{simulated annealing}
  & \xrot{differential evolution} & \xrot{scatter, tabu, cuckoo search} &
  \xrot{particle swarm optimization} & \xrot{other algorithms} &
  \xrot{mach./deep/transf. learning} \\ \hline }

\newcommand{\headerBa}{ \hline \T & & \multicolumn{5}{c}{Tasks} &
  \multicolumn{4}{c}{Measures} & \multicolumn{4}{c}{\xcell{Bayesian\\ methods}}
  & \multicolumn{3}{c}{\xcell{Monte\\ Carlo}} &
  \multicolumn{3}{c}{\xcell{Kalman\\ filter}} &
  \multicolumn{8}{c}{Model fitting} & {XLR} \\
  & & \xrot{identifi., observab., reachability} & \xrot{optimum experiment
    design} & \xrot{bifurcation analysis} & \xrot{inference, identification} &
  \xrot{sensitivity analysis} & \xrot{confidence/credible intervals} &
  \xrot{Akaike/Fisher/mutual info.} & \xrot{entropy} & \xrot{sum of squared
    errors} & \xrot{MAP, ML, likelihood} & \xrot{approximate Bayesian comput.}
  & \xrot{expectation-maximization} & \xrot{variational Bayesian inference} &
  \xrot{MCMC} & \xrot{Metropol./import. sampling} & \xrot{sequential MC,
    particle filters} & \xrot{Kalman filter} & \xrot{extended Kalman filter} &
  \xrot{unscented Kalman filter} & \xrot{LS and regression} & \xrot{genetic
    algorithms} & \xrot{optimization programming} & \xrot{simulated annealing}
  & \xrot{differential evolution} & \xrot{scatter, tabu, cuckoo search} &
  \xrot{particle swarm optimization} & \xrot{other algorithms} &
  \xrot{mach./deep/transf. learning} \\ \hline }

\newcommand{\headerBb}{ \hline \T & & \multicolumn{3}{c}{Measures} &
  \multicolumn{4}{c}{\xcell{Bayesian\\ methods}} &
  \multicolumn{3}{c}{\xcell{Monte\\ Carlo}} &
  \multicolumn{3}{c}{\xcell{Kalman\\ filter}} &
  \multicolumn{9}{c}{Model fitting} & {XLR} \\
  & & \xrot{confidence/credible intervals} &\xrot{Akaike/Fisher/mutual info.} &
  \xrot{entropy} & \xrot{sum of squared errors} & \xrot{MAP, ML, likelihood} &
  \xrot{approximate Bayesian comput.} & \xrot{expectation-maximization} &
  \xrot{variational Bayesian inference} & \xrot{MCMC} & \xrot{Metropol./import.
    sampling} & \xrot{sequential MC, particle filters} & \xrot{Kalman filter} &
  \xrot{extended Kalman filter} & \xrot{unscented Kalman filter} & \xrot{LS and
    regression} & \xrot{genetic algorithms} & \xrot{optimization programming} &
  \xrot{simulated annealing} & \xrot{differential evolution} & \xrot{searches}
  & \xrot{particle swarm optimiz.} & \xrot{other algorithms} &
  \xrot{mach./deep/transf. learning} \\ \hline }

\title{Comprehensive review of models and methods for inferences in
  bio-chemical reaction networks}

\author[\firstAuthorLast ]{\Authors}\address{}\correspondance{} 

\def\Emails{\{p.loskot, komlan.atitey\}@swan.ac.uk,
  l.s.mihaylova@sheffield.ac.uk}

\maketitle\vspace*{-1\baselineskip}

\begin{abstract}
  Key processes in biological and chemical systems are described by networks of
  chemical reactions. From molecular biology to biotechnology applications,
  computational models of reaction networks are used extensively to elucidate
  their non-linear dynamics. Model dynamics are crucially dependent on
  parameter values which are often estimated from observations. Over past
  decade, the interest in parameter and state estimation in models of (bio-)
  chemical reaction networks (BRNs) grew considerably. Statistical inference
  problems are also encountered in many other tasks including model
  calibration, discrimination, identifiability and checking as well as optimum
  experiment design, sensitivity analysis, bifurcation analysis and other.

  The aim of this review paper is to explore developments of past decade to
  understand what BRN models are commonly used in literature, and for what
  inference tasks and inference methods. Initial collection of about 700
  publications excluding books in computational biology and chemistry were
  screened to select over 260 research papers and 20 graduate theses concerning
  estimation problems in BRNs. The paper selection was performed as text mining
  using scripts to automate search for relevant keywords and terms. The outcome
  are tables revealing the level of interest in different inference tasks and
  methods for given models in literature as well as recent trends. In addition,
  a brief survey of general estimation strategies is provided to facilitate
  understanding of estimation methods which are used for BRNs.

  Our findings indicate that many combinations of models, tasks and methods are
  still relatively sparse representing new research opportunities to explore
  those that have not been considered - perhaps for a good reason. The most
  common models of BRNs in literature involve differential equations, Markov
  processes, mass action kinetics and state space representations whereas the
  most common tasks in cited papers are parameter inference and model
  identification. The most common methods are Bayesian analysis, Monte Carlo
  sampling strategies, and model fitting to data using evolutionary algorithms.
  The paper concludes by discussing future research directions including
  research problems which cannot be directly deduced from presented tables.

  \keyFont{ \section{Keywords:} automation, Bayesian analysis, biochemical
    reaction network, estimation, inference, modeling, survey, text mining}
\end{abstract}\vspace*{-0.5\baselineskip}

\toc{sc:intro}{Introduction}
\toc{sc:meth}{Methodology}
\toc{sc:estim}{Review of general estimation strategies}
\toc{sc:models}{Review of modeling strategies for BRNs}
\toc{sc:methods}{Review of parameter estimation strategies for BRNs}
\toc{sc:select}{Choices of models and methods for parameter estimation in
  BRNs}
\toc{sc:concl}{Conclusions}
\toc{sc:ref}{References}
\toc{sc:suppl}{Supplementary tables}

\section{Introduction}\label{sc:intro}

Biological systems are presently subject to extensive research efforts to
ultimately control underlying biological processes. The challenge is the level
of complexity of these systems with intricate dependencies on internal and
external conditions. Biological systems are inherently non-linear, dynamic as
well as stochastic. Their response to input perturbations is often difficult to
predict as they may respond differently to the same inputs. Furthermore,
biological phenomena must be considered at different spatio-temporal scales
from single molecules to gene-scale reaction networks.

Many biological systems can be conveniently represented either as biological
circuits \citep{sillero2011}, or as networks of biochemical reactions
\citep{ashyraliyev2009}. Common examples of biological systems described as
BRNs are: metabolic networks, signal transduction networks, gene regulatory
networks (GRNs), and more generally, networks of biochemical pathways.
Moreover, BRNs share similar characteristics with evolutionary and
prey-predatory networks in population biology, and disease spreading networks
in epidemiology. There are also synthetic bio-reactors and other types of
chemical reactors used in industrial production \citep{ali2015}.

Qualitative as well as quantitative observations of biological systems are
necessary to elucidate their functional and structural properties. Despite the
advent of high throughput experiments, biological phenomena are often only
partially observed. It means that internal system state cannot be fully or
directly observed, but it must be inferred from measurements. Such inferences
are possible due to dependency of observations on internal states and parameter
values \citep{frohlich2017}. Single molecule techniques are promising as they
enable more focused observations, however, their resolution and dimensionality
is still limiting.

Observations are often distorted and noisy. For instance, observations may be
time-averaged values. If measurements introduce distortion, we can assume
extended models \citep{ruttor2010}. Measurement noise may not be additive nor
Gaussian, and its variance may be dependent on values of parameters considered.
Parameter values may differ for in vitro and in vivo experiments
\citep{famili2005}. In systems comprising chemical reactions, the parameters of
interest are initial and instantaneous concentrations, reaction rates and
possibly other kinetic constants such as diffusion coefficients and drift
parameters. The number of chemical species is usually much smaller than the
number of chemical reactions.

In some cases, it may be necessary to estimate the number of reactions between
consecutive measurements \citep{reinker2006}. Structural identifiability of
chemical reaction systems can identify which reactions are occurring. Molecular
concentrations are usually measured directly, and they are functions of
reaction rates and other parameters which are inferred from measurements
\citep{frohlich2017}. The measurements and subsequent inferences of parameters
and states can be performed sequentially (online) or in batches (off-line)
\citep{arnold2014}.

Observations as longitudinal data are usually obtained at discrete time
instances which may not be equidistant. Observations can be used to create or
validate mathematical models. The rate of measurements is important
\citep{fearnhead2014}, since more frequent observations may be costly, and
affect observed biological processes. Processing large volumes of data is also
more computationally demanding. Observations and their processing are sometimes
combined to create so-called observers in order to replace high-cost sensors in
chemical reactors \citep{rapaport2005}. Such observers can yield interval
measurements of quantities with variable observation gain while allowing to
process discretized and delayed measurements \citep{vargas2014}. Average state
observers of large scale systems are considered in \citep{sadamoto2017}.
Observers can be classified as explanatory or predictive to describe existing
or future data \citep{ali2015}.

Biological phenomena can be studied by elucidating properties of their
mathematical models. Biological research problem dictates what physical and
chemical processes must be included in the model. It is usually more efficient
to only collect observations which are necessary to formulate and test some
biological hypothesis than to perform extensive time consuming and expensive
laboratory experiments. Such strategy is referred to as forward modeling
\citep{reinker2006}. On the other hand, finding parameter values to reproduce
observations is known as reverse modeling. Reverse modeling can be enhanced by
experiment design \citep{hagen2018}. Differences between forward and reverse
modeling are explained in \citep{ashyraliyev2009}.

The importance of modeling in biology is discussed in \citep{chevalier2011},
and general modeling strategies are described in \citep{banga2008}. Models of
biological systems are dependent on in vivo or in vitro experiments considered.
BRNs can be modeled as deterministic input-output non-linear transformations
which can, however, be locally linearized at given time scales and resolution.
Models can be modified using some transformation to facilitate their analysis.
There are also stochastic event-driven and probabilistic models of BRNs. For
large number of species, stochastic model converges to deterministic
description \citep{rempala2012}. However, models need to be unbiased in order
to avoid systematic errors. The same model may be used multiple times to
represent biological population \citep{woodcock2011}. Models can be
hierarchical or nested, and have parts interconnected with feedback loops
\citep{fernandez2013}. Since models must be usually evaluated many times, they
need to be computationally fast, and at the right level of coarse grain
description. For instance, microscopic stochastic models may be computationally
expensive whereas deterministic macroscopic description such as
population-average modeling may not be sufficiently accurate or have low
resolution.

Moreover, models can be multidimensional and have 100's or even 1000's of
parameters and states with multiple constraints and unknown initial conditions.
The development of large scale kinetic systems is one of the key tasks in
current computational biology \citep{penas2017}. Parameters estimation for
large scale reaction networks is considered in \citep{remli2017}.

Model analysis is performed to find transient responses of dynamic systems, to
obtain their behavior at steady-state or in equilibrium \citep{atitey2018a},
and to explore complex multi-dimensional parameter spaces. For many biological
models, viable parameter values form only a small fraction of overall parameter
space \citep{atitey2019}, so identifying this sub-volume by ordinary sampling
would be rather inefficient \citep{sillero2011}. Other challenges include size
of state space, unknown parameters, analytical intractability and numerical
problems. Evaluation of observation errors can both facilitate as well as
validate the analysis \citep{bouraoui2015}.

Most analytical and numerical methods can be used universally for different
model structures. However, efficiency of analysis can be considered in
statistical or computational sense. In statistical sense, analysis needs to be
robust against uncertainty in model structure and parameter values under noisy
and limited observations. Computational efficiency can be achieved by
developing algorithms which are prone to massively parallel implementation.

In this paper, we are concerned with parameter inference in biological and
chemical systems described by various BRN models. In literature, parameter
inference is also referred to as inverse problem \citep{eng2009}, point
estimation, model calibration and model identification. More recently, machine
learning methods became popular as an alternative strategy to learn not only
model parameters, but also model features from labeled or unlabeled
observations \citep{sun2012, schnoerr2017}. The key objective of parameter
inference is to define, and then minimize the estimation error while
suppressing the effect of measurement errors \citep{sadamoto2017}.

Parameter inference is affected by many factors. In particular, different
models have different degree of structural identifiability. The model
parameters cannot be identified or only partially identified, provided that
different parameter values or different inputs generate the same dynamic
response such as distributions of synthesized molecules. In some cases,
structural identifiability can be overcome by changing modeling strategy
\citep{yenkie2016}. Structural identifiability is a necessary but not
sufficient condition for overall identifiability \citep{gabor2017}. The
relationship between identifiability and observability is discussed in
\citep{baker2011}. Practical identifiability (also known as posterior
identifiability) is defined to assess whether there are enough data to overcome
measurement noise. It may be beneficial to test identifiability of parameters
which are of interest prior to their inference. For instance, parameters may
not be identifiable at given time scales, or data do not have sufficient
dimensionality (variability) or volume. A lack of suitable data makes the
inference problem to be ill conditioned. A crucial issue is then how well the
parameters need to be known in order to answer given biological question. In
all cases, it is important to validate the obtained estimates.

Sensitivity analysis can complement as well as support parameter estimation
\citep{saltelli2004, frohlich2016}. In particular, parameters can be ranked in
the order of their importance, from the most easy to the most difficult to
estimate. Parameters can be screened using a small amount of observations to
select those which are identifiable prior to their inference from a full data
set. Other tasks in sensitivity analysis include prioritizing and fixing
parameters, testing their independence, and identifying important regions of
their values. A survey of sensitivity analysis methods suitable for BRNs is
provided in \citep{saltelli2005}. Sensitivity profiles of 180 biological models
were compared and analyzed in \citep{erguler2011}.

In the rest of this paper, our main objective is to survey models and methods
which are used for parameter inferences in BRNs. After explaining our
methodology in \scref{sc:meth}, general estimation strategies are explained in
\scref{sc:estim}. Different modeling strategies for BRNs are outlined in
\scref{sc:models}. It is followed by a survey of parameter estimation methods
and related computational tasks in \scref{sc:methods}. Since the performance
and effectiveness of parameter estimation methods is crucially dependent on
specific models adopted, in \scref{sc:select}, we explore what methods are used
in literature for given models, and also, what parameter estimation methods are
used in given parameter estimation tasks. It enables us to point out future
research directions in \sucref{sc:select:1} including statistical inference
techniques which are used in other fields and which can likely be assumed for
BRNs. The paper is concluded in \scref{sc:concl}.

Our main contributions are the survey of general parameter estimation
strategies and additional 3 surveys with assessed levels of interest assumed in
the extensive list of references cited at the end of this paper. In particular,
the surveys constituting the contributions are:
\begin{enumerate}
\item estimation strategies for general systems;
\item models and modeling strategies for BRNs;
\item parameter estimation methods, strategies and related tasks for BRNs;
\item combinations of models and parameter estimation methods and tasks for
  BRNs.
\end{enumerate}

\section{Methodology}\label{sc:meth}

In order to explore how parameter estimation methods and tasks are used in
research literature with different models of BRNs, a large number of
representative or otherwise relevant papers had to be collected. The paper
collection resumed by keyword searches using the Google search portal. The
searches often led to Google Scholar which also provides information about
subsequent citing papers. These citing papers were added to the collection
provided that they are directly concerned about models or methods of parameter
inference in BRNs. We have also considered a number of graduate research theses
which can be publicly accessed online. However, books as well as textbooks were
mostly excluded due to difficulty in obtaining them in electronic form.
Overall, almost 700 electronic documents in portable document format (PDF)
format were collected during the first phase.

\begin{figure}[!t]
  \begin{center}
    \includegraphics[scale=1.0]{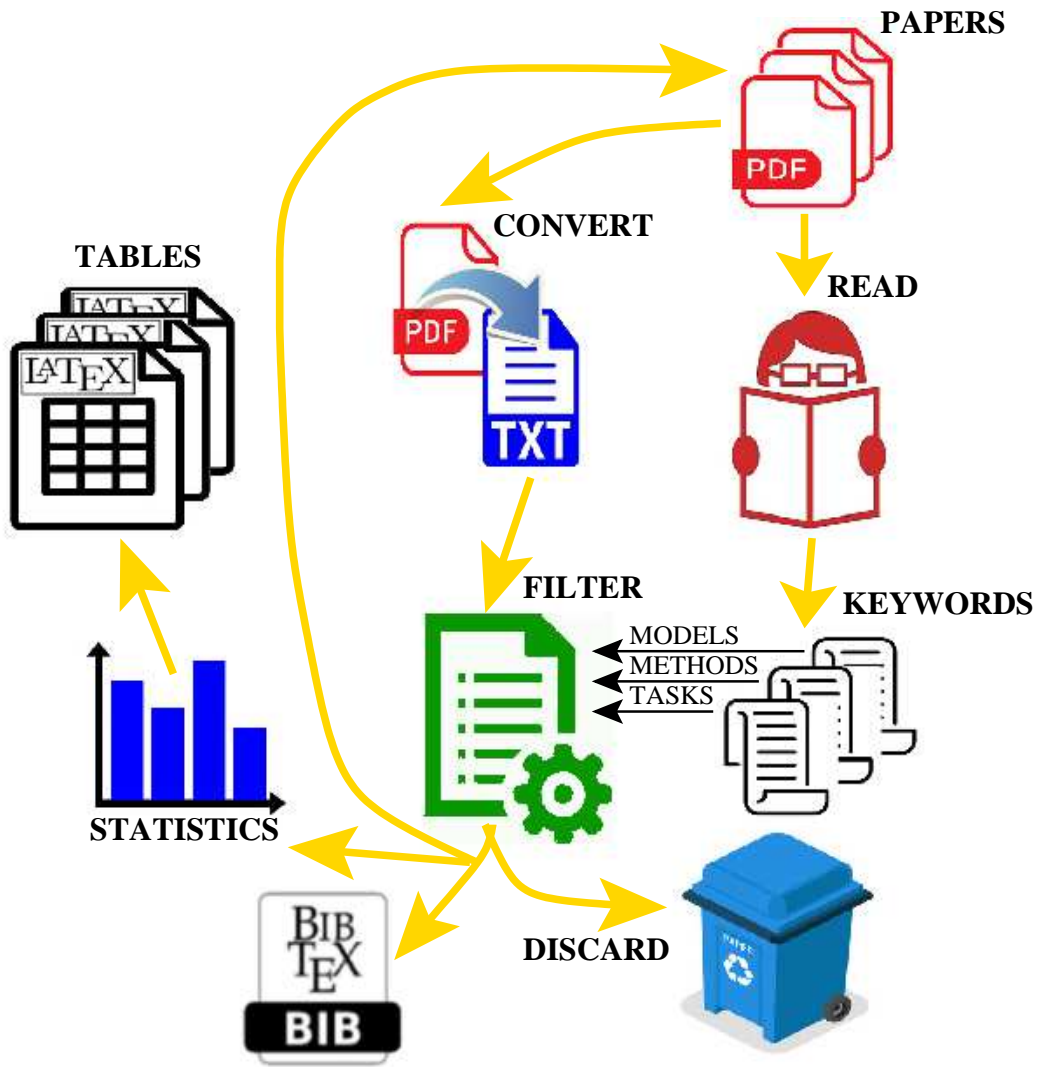}
  \end{center}
  \caption{A workflow of processing PDF files to automate production of
    \bibtex\ reference file, and tables with statistics in \LaTeX.}
  \label{fg:1}
\end{figure}

In the second phase, the collection of 700 documents was reduced to less than
300 by defining selection rules and using text mining. For the needs of this
review, text mining consists mostly of searchers for keywords using regular
expressions. The list of search patterns was built gradually as more documents
were explored. One list of keywords was created for BRN models, and another
list was created for estimation methods and tasks. The initial lists assumed
the keywords used in paper titles, before expanding the list with additional
keywords identified in abstracts, and elsewhere in the papers. We could then
count the number of occurrences of all keywords in the two lists in every
document considered. We observed that the larger the number of occurrences of
the keyword in a document, the larger the probability that this document is
concerned with given model, task or method. We can set the minimum number of
occurrences to be at least 5 in order to deem the document to be highly
relevant. If the number of occurrences is less than 5, it may indicate that the
keyword appeared mainly within the titles of cited references. In the end, we
obtained the collection of nearly 300 documents which are most relevant to our
review. The high-level view of processing workflow of PDF documents in our
study is shown in \fref{fg:1}.

The document contents are explored using a combination of automated text
processing and manual reading to extract desired information. We took advantage
of text processing capabilities readily available on Linux operating systems to
automate many tasks of text processing and analysis. In particular, all PDF
files were first converted to ordinary text files with ascii encoding of
characters (UTF-8) and with transliterated special characters in foreign
alphabets. The conversion was performed using standard \sf{pdftotext} utility
version 0.62 which is based on open source Poppler library for rendering PDF
files. The conversion is not and does not have to be 100\% accurate. For
example, the words with characters which are not recognized can be simply
omitted. Some words may be accidentally split into multiple parts during the
conversion from PDF to a text file. However, such undesirable cases can be
largely neglected for our purposes. It is also useful to remove the end-of-line
characters from within paragraphs while merging paragraphs which were split
across pages in order to enable searches for more complex text patterns.

The scripts to automate many text processing tasks were written in the BASH
interpreter version 4.4 running in a Linux terminal. The scripts extensively
employ standard Linux tools including \sf{grep}, \sf{sed} and \sf{awk}
programmable text filters. The scripts were used to automatically identify and
count relevant papers, generate \sf{LaTeX} tables with results, facilitate
semi-automated creation of bibliographic entries in the master \sf{BibTeX}
file, to obtain URL links for citing papers on Google Scholar in supplementary
Table S14 as well as to identify authors with publications concerning parameter
estimation in BRNs which are listed in Table S15. The keyword search within PDF
documents can assume multiple terms combined in sophisticated hierarchical
expressions with \sf{AND-OR} operators including conditions on the number of
occurrences, and sorting the matched documents as required.

Our procedure for identifying and selecting the most relevant papers has some
limitations. In particular, books are generally rather comprehensive sources of
information, but they are not included in our study. Unlike papers, books
should not be processed on per file basis, but on per chapter basis, especially
if the book is edited. This requires to identify page ranges for each chapter
to enable their extraction into separate files. On the other hand, research
theses have been included and evaluated in our study, although they were
considered separately from research papers. Moreover, paper selection and text
mining in our study is restricted to keyword searches using regular expressions
followed by manual reading of papers. A fully automated paper analysis with
minimum human intervention would require the use of methods of natural language
processing. Such capabilities are already available in some programming
languages, but this is outside the scope of this paper.

\section{Review of general estimation strategies}\label{sc:estim}

The objective of this section is to review general strategies of parameter
estimation, and highlight assumptions limiting their use \citep{kay93,kay98}.
We strive to minimize use of mathematical expressions, but assume the following
mathematical notations: $\E{\cdot}$ denotes expectation, $\bm{a}$ is a column
vector, $\bm{A}$ is a vector or matrix, $(\cdot)^T$ is vector or matrix
transpose, $\bm{a}_i$ is the $i$-th component of a vector, and $|\cdot|$ is the
Euclidean norm of a vector or matrix.

The measurements are noisy, and they are obtained at discrete time
instances. Important assumption for choosing estimation strategy is whether the
dependency of measurements on parameter values is known or only partially
known. This corresponds to knowing which chemical reactions are occurring.
Furthermore, concentrations, or equivalently, the number of molecules of each
chemical species in a constant spatial volume represent the system state.
Unobserved concentrations are then considered to be hidden states.

The basic estimation problem can be defined as follows. Let $\vK$ denote a
vector of reaction rates and other kinetic parameters to be estimated from
observed concentrations (or, equivalently, molecule counts) $\vC$. The
parameters are unknown, and assumed not to vary over time. The dependency of
concentrations at time $t$ on concentrations at time $(t-1)$ and constant
parameters is expressed as,
\begin{equation}\label{eq:10}
  \vC_t = f(\vC_{t-1},\vCt_{t-1},\vK,\vKt,\vw_t) = f(\vC_{t-1},\vCt_{t-1},
  \vK,\vKt) +\vw_t.
\end{equation}
Concentrations $\vC$ are measured whereas unobserved concentrations are denoted
as $\vCt$ in \eref{eq:10}. Note that model \eref{eq:10} of BRN assumes
dependency only between two successive concentration measurements. Such Markov
chain assumption is normally satisfied for all BRNs. Our aim is to estimate
kinetic parameters $\vK$ whereas parameters $\vKt$ in \eref{eq:10} are either
known or not of interest, despite affecting measured concentrations $\vC$. The
second equality in \eref{eq:10} indicates that measurement noise $\vw$ is
additive and independent from value of concentrations and kinetic parameters.
Such independence assumption is somewhat strong, and it needs to be carefully
examined for a given experimental procedure considered. Measurement noises are
assumed to have zero mean, and to be uncorrelated in time (a white noise
assumption), i.e.,
\begin{equation}\label{eq:20}
  \E{\vw_t}=\Zs,\quad\mbox{and}\quad \E{\vw_t\vw_l^T}=
  \begin{cases} \vQ_t & t=l \\ \Zs & t\neq l. \end{cases}
\end{equation}

It is tempting to assume that measurement noise is (approximately) Gaussian
distributed in order to facilitate parameter estimation. However,
concentrations $\vC$ and kinetic parameters $\vK$ are all non-negative, so
unconstrained Gaussian noise $\vw$ may produce negative values of $\vC$ in
\eref{eq:10}. Consequently, measurement noises are dependent on concentration
values which makes noises also correlated over time
\citep{karimi2013,lillacci2010,zimmer2010}. Such noise statistics may restrict
and significantly complicate parameter estimation. It is therefore important to
consider whether a given estimation strategy may be used when measurement
noises are neither additive, nor white and nor stationary.

At time $t$, parameter values are estimated from all measurements collected so
far, i.e.,
\begin{equation}\label{eq:30}
  \vKh_t = g(\vC_t,\vC_{t-1},\ldots,\vC_1).
\end{equation}
Note that, unlike Markovian assumption adopted in system model \eref{eq:10},
all measurements may affect estimated values. Online estimation updates
estimates regularly with every new measurement whereas offline estimation
processes measurements in batches. If measurement noises are relatively small,
we can simply solve the set of equations \eref{eq:10} (e.g., numerically) for
unknown values $\vK$. This may, however, amplify noise and render the estimates
unreliable, especially when measurement noise cannot be neglected. In such
case, more sophisticated strategies are required to suppress measurement noise.

In general, all estimators are designed to minimize some penalty function
$q(\vKh,\vK)$. The choice of penalty function depends on specific estimation
problem considered, and whether the parameters to be estimated have continuous
or discrete values. If the number of molecules of chemical species is large, we
can use continuous approximation, and assume their concentrations to be
non-negative real values. Such approximation may introduce significant modeling
error when the number of molecules becomes small. Another simplifying
approximation is to assume that concentrations are continuous functions of
time, even though reactions are occurring at discrete time instances.

There are several criteria for selecting a suitable estimator. First, unknown
parameters $\vK$ may be considered random, provided that their prior
probability density function (PDF) is known. Estimators exploiting knowledge of
the prior parameter distribution are known as Bayesian estimators. If such
knowledge is not available, we may assume a uniform distribution over some
range of sensible values. This allows to assume Bayesian estimation approaches.
Since kinetic parameters in BRNs are determined by physical laws, their values
are the same under the same experimental conditions. In this case, it is common
to assume that kinetic parameters as deterministic but unknown constants which,
however, excludes Bayesian estimators from consideration.

Second, system model \eref{eq:10} may be only partially known. For instance,
there may be uncertainty which reactions should be included in BRN. Adding more
reactions to BRN may make model \eref{eq:10} to be mathematically intractable,
or non-identifiable. Estimators which cannot be expressed using closed form
mathematical expression are still solvable numerically. Third, parameters may
be time-varying, so they can be treated as random or non-random processes. For
instance, measured concentrations in \eref{eq:10} are random processes whereas
reaction rates are random or non-random constants.

Main strategies for parameter estimation are summarized in \tref{tb:1}. If the
prior distribution $p(\vK)$ of parameters is known, we can assume minimum mean
square error (MMSE) or maximum a posterior (MAP) estimators. Assuming that
parameters are deterministic constants, we can use minimum variance unbiased
(MVUB) or maximum likelihood (ML) estimators. Particularly ML estimator is
attractive and frequently used, since it always exists, and it is
asymptotically unbiased, unlike MVUB estimator which may not exist or is
difficult to obtain. Unbiased estimator of unknown deterministic parameter with
the smallest possible variance given by the Cramer-Rao bound is said to be
efficient. Estimator is said to be unbiased, provided that, at any time $t$,
$\E{\vKh_t}=\vK$, and it is consistent, provided that the estimation error,
$e_t=\vKh_t-\vK$, becomes negligible with enough measurements, i.e.,
$\lim_{t\to\infty}\vKh_t=\vK$.

In many scenarios, it is difficult or not possible to find distribution of
measured concentrations. Provided that we find approximation
$\vC\approx g(\vK)$ (a meta-model) of BRN considered, we can assume different
types of least square (LS) estimators also known as generalized linear
regression (GLR). The estimated parameter values are those that best fit the
measurements. High-dimensional parameter fitting to measurements is
conveniently carried out numerically. The aim is to improve the fitness of
meta-model (i.e., minimize loss function). Various types of evolutionary
algorithms are considered for these type of problems such as genetic algorithm,
simulated annealing, ant colony optimization and particle swam optimization.
The main issue of these numerical methods is how to determine initial
estimates, the speed of convergence, how easy they can be implemented, and if
they are derivative free. Moreover, various strategies are used to enable
search for a globally optimum solution. Furthermore, if measurements are
stationary, for instance, once BRN reaches a steady-state, MM estimator can be
used to find parameter values matching selected moments of measurements. These
moments are determined empirically as,
$\E{\vC^n}\approx \frac{1}{N}\sum_{i=1}^N \vC^n_i$.

\begin{instab}{ccccl}{tb:1}{Main strategies for parameter estimation.}
  \hline \textbf{Strategy} & \textbf{Assumptions} & \textbf{Objective} &
  \textbf{Estimator} & \textbf{Notes} \\ \hline MMSE & $p(\vK)$, $p(\vC|\vK)$ &
  $\E{|\vKh-\vK|^2}$ & $\EE{\vK}{\vK|\vC}$
  & \xcell[l]{unbiased, estimation error\\ orthogonal to $\vKh$ and $\vC$}\\
  MAP & $p(\vK)$, $p(\vC|\vK)$ & $\Prob{|\vKh-\vK|>\Delta}$ &
  $\argmax_{\vK} p(\vK|\vC)$ & solve $\frac{\df}{\df\vK}p(\vC|\vK)p(\vK)=0$\\
  MVUB & \xcell[c]{$p(\vC|\vK)$\\$p(\vK)$ unknown}
  &  $\var{\vKh}$ & \xcell[l]{closed solution if\\
    $p(\vC)$ exponential\\ in
    $\vK$, otherwise\\ may not exist} & \xcell[l]{Cramer-Rao bound: \\
    $\var{\vKh}\geq J^{-1}(\vK)$ \\ (Fisher
    information)} \\
  ML & \xcell[c]{$p(\vC|\vK)$\\$p(\vK)$ unknown} & \xcell[c]{maximize
    \\likelihood} & $\argmax_{\vK}p(\vC|\vK)$ & \xcell[l]{always exist,
    asymptotically\\ unbiased \& efficient, may\\ not be minimum
    variance\medskip} \\
  LS & \xcell[c]{meta-model\\ $\vC\approx g(\vK)$} & $\var{\vC}$ &
  $\argmin_{\vK} | \vC-g(\vK) |^2$ & \xcell[c]{best fit to data, $g(\cdot)$
    usually\\ linear combination with \\transformation to make\\ measure.
    errors Gaussian} \\
  MM & \xcell[c]{measurements\\stationary,\\ invertible $g_n(\cdot)$} &
  $g_n(\vK)=\E{\vC^n}$ & $\vKh = g^{-1}_n\left( \frac{1}{N}\sum\limits_{i=1}^N
    \vC_i^n\right)$ & asymptotically unbiased\\
  \hline \multicolumn{5}{l}{\renewcommand{\arraystretch}{1}\footnotesize
  \begin{tabular}{lll}
  MMSE: minimum mean square error &
  MVUB: minimum variance unbiased &
  LS: least squares \\
  MAP: maximum a posterior probability &
  ML: maximum likelihood &
  MM: method of moments \\
  \end{tabular}}
\end{instab}

There are other estimators which are not included in \tref{tb:1}. In
particular, linear MMSE, BLUE (best linear unbiased) and Kalman filter (KF) are
linear filters. Since both BLUE and KF also require linear model of
measurements, none of these estimators is suitable for inference in non-linear
BRN. KF further assumes that measurement and process noises are Gaussian,
although they can be non-stationary and non-white. In order to exploit fast
tracking properties of KF and extend its use to non-linear state estimation
problems, several linearization strategies were proposed in literature.
Extended KF (EKF) uses first-order linearization about predicted value of
parameter. Linearization and the need to also estimate covariance matrix
significantly increases numerical complexity. Accuracy of EKF depends on
accuracy of linearization, and it is not guaranteed to be unbiased and may even
diverge. Augmented KF (AKF) uses second order linearization, but it is not as
popular as KF. Unscented KF (UKF) represents distribution of parameter to be
estimated by a group of random samples followed by unscented transformation to
make them Gaussian. It improves linearization by providing better estimates of
mean and covariance, and there is no need to obtain or calculate derivatives
(i.e., Jacobian), so this estimator is more robust than EKF.

Markov chain Monte Carlo (MCMC) is another type of sampling estimator. It uses
random or semi-random walk sampling of posterior distribution of estimated
parameter. MCMC sampling requires a transition before converging to the desired
sampling distribution at equilibrium. More general sampling strategies of
posterior distribution are known as sequential MC (SMC) estimators or particle
filter (PF). The main advantage of these methods is that they are very
universal and require no assumptions about system model or its parameters. They
use genetic sampling with mutations, sample selection, and resampling. However,
in general, all sampling based estimators such as UKF, MCMC, SMC and PF are
negatively affected by non-smooth non-linearities and systems involving large
number of dimensions. All these estimators iterates between parameter
prediction and updating steps, and they are often used to estimate hidden
(i.e., unobserved) states in dynamical systems.

Alternative method for iteratively calculating posterior distribution (i.e.,
MAP estimate) or likelihood (i.e., ML estimate) of parameter is
expectation-maximization (EM). This method is suitable for jointly estimating
parameters and unobserved states in hidden Markov models and mixed distribution
models. Different implementations of EM method may involve naive Bayes
strategy, and Baum-Welch or inside-outside algorithm. The expectation step to
predict parameter value can be obtained by any estimator such as KF. The
maximization step updates the predicted parameter value.

Finally, if enough labeled data are available, we can use supervised or
semi-supervised methods of machine learning. However, the use of these methods
have been explored only briefly in our review paper.

\section{Review of modeling strategies for BRNs}\label{sc:models}

Mathematical models describe dependencies of observations on model parameters.
A general procedure for constructing mathematical models of biological systems
is described in \citep{chun2009}. Bio-reactors are mathematically described in
\citep{bouraoui2016, vargas2014, ali2015}. Model building is an iterative
process which is often combined with optimum experiment design
\citep{fernandez2005}. Model structure affects selection as well as performance
of parameter estimators. Structural identifiability and validity of multiple
models together with parameter sensitivity was considered in
\citep{jaqaman2006}. Parameter estimation can be assumed together with
discriminating among several competing models, for instance, when the model
structure is only partially known. Model structure and its parameter values to
achieve desired dynamics can be derived using statistical inference
\citep{barnes2018}. Synthesizing parameter values for BRNs is also considered
in \citep{ceska2017}. Probabilistic model checking can be used to facilitate
robustness analysis of stochastic biochemical models \citep{ceska2014}.
Iterative, feedback dependent modularization of models with parameters
identification was devised in \citep{lang2016}. Selection among hierarchical
models assuming Akaike information was studied in \citep{fernandez2013}.

Different modeling strategies have been considered for BRNs. Two main physical
laws considered for modeling BRNs are the rate law and the mass action law.
These laws relate reactant concentrations to reactions rates in chemical
equilibrium. Mechanistic models are generally derived from physical laws of
system components. Most commonly assumed random processes in models of BRNs are
several basic variants of Markov process, and also birth-death process.
Majority of mathematical models describing BRNs involve some form of ODEs, PDEs
and SDEs. These models are often mathematically intractable, and must be
analyzed numerically. Some of these models are derived from CME or one of its
several approximations. CME approximations can be deterministic or stochastic,
and their accuracy depends on BRN structure as well as parameter values.

Chemical reactants in dynamic equilibrium are governed by the law of mass
action whereas kinetic properties of BRNs are described by the rate laws
\citep{schnoerr2017}. The reaction kinetics can be considered at steady-state
or in transition to steady-state. There are also other kinetic models such as
Michaelis-Menten kinetics for enzyme-substrate reactions
\citep{rumschinski2010}, Hill kinetics for cooperative ligand binding to
macromolecules \citep{fey2010}, kinetics for logistic growth models in GRNs
\citep{ghusinga2017}, kinetics for birth-death processes \citep{daigle2012},
and stochastic Lotka-Volterra kinetics associated with prey-predatory networks
\citep{boys2008}.

Single molecule stochastic models describe BRN qualitatively by generating
probabilistic trajectories of species counts. BRNs can be modeled as a sequence
of reactions occurring at random time instances \citep{amrein2012}. Such
stochastic kinetics mathematically correspond to a Markov jump process with
random state transitions between the species counts \citep{andreychenko2012}.
Alternatively, sequence of chemical reactions can be viewed as a hidden Markov
process \citep{reinker2006}. Markov jump process can be exactly simulated using
the classical Gillespie algorithm, so that competing reactions are selected
assuming a Poisson process with the intensity proportional to the species
counts \citep{golightly2012, kugler2012}. Random occurrences of reactions can
be also described using a hazard function \citep{boys2008}. Non-homogeneous
Poisson processes can be simulated by the thinning algorithm of Lewis and
Shedler \citep{sherlock2014}.

The number of species and their molecule counts can be large, so state space of
continuous time Markov chain (CTMC) model can be huge \citep{angius2011}. Large
state space can be truncated by considering only states significantly
contributing to the parameter likelihood \citep{singh2005}. Parameter
likelihoods can be updated assuming increments and decrements of the species
counts \citep{lecca2009}. Probabilistic state space representation of BRNs as
dynamic systems was considered in \citep{andreychenko2011, gupta2014,
  mcgoff2015, schnoerr2017}. Augmented state space representation of BRN from
ordinary differential equations (ODEs) is obtained in \citep{baker2013}.

More generally, mechanistic models are obtained by assuming that biological
systems are built up from actual or perceived components which are governed by
physical laws \citep{frohlich2017, hasenauer2013th, pullen2014, white2016}. It
is a different strategy to empirical models which are reverse engineered from
observations \citep{bronstein2015, dattner2015, geffen2008}. Black-box modeling
can be used with some limitations when there is little knowledge about the
underlying biological processes \citep{chun2009}.

\subsection{Modeling BRNs by differential equations}\label{sc:metho:1}

Time evolution of states with probabilistic transitions is described by
chemical master equation (CME) \citep{andreychenko2011, weber2017}. It is a set
of coupled first-order ODEs or partial differential equations (PDEs)
\citep{teijeiro2017, penas2017, fearnhead2014} representing a continuous time
approximation and describing BRN quantitatively. ODE model of BRN can be also
derived as a low-order moment approximation of CME \citep{bogomolov2015}. For
model with stochastic differential equation (SDE), it is often difficult to
find transition probabilities \citep{fearnhead2014, sherlock2014, karimi2013}.
The PDE approximation can be obtained assuming Taylor expansion of CME
\citep{schnoerr2017}. The error bounds for numerically computed stationary
distributions of CME are obtained in \citep{kuntz2017}. CME for hierarchical
BRNs consisting of dependent and independent sub-networks is solved
analytically in \citep{reis2018}. Path integral form of ODEs has been
considered in \citep{liu2014, weber2017}. Models with memory described by delay
differential equations (DDEs) are investigated in \citep{zhan2014}.
Mixed-effect models assume multiple instances of SDE based models to evaluate
statistical variations between and within these models \citep{whitaker2017}.

Comprehensive tutorial on ODE modeling of biological systems is provided in
\citep{gratie2013}. ODE models can be solved numerically via discretization.
For instance, the method of finite differences (FDM) can be used to obtain
difference equations \citep{frohlich2016}. However, algorithms for numerically
solving deterministic ODE models or simulating models with SDEs may not be
easily parallelizable, or they have problems with numerical stability. ODE
models are said to be stiff, if they are difficult to solve or simulate, for
example, if they contain multiple multiple processes at largely different time
scales \citep{sun2012, kulikov2017, cazzaniga2015}. Alternatively, BRN
structure can be derived from its ODE representation \citep{fages2015}. Similar
strategy is assumed in \citep{plesa2017} where BRN is inferred from
deterministic ODE representation of time series data.

A survey of methods for solving CME of gene expression circuits is provided in
\citep{veerman2018}. These methods involve propagators, time scale separation,
and generating functions \citep{schnoerr2017}. For instance, time scale
separation can be used to robustly decompose CME into a hierarchy of models
\citep{radulescu2012}. Reduced stochastic description of BRN exploiting time
scale separation is studied in \citep{thomas2012}.

If deterministic ODEs cannot be solved analytically, one can use Langevin and
Fokker-Planck equations as stochastic diffusion approximations of CME
\citep{schnoerr2017, hasenauer2013th}. Fokker-Planck equation can be solved to
obtain deterministic time evolution of system state distribution
\citep{kugler2012, liao2015, schnoerr2017}. Deterministic and stochastic
diffusion approximations of stochastic kinetics are reviewed in
\citep{mozgunov2018}. Chemical Langevin equation (CLE) is a SDE consisting of
deterministic part describing slow macroscopic changes, and stochastic part
representing fast microscopic changes \citep{golightly2012, dey2018, cseke2016}
which are dependent on the size of deterministic part. In the limit, as
deterministic part increases, random fluctuations can be neglected, and
deterministic kinetics of the Langevin equation becomes reaction the rate
equation (RRE) \citep{bronstein2015, frohlich2016, loos2016}.

\subsection{Modeling BRNs by approximations}\label{sc:method:2}

A popular strategy to obtain computationally efficient models is to assume
approximations, for example, using meta-heuristics and meta-modeling
\citep{sun2012,cedersund2016}. Quasi-steady state (QSS) and quasi-equilibrium
(QE) approximations of BRNs are assumed in \citep{radulescu2012}. Modifications
of QSS model are investigated in \citep{wong2015}. It is also common to
approximate system dynamics by continuous ODEs or SDEs \citep{fearnhead2014}.
Thus, when the number of molecules is small, the SDE model is preferred, since
deterministic ODE model may be inaccurate \citep{gillespie2014}. It is
generally difficult to quantify approximation error for diffusion approximation
models. Forward-reverse stochastic diffusion with deterministic approximation
of propensities by observed data was considered in \citep{bayer2015}.

Mass action kinetics can be used to obtain deterministic approximation of CME.
Corresponding deterministic ODEs can accurately describe system dynamics,
provided that molecule counts of all species are sufficiently large
\citep{sherlock2014, yenkie2016}. Other CME approximations assume finite state
projections, system size expansion, and moment closure methods
\citep{chevalier2011, schnoerr2017}. These methods are popular, since they are
easy to implement, efficient computationally, do not require complete
statistical description, and also achieve good accuracy if species appear in
large copy numbers \citep{schnoerr2017}. Moment closure methods leading to
coupled ODEs can approach CME solution with a low computational complexity
\citep{frohlich2016, bogomolov2015, schilling2016}. Specifically, the $n$-th
moment of population size depends on its $(n+1)$ moment. In order to close the
model, the $(n+1)$-th moment is approximated by a function of lower moments
\citep{ruess2011, ghusinga2017}. Only the first several moments can be used to
approximate deterministic solution of CME \citep{schnoerr2017}. Limitations of
moment closure method are analyzed in \citep{bronstein2017}. Multivariate
moment closure method is developed in \citep{lakatos2015} to describe nonlinear
dynamics of stochastic kinetics. General moment expansion method for stochastic
kinetics is derived in \citep{ale2013}. Approximation of state probabilities by
their statistical moments can be used to perform efficient simulations of
stochastic kinetics \citep{andreychenko2014}.

The leading term of CME approximation in system size expansion (SSE) method
corresponds to linear noise approximation (LNA). It is the first order Taylor
expansion of deterministic CME with a stochastic component where transition
probabilities are additive Gaussian noises. Other terms of the Taylor expansion
can be included in order to improve modeling accuracy \citep{frohlich2016}. In
\citep{sherlock2014}, LNA is used to approximate fast reactions as continuous
time Markov process (CTMP) whereas slow reactions are represented as Markov
jump process with time-varying hazards. There are other variants of LNA such as
a restarting LNA model \citep{fearnhead2014}, LNA with time integrated
observations \citep{folia2018}, and LNA with time scale separation
\citep{thomas2012}. LNA for reaction-diffusion master equation (RDME) is
computed in \citep{lotstedt2018}. The impact of parameter values on stochastic
fluctuations for LNA of BRN is investigated in \citep{pahle2012}.

S-system model is a set of decoupled non-linear ODEs in the form of product of
power-law functions \citep{chou2006, liu2012, meskin2011, iwata2014}. Such
model is justified by multivariate linearization in logarithmic coordinates. It
provides good trade-off between flexibility and accuracy, and offers other
properties particularly suitable for modeling complex non-linear systems.
S-system modeling with additional constraints is discussed in \citep{sun2012}.
S-system model representing biological pathways is investigated in
\citep{mansouri2015}. S-system model assuming weighted sum of kinetic orders is
obtained in \citep{liu2008}. Bayesian inference for S-system models is
investigated in \citep{mansouri2014}.

Polynomial models of biological systems are investigated in \citep{fey2010,
  dattner2015, kuepfer2007a, vrettas2011}. Rational models as fractions of
polynomial functions are examined in \citep{villaverde2016, fey2010,
  eisenberg2014}. Methods for validating polynomial and rational models of BRNs
are studied in \citep{rumschinski2010}. Eigenvalues are used in
\citep{hori2013} to obtain a low order linear approximation of time series
data. More generally, differential-algebraic equations (DAEs) are considered in
\citep{ashyraliyev2009, deng2014, fernandez2013, michalik2009}. These models
have different characteristics than ODE models, and are often more difficult to
solve. Review of autoregressive models for parameter inference including
stability and causality issues is given in \citep{michailidis2013}.

\subsection{Other models of BRNs}\label{sc:method:3}

There are many other types of BRN models considered in literature. Birth-death
process is a special case of CTMP having only two states \citep{paul2014th,
  daigle2012, zechner2014th}. It is closely related to telegraph process
\citep{veerman2018}. Computationally efficient tensor representation of BRNs to
facilitate parameter estimation and sensitivity analysis is devised in
\citep{liao2015}. Other computational models for qualitative description of
interactions and behavioral logic in BRNs involve Petri nets
\citep{mazur2012th, sun2012, schnoerr2017}, probabilistic Boolean networks
\citep{mizera2014, liu2012, mazur2012th}, continuous time recurrent neural
networks \citep{berrones2016}, and agent based model (ABM) \citep{hussain2015}.
Hardware description language (HDL) which is normally used to describe logic of
electronic circuits is adopted for the case of spatially-dependent biological
systems described by PDEs in \citep{rosati2018}. Multi-parameter space was
mapped to 1D manifold in \citep{zimmer2010}.

Many models containing multiple unknown parameters are poorly constrained. Even
though such models may be still fully identifiable, they are ill-conditioned,
and often referred to as being sloppy \citep{erguler2011, toni2009, white2016}.
Parameter estimation and experimental design for sloppy models are evaluated in
\citep{mannakee2016} where it is shown that dynamic properties of sloppy models
usually depend only on several parameters with the remaining parameters being
largely unimportant. A sequence of hierarchical models of increasing complexity
was proposed in \citep{white2016} to overcome complexity and sloppiness of
conventional models.

Main modeling strategies discussed in this section are summarized in
\tref{tb:2}. They are categorized as physical laws, random processes,
mathematical models, interaction models and CME based models. Models in four of
these categories are mostly quantitative whereas interaction models are
qualitative. However, \tref{tb:2} does not consider model properties such as
sloppiness, and model structure which may be hierarchical, modular or
sequential. Hybrid models which are excluded from \tref{tb:2} combine different
modeling strategies in order to mitigate various drawbacks \citep{babtie2017,
  mikeev2012, sherlock2014}. For example, a hybrid model can assume
deterministic description of large species populations with stochastic
variations of small populations \citep{mikeev2012}. Hybrid model consisting of
parametric and non-parametric sub-models can offer some advantages over
mechanistic models \citep{stosch2014}.

\begin{instab}[1.5]{ll}{tb:2}{Overview of main modeling strategies for BRNs.} 
  \hline
  \textbf{\TT\BB Strategy} & \textbf{Motivation and key papers} \\
  \hline \TT \textbf{Physical laws} & \xcell{reaction rates in dynamic
    equilibrium are functions of \\reactant concentrations} \\
  \iitem kinetic rate laws & \xcell{\scite{eng2009, chun2009, villaverde2012};
    \\ \scite{baker2011, voit2013, joshi2006}} \\
  \iitem mass action kinetics & \xcell{\scite{wong2015, smith2018};\\
    \scite{linder2014, angius2011}} \\
  \iitem mechanistic models & \xcell{\scite{white2016, stosch2014};
    \\ \scite{chun2009, pullen2014}} \\
  \TT \textbf{Random processes} & \xcell{probabilistic behavioral description
    of chemical reactions} \\
  \iitem Markov process & \xcell{\scite{goutsias2012, andrieu2010};\\
    \scite{septier2016, weber2017}} \\
  \iitem Poisson process & \xcell{\scite{weber2017, reis2018};\\
    \scite{bronstein2017, daigle2012}} \\
  \iitem birth-death process & \xcell{\scite{daigle2012, weber2017};\\
    \scite{wang2010, mikelson2016}} \\
  \iitem telegraph process & {\scite{veerman2018, weber2017}}\\
  \TT\TT\textbf{Mathematical models} & adopted models for dynamic systems \\
  \iitem quasi-state models & \xcell{\scite{wong2015, srivastava2012th,
      schnoerr2017};\\ \scite{liao2017th, thomas2012, radulescu2012}} \\
  \iitem state space representation & \xcell{\scite{andrieu2010, weber2017};\\
    \scite{brim2013, andreychenko2011}} \\
  \iitem ODEs, PDEs, SDEs, DDEs & \xcell{\scite{fages2015,
      weber2017};\\\scite{ramsay2007, jia2011};\\
    \scite{liu2014, teijeiro2017}} \\
  \iitem path integral form of ODEs & {\scite{weber2017}} \\
  \iitem rational model & \xcell{\scite{hussain2015, vanlier2013};
    \\\scite{villaverde2016, sun2012}} \\
  \iitem differential algebraic eqns. & \xcell{\scite{michalik2009, deng2014};
    \\\scite{ashyraliyev2009, ramsay2007}} \\
  \iitem tensor representation & {\scite{liao2015, wong2015, smith2018}} \\
  \iitem S-system model & \xcell{\scite{ voit2013, chun2009, liu2012};\\
    \scite{meskin2011, kutalik2007}} \\
  \iitem polynomial model & \xcell{\scite{vrettas2011, ceska2017};
    \\\scite{weber2017, kuntz2017}} \\
  \iitem \BB manifold map & \xcell{\scite{mannakee2016, radulescu2012};
    \\\scite{white2016, septier2016}} \\
  \hline
\end{instab}

\begin{instabcont}[1.5]{ll}{tb:2c}{Overview of main modeling strategies for
    BRNs. (cont.)} \hline
  \textbf{\TT\BB Strategy} & \textbf{Motivation and key papers} \\
  \hline \TT \textbf{Interaction models} & qualitative modeling of chemical
  interactions \\
  \iitem Petri nets & {\scite{voit2013, chun2009, liu2012}} \\
  \iitem Boolean networks & {\scite{streib2012, chun2009}} \\
  \iitem neural networks & \xcell{\scite{stosch2014, camacho2018};
    \\\scite{ali2015, goutsias2012}} \\
  \iitem agent based models & \xcell{\scite{carmi2013, hussain2015};\\
    \scite{jagiella2017, goutsias2012}} \\
  \TT \textbf{CME based models} & stochastic and deterministic approximations
  of CME \\
  \iitem Langevin equation & \xcell{\scite{thomas2012, septier2016};
    \\ \scite{schnoerr2017, goutsias2012};\\ \scite{smith2018, weber2017}} \\
  \iitem Fokker-Planck equation & {\scite{weber2017, schnoerr2017, liao2015}}
  \\
  \iitem reaction rate equation & \xcell{\scite{liu2014,
      loos2016};\\\scite{linder2014, koeppl2012}} \\
  \iitem moment closure & \xcell{\scite{schnoerr2017, bronstein2017};\\
    \scite{schilling2016, ruess2011};\\\scite{andreychenko2014, lakatos2015}}
  \\
  \iitem linear noise approximation & \xcell{\scite{thomas2012,
      fearnhead2014};\\ \scite{golightly2015, golightly2012}; \\
    \scite{whitaker2017, schnoerr2017}} \\
  \iitem system size expansion & {\scite{schnoerr2017, frohlich2016}}\\
  \hline
\end{instabcont}

\begin{figure}[!t]
  \begin{center}
    \includegraphics[scale=0.8]{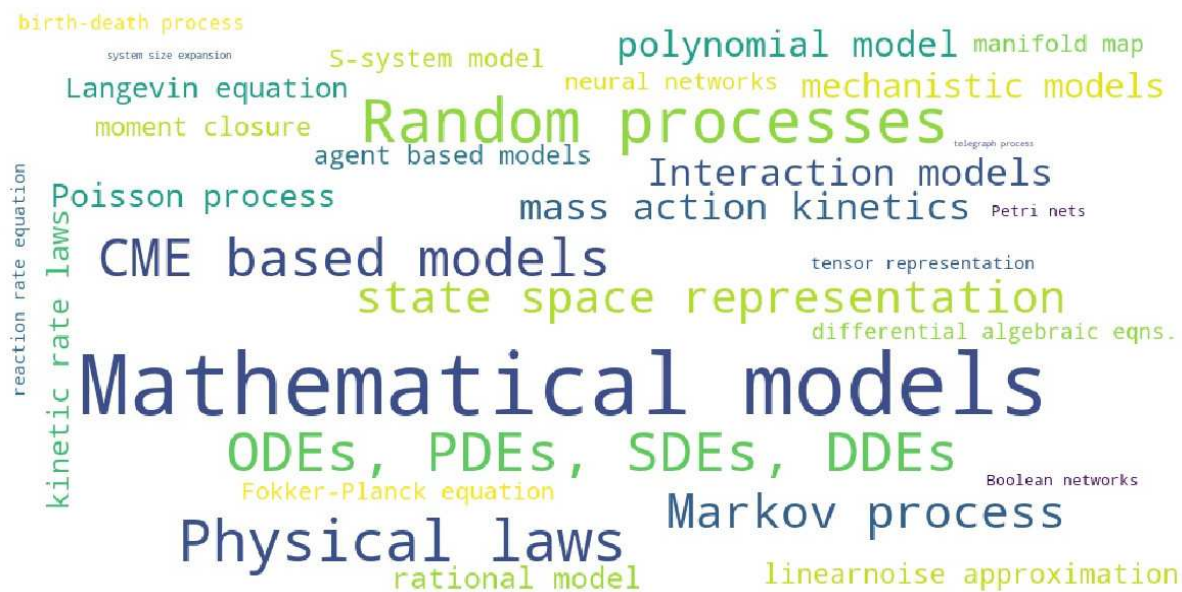}
  \end{center}
  \caption{A word cloud visualizing the level of interest in different models
    of BRNs.}
  \label{fg:2}
\end{figure}

In order to assess the level of interest of different BRN models in literature,
supplementary Table S12 presents the occurrences of 25 main modeling strategies
for all references cited in this paper. The summary of Table S12 is reproduced
in \tref{tb:6}, and further visualized as a word cloud in \fref{fg:2}. We
observe that differential equations are the most commonly assumed models of
BRNs in literature. About half of the cited papers consider Markov chain models
or their variants, since these models naturally and accurately represent time
sequence of randomly occurring reactions in BRN. State space representation is
assumed in over one third of the cited papers. Other more common models of BRNs
include mass action kinetics, mechanistic models, and models involving
polynomial functions.

\begin{table}[!t]
\centering\footnotesize
\caption{Coverage of modeling strategies for BRNs.}
\label{tb:6}
\setlength{\tabcolsep}{2pt} \renewcommand{\arraystretch}{1.1}
\begin{tabular}{l*{25}{c}}
\headerA{}
  {\footnotesize\bf \# papers} &59 &104 &82 &166 &72 &22 &2 &150 &216 &58 &27
  &19 &39 &89 &35 &13 &13 &36 &43 &55 &35 &19 &45 &50 &4 \\\hline
\end{tabular}
\end{table}

Another viewpoint is to consider publication years of papers concerning
different modeling strategies. \tref{tb:10} shows the number of papers for
given modeling strategy in given year starting from 2005. We observe that the
interest in some modeling strategies remain stable over a decade, for example,
for models with state space representation and models involving differential
equations. The number of cited papers is the largest for years 2013 and 2014.
The paper counts in \tref{tb:10} are indicative that the interest in
computational modeling of BRNs has been steadily increasing over the past
decade.

\begin{table}[!t]
\centering\footnotesize
\caption{Number of papers concerning models of BRNs in given years.}
\label{tb:10}
\setlength{\tabcolsep}{2pt} \renewcommand{\arraystretch}{1.1}
\begin{tabular}{p{1cm}*{26}{c}}
\headerA{Year}
2005 &3 &3 &1 &2 &2 &. &. &4 &5 &. &2 &. &. &. &. &. &. &. &. &1 &2 &. &. &. &.  \\
2006 &2 &4 &2 &2 &. &. &. &2 &3 &4 &1 &. &1 &3 &. &. &. &2 &. &. &. &. &. &. &.  \\
2007 &. &4 &1 &3 &2 &. &. &2 &6 &1 &1 &. &2 &4 &2 &. &. &1 &1 &1 &. &. &. &. &.  \\
2008 &1 &4 &2 &2 &1 &. &. &6 &6 &1 &1 &. &2 &2 &1 &. &. &2 &. &. &. &1 &. &. &.  \\
2009 &4 &7 &2 &5 &1 &. &. &6 &11 &1 &2 &1 &4 &2 &2 &2 &2 &2 &3 &1 &1 &. &. &1 &.  \\
2010 &7 &11 &5 &12 &3 &1 &. &8 &13 &6 &2 &. &4 &5 &5 &1 &1 &2 &5 &7 &2 &1 &. &2 &.  \\
2011 &5 &4 &5 &11 &4 &. &. &10 &13 &2 &. &. &3 &4 &1 &. &. &2 &2 &4 &2 &2 &2 &2 &.  \\
2012 &6 &11 &6 &21 &11 &3 &. &14 &19 &5 &4 &1 &5 &6 &3 &2 &3 &6 &6 &9 &6 &1 &4 &9 &.  \\
2013 &7 &9 &12 &16 &8 &3 &. &17 &26 &9 &3 &1 &7 &12 &4 &2 &. &3 &7 &6 &2 &2 &4 &4 &.  \\
2014 &8 &13 &14 &33 &11 &4 &. &26 &33 &7 &4 &2 &5 &14 &2 &1 &3 &7 &6 &7 &5 &5 &10 &9 &.  \\
2015 &6 &10 &8 &15 &5 &2 &. &20 &24 &5 &1 &2 &3 &10 &4 &2 &1 &1 &2 &4 &4 &. &6 &5 &.  \\
2016 &4 &8 &13 &19 &5 &2 &. &14 &23 &4 &1 &2 &3 &10 &5 &1 &2 &4 &7 &6 &3 &3 &7 &8 &2  \\
2017 &4 &8 &8 &13 &10 &6 &1 &12 &18 &7 &4 &7 &. &10 &6 &1 &1 &3 &3 &5 &5 &2 &8 &5 &2  \\
2018 &2 &8 &3 &11 &8 &1 &1 &7 &13 &5 &. &3 &. &5 &. &1 &. &1 &1 &4 &3 &2 &4 &5 &.  \\  \hline
\end{tabular}
\end{table}

\section{Review of parameter estimation strategies for BRNs}\label{sc:methods}

Parameter estimation appears in many computational problems including model
identification \citep{banga2008}, model calibration \citep{zechner2011}, model
discrimination \citep{kuepfer2007a}, model identifiability \citep{geffen2008},
model checking \citep{cseke2016}, sensitivity analysis \citep{erguler2011},
optimum experiment design \citep{ruess2015}, bifurcation analysis
\citep{eng2009}, reachability analysis \citep{tenazinha2011}, causality
analysis \citep{carmi2013}, stability analysis \citep{dochain2003}, network
inference \citep{smet2010}, and control of BRN \citep{venayak2018}. A survey of
parameter estimation methods for chemical reaction systems can be found, for
example, in \citep{gupta2013th, baker2015, chun2009, mcgoff2015}. Other review
papers on parameter estimation in BRNs and other dynamic systems are listed in
\tref{tb:3}.

\begin{instab}[1.5]{ll}{tb:3}{Review papers on parameter estimation in
    BRNs and other dynamic systems.}
  \hline
  \textbf{\TT\BB Reference} & \textbf{Focus} \\ \hline
  \citep{banga2008} & \xcell{model calibration using global optimization
    methods\\ supported by maximum information experiment design} \\
  \citep{chun2009} & \xcell{very comprehensive survey of available optimization
    methods\\ for parameter estimation and model-free and model-based
    \\structure identification from data} \\
  \citep{ashyraliyev2009} & \xcell{a priori and a posteriori model
    identifiability and survey of\\ parameter space search strategies} \\
  \citep{smet2010} & \xcell{methods for underdetermined inferences of BRNs from
    data} \\
  \citep{tenazinha2011} & \xcell{integrated models of BRNs reflecting
    availability of omics\\ data assuming chemical organization theory,
    flux-balance\\ analysis, logical discrete modeling, Petri nets, kinetic
    models, \\ stochastic models, and hybrid models} \\
  \citep{goutsias2012} & \xcell{comprehensive review of analytical methods for
    evaluating\\ dynamics of Markov reaction networks} \\
  \citep{streib2012} & \xcell{systematic and conceptual overview of methods for
    inferring\\ gene regulatory networks from gene expression data; survey\\
  of strategies to compare performance of inference methods} \\
  \citep{sun2012} & \xcell{survey of metaheuristic methods applied to
    reliability and\\ identifiability of biochemical model parameters
    including\\ optimum experiment design} \\
  \citep{kuwahara2013} & \xcell{scalable framework for parameter estimation in
    genetic\\ circuits assuming mean time evolution of gene products} \\
  \citep{voit2013} & \xcell{review of biological system models and methods
    for their\\ analysis as well as design} \\
  \citep{baker2015} & \xcell{general framework to deal with non-identifiable
    parameters\\ in BRNs using constrained parameter estimation} \\
  \citep{mcgoff2015} & \xcell{mathematical survey of statistical methods for
    parameter\\ inference in general non-linear dynamical systems} \\
  \citep{weiss2016} & survey of transfer learning methods \\
  \citep{schnoerr2017} & \xcell{a comprehensive survey of deterministic and
    stochastic models\\ of BRNs followed by introduction to Bayesian
    parameter\\ inference from data} \\
  \citep{camacho2018} & \xcell{application of machine learning techniques to
    computational\\ problems in biological networks} \\
  \citep{smith2018} & \xcell{review of spatial stochastic kinetics including
    reaction-diffusion\\ master equation and models involving Brownian
    dynamics} \\ 
  \hline
\end{instab}

A survey of tasks in modeling and system identification is provided in
\citep{chun2009}. Model identifiability determines which parameter values can
be estimated from observations \citep{villaverde2016}. It is known as
structural identifiability, and it is inspired by the concept of system
observability. Structural identifiability is normally evaluated prior to
estimating parameters. There is also practical identifiability which accounts
for quality and quantity of observations, i.e., whether it is possible to
obtain good parameter estimates from noisy and limited data. Theory and tools
for model identifiability and closely related concepts such as sensitivity to
parameter perturbations, observability, distinguishability and optimum
experiment design are reviewed in \citep{villaverde2016a}. Models which are not
identifiable can be modified or simplified to make them identifiable
\citep{baker2015, villaverde2016, villaverde2016a}. Model identifiability
formulated as observability was considered in \citep{geffen2008} to replace
traditional analytical approaches which often require model simplifications
with more deterministic empirical methods. Changes in structural and practical
identifiability of models with availability of new knowledge and data is
studied in \citep{babtie2017}. Global observability and detectability of
reaction systems was studied in \citep{moreno2005}. Parameter identifiability
of power law models is investigated in \citep{srinath2010} and of linear
dynamic models in \citep{li2013}. Parameters can be mutually dependent
\citep{fey2008}. Parameter dependencies measured by correlations and other
higher order moments are exploited to determine structural and practical
identifiability in \citep{li2015}. Intrinsic noise in species counts can be
exploited to overcome structural non-identifiability within a deterministic
framework as shown in \citep{zimmer2010}. Chemical reaction optimization (CRO)
is used to maximize production of a bio-reactor in \citep{abdullah2013a}.

Many different parameter estimation strategies have been devised in literature
for BRNs and dynamic systems. All parameter estimation problems lead to
minimization or maximization of some fitness function. Deriving optimum value
analytically is rarely possible whereas numerical search for the optimum in
high-dimensional parameter spaces can be ill-conditioned when objective or
fitness function is multi-modal. If there is large flat surface about the
minimum, the obtained solution cannot be trusted \citep{fernandez2006,
  srinivas2007}. Moreover, the optimum values can change over order of
magnitude under different implicit or explicit constraints which is often the
case for biological systems. Numerical algorithms for non-convex optimization
problems need to be stable and provide convergence guarantees. Other aspects
include scalability, computational efficiency, numerical stability and
robustness, and all methods need to be also statistically validated. All search
strategies experience trade-off between efficiency and robustness.

Measurements can be produced from heterogeneous sources (omics data), and from
heterogeneous populations \citep{zechner2011}. In deterministic models,
parameter estimation is often carried out by fitting model to data. Parameter
uncertainty analysis can be used to assess how well the model explain
experimental data \citep{vanlier2013}. Stochastic models require more
sophisticated strategies to perform parameter estimation \citep{zimmer2012}.
Multiple-shooting method for stochastic systems is used in \citep{zimmer2016}
to calculate the Fisher information matrix. In literature, deterministic
methods appear to be assumed much more often than stochastic methods
\citep{daigle2012}. Since the mean approximation of SDEs may differ from the
solution obtained for deterministic ODEs, parameter estimation assuming
stochastic rather than deterministic models is preferable when some species
counts are relatively small \citep{andreychenko2012}.

Parameter estimation in transient and steady states are quite different
\citep{ko2009}. At steady-state, small perturbations are sufficient to observe
system responses whereas at transient state, experiment design for model
identification is more complicated. In particular, quick transient response
after external perturbation limits information content of measurements
\citep{zechner2012}. Sensitivity analysis can be used to improve computational
efficiency of parameter estimation \citep{frohlich2017}. The parameter space
boundaries can be estimated by sampling \citep{fey2010}. Confidence and
credible intervals can be obtained also for stiff and sloppy models assuming
inferability, sensitivity and sloppiness \citep{erguler2011}. Furthermore,
observers design may be different for systems with and without inputs
\citep{singh2005}.

Scalability of parameter estimation can be resolved by decoupling rate
equations and by assuming mean-time evolution of species counts
\citep{kuwahara2013}. However, exploring large parameter spaces can be
complicated if the estimation problem is ill-conditioned and multi-modal
\citep{liu2009}. State-dependent Markov jump processes are difficult to
estimate at large scale, especially when these processes are faster than the
rate of observations \citep{fearnhead2014}.

Parameter estimation can be facilitated by grouping parameters and identifying
which are uncorrelated \citep{gabor2017}. Parameter estimation in groups can
provide robustness against noisy and incomplete data \citep{jia2011}. Only
parameters which are consistent with measured data can be selected and jointly
estimated \citep{hasenauer2010}. Parameter clustering can also improve model
tractability and identifiability, since changes in some parameters could be
compensated by changes in other parameters \citep{nienaltowski2015}. Grouping
of parameters to elucidate dynamics of genetic circuit is assumed in
\citep{atitey2019}. Parameters can be assumed hierarchically to gradually
estimate their values starting from a minimum set \citep{shacham2014}. A hybrid
hierarchical parameter estimation prone to parallel implementation is devised
in \citep{sosonkina2004}.

Incremental parameter estimation usually requires data smoothing which can
create estimation bias \citep{liu2014}. Such bias can be mitigated by
estimating independent parameters before dependent ones. Parameter inference
can be paired with hypothesis testing or model selection \citep{fernandez2013}.
Joint model and parameter identification with incremental one-at-a-time
parameter estimation and model building is performed in \citep{gennemark2007}.
Unobserved states, latent variables and parameters in BRNs can be estimated
jointly by sequential processing of measurements \citep{zimmer2012,
  arnold2014}, by using sliding window observers \citep{liu2006}, and by other
numerical methods \citep{kamaukhov2007}. Estimation of kinetic rates in BRNs is
transformed into a problem of state estimation in \citep{fey2010}. Parameter
estimation and state reconstruction are linked via extended models in
\citep{busetto2009}. Unobservable sub-spaces can be excluded to only consider
model parts which can be identified reliably \citep{singh2005}. Unknown
parameters which are not of interest can be margninalized
\citep{bronstein2015}. Another strategy is to reconstruct states prior to
parameter estimation \citep{fey2008}.

Information theoretic metrics can be used to infer BRN structure
\citep{villaverde2014}, and to perform identifiability analysis of parameters
\citep{nienaltowski2015}. Akaike information can be used to assess quality of
statistical model given observations, so the best model is selected
\citep{gosalbez2013, pullen2014}. However, in order to avoid overfitting and
constrain model complexity, there is a penalty being simply the number of model
parameters to estimate. Model overfitting leads to poor generalization
capability. Overfitting can be resolved by model reduction techniques
\citep{sadamoto2017,srivastava2012th}. For instance, only essential chemical
reactions can be considered \citep{sillero2011}. Simplified modeling with the
reduced number of parameters and parameter subset selection is used in
\citep{eghtesadi2014} to avoid overfitting noisy data. On the other hand,
under-determined models may yield several or infinitely many solutions of
fitting data in which case they are not identifiable. In such cases, data
fitting can be performed subject to additional constraints. There can also be
cases where multiple models all fit measured data well. However, a model with
the best fit to data may not necessarily provide a satisfactory biological
explanation \citep{slezak2010}.

Simultaneous estimation of parameters and structure of BRN as a mixed binary
dynamic optimization problem with Akaike information is formulated in
\citep{gosalbez2013} to trade-off estimation accuracy and evaluation
complexity. Fisher information is given by the mean amount of information
gained from observed data. It is often used when estimating non-random
parameters, for instance, using maximum likelihood (ML) \citep{fernandez2005,
  kyriakopoulos2014}. It can be also exploited to perform sensitivity,
robustness and identifiability of parameters. It is especially useful when
measurements and parameters are correlated \citep{komorowski2011}. Fisher
information can be used to improve parameter estimation \citep{transtrum2012},
to design optimum experiments \citep{kyriakopoulos2014, zimmer2016}, and to
select a subset of identifiable parameters \citep{eisenberg2014}. Mutual
information can be used as a similarity measure which statistically outperform
correlation measure in canonical correlation analysis (CCA)
\citep{nienaltowski2015}. Other uses of mutual information are outlined in
\citep{mazur2012th}, and for parameter estimation in \citep{streib2012}.

Cross-entropy methods can be used with stochastic simulations
\citep{revell2018}, and to improve computational efficiency of parameter
estimation \citep{daigle2012}. Maximum entropy sampling (MES) methods for
experiment design and parameter estimation are discussed in \citep{mazur2013}.
Maximum entropy principle to reconstruct probability distributions is described
in \citep{schnoerr2017}. Relative entropy rate is assumed in
\citep{pantazis2013} to perform sensitivity analysis of BRNs. Kantorovich
distance between two probability measures is used in \citep{koeppl2010} to
estimate model parameters of BRNs.

Sum of squared errors (SSE) is often assumed to obtain regression estimators
\citep{chou2006}, and to assess goodness of fit and quality of estimators
\citep{iwata2014, kimura2015, nim2013}. The SSE acronym should not be confused
with system size expansion (SSE) which is modeling strategy discussed
previously \citep{frohlich2016, schnoerr2017}.

In general, many algorithms for parameter estimation and other related problems
have been considered in literature. These algorithms are often modifications of
several fundamental estimation strategies, and they are adopted for specific
models and availability and quality of measurements. Since graduate research
theses usually contain more or less comprehensive and up to date surveys of
relevant literature, the theses concerned with parameter estimation in BRNs are
summarized in \tref{tb:4}.

\begin{instab}[1.5]{ll}{tb:4}{Selected research theses concerning parameter
    estimation and related problems in BRNs.}
  \hline
  \textbf{\TT\BB Thesis} & \textbf{Main research problems considered} \\ \hline
  \citep{dargatz2010th} & {Bayesian inference for biochemical models involving
    diffusion} \\
  \citep{mu2010th} & rate and state estimation in S-system and linear fractional
  model (LFM) \\
  \citep{palmisano2010th} & software tools for modeling and parameter estimation
  in BRNs \\
  \citep{mazur2012th} & inference via stochastic sampling and Bayesian learning
  framework \\
  \citep{srivastava2012th} & \xcell{stochastic simulations of BRNs combined with
    likelihood based\\ parameter estimation, confidence intervals, sensitivity 
    analysis} \\
  \citep{gupta2013th} & \xcell{parameter estimation in deterministic and
    stochastic BRNs, inference \\with model reduction, mostly MCMC methods} \\
  \citep{hasenauer2013th} & \xcell{Bayesian estimation and uncertainty analysis
    of population \\ heterogeneity and proliferation dynamics} \\
  \citep{linder2013th} & \xcell{penalized LS algorithm and diffusion and linear
    noise approximations\\ and algebraic statistical models} \\
  \citep{flassig2014th} & model identification for large scale gene regulatory
  networks \\
    \citep{liu2014th} & approximate Bayesian inference methods and sensitivity
  analysis\\
  \citep{moritz2014th} & \xcell{structural identification and parameter
    estimation for modular\\ and layered type of modes} \\
  \citep{paul2014th} & analysis of MCMC based methods \\
  \citep{ruess2014th} & \xcell{optimum estimation and experiment design assuming
    ML and \\Bayesian inference and Fisher information} \\
  \citep{schenkendorf2014th} & \xcell{quantification of parameter uncertainty,
    optimal experiment design for\\ parameter estimation and model selection}\\
  \citep{smadbeck2014th} & \xcell{moment closure methods, model reduction,
    stability and\\ spectral analysis of BRNs} \\
  \citep{schnoerr2016th} & \xcell{Langevin equation, moment closure
    approximations, representations\\ of stochastic RDME} \\
  \citep{zechner2014th} & inference from heterogeneous snapshot and
    time-lapse data \\  
  \citep{galagali2016th} & \xcell{Bayesian and non-Bayesian inference in BRNs,
    adaptive MCMC\\ methods, network-aware inference, inference for
    approximated BRNs} \\
  \citep{hussain2016th} & \xcell{sequential probability ratio test, Bayesian
    model checking, \\ automated and formal verification, parameter
    discovery} \\
  \citep{lakatos2017th} & multivariate moment closure and reachability analysis
  \\
  \citep{liao2017th} & tensor representation and analysis of BRNs \\ \hline
\end{instab}

In the rest of this section, we survey the algorithms for parameter estimation
in models of BRNs or dynamic systems in the following 4 subsections: Bayesian
methods, Monte Carlo methods, other statistical methods including Kalman
filtering, and model fitting methods.

\subsection{Bayesian methods}\label{sc:methods:1}

Fundamental premise of Bayesian estimation methods is that prior probabilities
or distributions of parameters are known. The objective is then to obtain
posterior probabilities of parameters to be estimated. It is often sufficient
to find the maximum value of posterior distribution corresponding to the
maximum a posterior (MAP) estimate. The value of this maximum can be also used
to select among several competing models \citep{andreychenko2012} and to design
optimum experiments \citep{mazur2012th}. Model checking via time-bounded path
properties is represented as a Bayesian inference problem in
\citep{milios2018}. Biological models often assume conjugate priors to perform
Bayesian inference \citep{galagali2016th, mazur2012th, boys2008, murakami2014}.
Bayesian inference for low copy counts can be improved by separating intrinsic
and extrinsic noises \citep{koeppl2012}. Bayesian analysis is facilitated by
separating slow and fast reactions in \citep{sherlock2014}. Bayesian inference
strategies for biological models involving diffusion processes are investigated
in \citep{dargatz2010th}.

In many cases, determining exact posterior distribution using Bayesian
framework may be intractable. Approximate Bayesian computation (ABC) method is
a strategy to estimate posterior distribution, or more specifically, to
estimate likelihood function \citep{tanevski2010}. A survey of ABC methods can
be found in \citep{drovandi2016}. The basic idea is to find parameter values
which generate the same statistics as the observed data. ABC method can be
performed sequentially, and it can be coupled with sensitivity analysis
\citep{liu2014th}. Parameter estimation and model selection using ABC framework
is studied in \citep{liepe2014, murakami2014}. Non-identifiability of
parameters having flat shape posterior followed by ABC inference is studied in
\citep{murakami2014} assuming conjugate priors. Efficient method to generate
summary statistics for ABC is presented in \citep{fearnhead2012}. A piece-wise
ABC sampling to estimate posterior density for Markov models is proposed in
\citep{white2015}. Parallel implementations of ABC and SMC methods are
introduced in \citep{jagiella2017}.

Expectation-maximization (EM) is a popular implementation of MAP estimator
where there are some other unobserved or unknown parameters \citep{bayer2015,
  karimi2014a, daigle2012}. It can be combined with Monte Carlo (MC) sampling,
and such method is known as MC expectation-maximization (MCEM)
\citep{angius2011}. Computationally efficient method for obtaining ML estimates
by MCEM with Modified Cross-Entropy Method (MCEM2) is developed in
\citep{daigle2012}. Approximate EM algorithm is devised in \citep{karimi2013}
which is robust against unknown initial estimates, and which is useful for
online state estimation during process monitoring. Another parameter estimation
strategy with the same structure as EM estimator is known as variational
Bayesian inference \citep{vrettas2011, weber2017}. It is more general than EM
estimation, as it can also yield posterior distribution in addition to
parameter estimates by exploiting analytical approximation of posterior
density. For instance, posterior density is approximated by radial basis
functions (RBFs) in \citep{frohlich2014a} to reduce the number of model
evaluations. Variational approximate inference with continuous time
constraints, and model checking problem are investigated \citep{cseke2016}.

ML estimation is a popular strategy for parameter inference, provided that the
likelihood of observed data can be computed efficiently for given model. Survey
of ML based methods for parameter estimation in BRNs is provided in
\citep{daigle2012}. Likelihood function can be approximated analytically using
Laplace and B-spline approximations \citep{karimi2014}, or numerically
including its derivatives \citep{mikeev2012}. Likelihood function is obtained
by simulations in \citep{tian2007}. Moment closure is used for fast
approximation of parameter likelihood in \citep{milner2013}. Stoachastic
simulations can be avoided by approximating transition distributions by
Gaussian distribution in the parameter likelihood \citep{zimmer2014}. ML of
transition probabilities is assumed in \citep{chen2017} to devise a new
estimation algorithm which can improve variational Bayesian methods with
summary statistics. ML estimation combined with regularization penalizing
complexity is investigated in \citep{jang2016}. ML estimation for BRN model
with concentrations increments and decrements is studied in \citep{lecca2009}.

\subsection{Monte Carlo methods}\label{sc:methods:2}

Motivation of MC methods is to represent probabilities and density functions as
relative frequencies of samples or particles in order to overcome mathematical
intractability of Bayesian inference. However, even sampling methods can be
computationally overwhelming due to frequent model evaluations. Markov chain
Monte Carlo (MCMC) methods are the most often used sampling strategies to
simulate conditional trajectories of system states. MCMC sampling with good
mixing properties requires carefully chosen proposal distribution and good
selection of initial samples in order to avoid sample degeneracy and
instability problems. The most well known sampling MCMC procedure is Metropolis
or Metropolis-Hastings algorithm \citep{golightly2011, sillero2011,
  mazur2012th, galagali2016th}. Another strategy for dealing with
high-dimensional sampling problems is to combine particle filters and MCMC
methods to obtain sequential MCMC (SMCMC) algorithms \citep{septier2016}. An
overview of particle filtering and MCMC methods for spatial objects is
presented in \citep{mihaylova2014}. MCMC methods for causality reasoning are
introduced in \citep{carmi2013}. Design of proposal distributions for MCMC and
SMC methods assuming large number of correlated variables is studied in
\citep{andrieu2010}.

Since the convergence rate of MCMC can be rather slow for heavy tail
distributions, factorization and approximation of posterior can improve MCMC
performance \citep{frohlich2014a}. MCMC methods can be made adaptive to improve
their convergence properties as shown in \citep{hasenauer2013th,
  galagali2016th, mazur2012th, muller2011}. Interpolation of observed data with
MCMC sampling is used in \citep{golightly2005a} to jointly estimate unobserved
states and reaction rates. MCMC sampling can be combined with importance
sampling to reduce computational complexity and simulation times
\citep{golightly2015}. Conditional density importance sampling (CDIS) is
introduced in \citep{gupta2014} as an alternative to MCMC parameter estimation.
MCMC methods for high-dimensional systems are compared in \citep{septier2016}.

Bayesian inference via MC sampling with stochastic gradient descent is studied
in \citep{wang2010}. The likelihood function of parameters is calculated by
combining MC global sampling with locally optimum gradient methods in
\citep{kimura2015}. Nested Bayesian sampling is used in \citep{pullen2014} to
compute marginal likelihoods to compare or rank several competing models. MCMC
sampling for mixed-effects SDE models is considered in \citep{whitaker2017}. In
order to overcome ill-conditioned least squares (LS) data fitting and numerical
instability, bootstrapped MC procedure based on diffusion and LNA was studied
in \citep{linder2014}. Particle filter assumes specific type of random
processes to identify posteriors while bounding computational complexity for
models with large number of parameters is considered in \citep{mikelson2016}.
Population MC (PMC) sampling framework for Bayesian inference in
high-dimensional models was developed in \citep{koblents2011}.

Sequential MC (SMC) method represents posterior distribution in Bayesian
inference by a set of samples referred to as particles
\citep{gordon1993,doucet2001,tanevski2010, yang2014}, so it is also known as
particle filter \citep{gordon1993,doucet2001,lillacci2012, golightly2015}. SMC
methods for joint state and parameter estimation are proposed in
\citep{nemeth2014}. The degeneracy phenomenon in particle filters can be
mitigated by more efficient sampling strategies \citep{golightly2017}.
Parallelization of SMC computations is devised in \citep{mihaylova2012}.
Further modifications of creating and processing particles to improve
computational efficiency is investigated in \citep{golightly2018}.

The computational complexity of particle filter can be reduced by particle MCMC
(pMCMC) method \citep{koblents2014}. The pMCMC method can be combined with
diffusion approximation \citep{golightly2011}, and further refined to improve
its scalability \citep{golightly2017}. A proposal distribution for Bayesian
analysis is obtained by pMCMC sampling in \citep{sherlock2014}. Proposal
samples to calculate marginal likelihoods are obtained for CLE and LNA
approximations in \citep{golightly2015}. Particle filter is validated and shown
to be more robust than LS data fitting by exploiting noise statistics of data
in \citep{lillacci2012}.

\subsection{Other statistical methods}\label{sc:methods:3}

The key assumption of using standard Kalman filter is linearity of
measurements. Kalman filter is used with CME approximations in \citep{dey2018}
while estimating noise covariance and allowing for dependency of noise on
states and parameter values. Kalman filter is used to obtain initial guess of
parameter values for data fitting parameter estimation in \citep{lillacci2010}.
Classical Kalman filter can be merged with particle filtering methods in
stochastic \citep{vrettas2011} and deterministic systems \citep{arnold2014}.
Kalman filter for time integrated observations is assumed in \citep{folia2018}.

Since BRNs are generally highly non-linear, extended and unscented Kalman
filters (EKFs and UKFs) have been developed \citep{baker2011}. EKF was modified
for stiff ODEs in \citep{kulikov2015, kulikov2017}. Joint estimation of
parameters and states by EKF is investigated in \citep{sun2008, ji2009}. EKF is
combined with moment closure method in \citep{ruess2011}, and it is modified
for parameter estimation in S-system models in \citep{meskin2011}. Modification
of EKF to penalize modeling uncertainty due to linearization in order to
improve estimation accuracy is proposed in \citep{xiong2013}. Square-root UKF
achieves good numerical stability, and it can be modified to deal with state
estimation constraints \citep{baker2013, baker2015}. For infrequent sampling or
sparse observations, UKF and cubature Kalman filter outperform EKF
\citep{kulikov2015a, kulikov2017}.

There are other less commonly used inference strategies which have not been
mentioned. In particular, Gaussian smoothing to compensate for missing and
noisy data is used in \citep{sun2012}. Parameter estimation assuming non-linear
ODE model combined with data smoothing was investigated in \citep{ramsay2007}.
Inference of state distributions via optimized histograms and statistical
fitting is performed in \citep{atitey2018}. Formal verification and sequential
probability ratio test for parameters estimation are considered in
\citep{hussain2016th}. The moment closure modeling is combined with stochastic
simulations for parameter estimation in \citep{bogomolov2015}. Classical
bootstrapping with data replication and resampling to enable repeated
estimations is described in \citep{vanlier2013}. Confidence intervals of
parameter estimates can be obtained using bootstrapping \citep{joshi2006,
  srivastava2014}. Bootstrapping can be used to improve efficiency in
recomputing model trajectories \citep{linder2014}. Bootstrap filter can
outperform EKF \citep{gordon1993}. Generalized method of moments with empirical
sample moments is performed in \citep{kugler2012, luck2016} whereas moment
based methods for parameters inference and optimum experiment design are
considered in \citep{ruess2015}. Expectation propagation (EP) for approximate
Bayesian inference is studied in \citep{cseke2016}.

\subsection{Model fitting methods}\label{sc:methods:4}

Parameter estimation by fitting measured data appears to be by far the most
common method used in literature. The main reason is that, unlike other
estimation strategies, it is relatively straightforward to formulate the
underlying optimization problem with minimum knowledge and assumptions. Various
continuous and discrete fitness functions are explored in \citep{deng2014}.
Multiple fitness functions may be also considered. Fitness function can be
derived assuming the likelihood \citep{fernandez2006}. Fitting of the
approximated likelihood function is considered in \citep{srivastava2014}.
Observations are interpolated with spline functions in \citep{nim2013}, so that
derivatives can be used to estimate production and consumption of molecules.
Such strategy decomposes a high-dimensional problem into the product of
low-dimensional factors. Fitness function is interpolated by spline functions
in \citep{zhan2011}.

The challenge is to develop numerically efficient methods to solve
high-dimensional problems with possibly many constraints. Even though
derivative free methods are easier to implement, gradient based methods have
faster, though only local convergence. For instance, gradient based
optimization with sensitivity analysis assuming finite differences is
investigated in \citep{loos2016}. Derivative free methods are necessary for
combinatorial and integer constrained problems \citep{cedersund2016,
  gabor2017}. Data fitting is generally more computationally demanding for
stochastic than for deterministic models, but the former are more likely to
find a global solution \citep{fernandez2005}.

Since many practical problems are non-convex, global optimization methods are
generally preferred. They can be implemented as multi-start local methods, or
by selecting a subset of parameters to be estimated. Sensitivity to initial
values can be reduced by methods tracking multiple solutions. Many of these
methods can be parallelized to overcome computational burden
\citep{mancini2015, teijeiro2017}. Parallel implementation of data fitting
algorithms employing Spark, MapReduce and MPI messaging are considered in
\citep{teijeiro2017}. Computational complexity of global methods can be
mitigated by incremental identification strategies \citep{michalik2009}. Global
methods also require proper selection of search parameters which is usually
achieved by multiple initial exploratory runs \citep{penas2017}. Another
strategy for global search is to assume transformations followed by non-uniform
sampling \citep{kleinstein2006}. Hybrid strategies switch between global and
local searches \citep{fernandez2005, fernandez2006, ashyraliyev2009}.

Majority of data fitting methods are rooted in simple LS estimation or
regression, or non-linear least squares (NLSQ) problem \citep{baker2011}.
Alternating regression (AR) reformulates non-linear fitting as iterative linear
regression \citep{chou2006}. Non-linear regression is converted into non-linear
programming problem which is solved by random drift PSO in \citep{sun2014}.
Asymptotic properties of LS estimation were evaluated in \citep{rempala2012}.
Iterative linear LS for systems described by ratio of linear functions is
considered in \citep{tian2010}. Regularization of optimization problems is a
strategy to deal with ill-conditioned problems due to insufficient or noisy
data for a given number of parameters to be estimated \citep{gabor2014,
  gabor2017}. In particular, regularization introduces additional constraints
to penalize complexity or constraints on parameters values using prior
knowledge which can trade-off estimator bias with its variance while not
over-fitting model \citep{kravaris2013, jang2016, liu2012}. Alternatively,
perturbation method has been developed to for fitting data in
\citep{shiang2009}.

Evolutionary algorithms (EAs) are the most frequently used methods for solving
high-dimensional constrained optimization problems. They require no particular
assumptions, and can be used even for problems of very large dimension. EAs
adopt heuristic strategies to find the optimum assuming a population of
candidate solutions which are iteratively improved by reproduction, mutation,
crossover or recombination, selection and other operations until fitness or
loss function reaches the desired value. Specific EAs commonly used in
literature for identification of BRNs and other dynamic systems are summarized
in \tref{tb:5}.

\begin{instab}[1.5]{ll}{tb:5}{Common evolutionary algorithms for parameter
    estimation in BRNs and dynamic systems.}
  \hline \textbf{\TT\BB Algorithm} & \textbf{Motivation and selected papers}
  \\ \hline
  Genetic algorithms (GAs)& \xcell{largest class of EAs, inspired by evolution
    and natural selection, \\ often near optimum solution} \\
  &  \xcell{\scite{sun2012, chun2009, besozzi2009};\\ \scite{tian2007, liu2012,
      matsubara2006}} \\
  Genetic programming (GP) & \xcell{evolution of computer programs towards
    improving their fitness\\ to solve a given task} \\
  & \scite{nobile2013, chun2009, sun2012} \\
  Evolutionary programming (EP) & \xcell{parameters of computer program evolve
    towards improving its\\ fitness to solve a given task} \\
  & \scite{baker2010, sun2012, revell2018} \\
  Simulated annealing (SA) & \xcell{probabilistic search combining sampling with
    random but\\ controlled acceptance of candidate solutions} \\
  & \xcell{\scite{hussain2015, sun2012, ashyraliyev2009}; \\
    \scite{dai2010, chun2009, cedersund2016}} \\  
  Differential evolution (DE) & \xcell{derivative free method, linearly
    combining randomly selected\\ candidate solutions to obtain iterative
    improvements} \\ & \xcell{\scite{teijeiro2017, liu2009, sun2012}; \\
    \scite{chong2014, srinivas2007, chong2012}} \\ 
  Scatter search (SS) & \xcell{often combined with tabu search, it is local
    search with temporarily\\ accepting worse solutions and avoiding already
    visited regions} \\ 
  & \xcell{\scite{penas2017, villaverde2012, remli2017}; \\
    \scite{fernandez2006, cedersund2016}} \\
  Particle swarm optimiz. (PSO) & \xcell{derivative free method, moving
    particles (i.e., samples or candidate\\ solutions) towards better solution}
  \\ & \xcell{ \scite{nobile2016, sun2014, besozzi2009};\\
    \scite{cazzaniga2015, abdullah2013, tangherloni2016}} \\
  \hline
\end{instab}

Cuckoo search employs random sub-populations which can be discarded to improve
the solution \citep{rakhshani2016}. Optimization programs include non-linear
simplex method \citep{cazzaniga2015}, non-linear programming (NLP)
\citep{moles2003, fernandez2013, sun2012, zhan2011}, semi-definite programming
\citep{kuepfer2007a, rumschinski2010}, and quadratic programming
\citep{gupta2013th}. Nelder-Mead method (also known as downhill simplex method)
maintains a simplex of test points which evolve to find the data fit
\citep{abdullah2013b}. Quantifier elimination (QE) is used to simplify
constrained optimization problems \citep{anai2006}. Other examples of nature
inspired algorithms include firefly algorithm (FA) \citep{abdullah2013a,
  abdullah2013b} and artificial bee colony (ABC) algorithm \citep{chong2014}.
Neural networks are becoming popular especially due to multi-layer deep
learning methods. Other works which are concerned with problems of traditional
neural networks consider training, overfitting, and smoothing as a mean value
approximation \citep{matsubara2006, chun2009, ali2015, berrones2016}. Parallel
implementation of scatter search for large scale systems are devised in
\citep{villaverde2012, penas2017}.

Benefits of individual optimization methods can be exploited by adaptively
combining different algorithms. For instance, DE is combined with tabu search
in \citep{srinath2010}, and another hybrid DE method is considered in
\citep{liu2008a}. Genetic programming and PSO are combined in
\citep{nobile2013} whereas fuzzy logic based PSO is developed in
\citep{nobile2016}. Regularization, pruning and continuous genetic algorithm
(CGA) are combined in \citep{liu2012}.

Machine learning (MLR) methods can be very effective provided that there are
enough training data drawn from some fixed distribution \citep{pan2010}. If
there is not enough labeled data, or the generating distribution changes,
transfer learning (TLR) can exploit data from multiple domains \citep{pan2010,
  azab2018, weiss2016}. A primer on MLR and deep learning (DLR) methods for
biological networks is provided in \citep{camacho2018}.

\begin{figure}[!t]
  \begin{center}
    \includegraphics[scale=0.8]{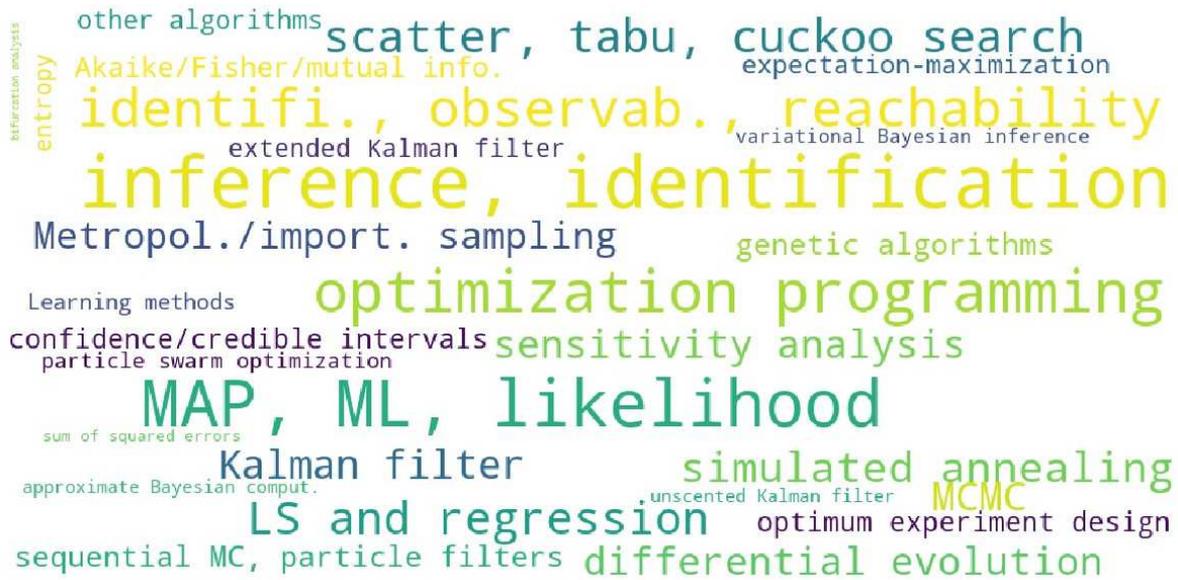}
  \end{center}
  \caption{A word cloud visualizing the level of interest in different
    parameter estimation methods in BRNs.}
  \label{fg:3}
\end{figure}

A survey of 5 estimation tasks and 23 estimation methods for BRNs considered in
references cited in this paper is provided in supplementary Table S13. This
table is summarized in \tref{tb:7} for convenience, and the corresponding word
cloud is shown in \fref{fg:3}. All cited references consider some parameter
estimation or identification task, since this was the primary objective of our
paper. Other common tasks in literature appear to be model identifiability,
parameter observability, and reachability analysis. Information theoretic
measures are used relatively often as alternative to probabilistic measures in
order to define rigorous inference problems. Parameter identification by model
fitting appears to be the most common strategy in literature. Bayesian analysis
which accounts for prior and posterior statistical distributions of parameters
is often performed numerically using MCMC and other statistical sampling
methods.

\begin{table}[!t]
\centering\footnotesize
\caption{Coverage of parameter estimation methods for BRNs.}
\label{tb:7}
\setlength{\tabcolsep}{2pt} \renewcommand{\arraystretch}{1.1}
\begin{tabular}{l*{28}{c}}
\headerB{}
  {\footnotesize\bf \# papers}
  &149 &61 &9 &288 &81 &63 &68 &50 &21 &233 &30 &57 &33 &82 &78 &77 &87 &53 &30
  &110 &77 &191 &97 &83 &113 &40 &56 &45 \\ \hline
\end{tabular}
\end{table}

In order to view a timeline of interest in different parameter estimation
methods, \tref{tb:11} contains the number of cited papers concerning estimation
tasks and methods in given years. As for methods in \tref{tb:10}, the number of
selected references peaked in 2014. However, we can again observe increasing
interest in parameter estimation problems for BRNs over past decade. This shows
that parameter estimation strategies are closely related to modeling strategies
as discussed previously. Moreover, from \tref{tb:4}, we can observe that the
largest number of research theses involving parameter estimation problems in
BRNs were again produced in 2014.

\begin{table}[!t]
\centering\footnotesize
\caption{Number of papers concerning estimation tasks and methods for BRNs in
given years.}
\label{tb:11}
\setlength{\tabcolsep}{2pt} \renewcommand{\arraystretch}{1.1}
\begin{tabular}{p{1cm}*{28}{c}}
\headerB{Year}
2005 &3 &2 &. &7 &1 &1 &1 &. &. &3 &. &. &. &2 &2 &1 &2 &. &. &1 &1 &3 &1 &1 &. &. &. &.  \\
2006 &6 &1 &. &10 &1 &4 &2 &. &2 &7 &. &1 &1 &. &. &. &. &. &. &4 &3 &4 &4 &6 &1 &. &2 &1  \\
2007 &2 &2 &1 &8 &1 &2 &1 &. &2 &5 &. &. &. &1 &1 &2 &2 &1 &1 &4 &4 &5 &4 &3 &1 &1 &2 &2  \\
2008 &3 &3 &. &9 &1 &1 &1 &. &1 &6 &. &2 &. &1 &1 &1 &2 &1 &1 &5 &3 &7 &3 &5 &4 &. &2 &1  \\
2009 &8 &4 &1 &13 &4 &4 &4 &2 &1 &11 &1 &. &2 &3 &3 &2 &4 &1 &2 &7 &8 &9 &7 &7 &6 &3 &4 &1  \\
2010 &13 &5 &. &22 &8 &4 &8 &3 &2 &18 &3 &4 &2 &5 &8 &6 &5 &3 &1 &7 &7 &16 &10 &8 &6 &2 &3 &3  \\
2011 &9 &5 &. &18 &6 &3 &2 &1 &. &15 &1 &4 &2 &6 &7 &4 &4 &4 &2 &4 &4 &11 &6 &4 &6 &2 &2 &.  \\
2012 &14 &7 &3 &25 &10 &5 &10 &6 &. &21 &2 &9 &3 &8 &12 &9 &5 &2 &2 &11 &8 &17 &10 &10 &12 &4 &7 &4  \\
2013 &19 &6 &2 &30 &11 &6 &11 &8 &3 &28 &3 &4 &4 &12 &6 &10 &14 &11 &2 &14 &12 &21 &9 &7 &15 &10 &7 &8  \\
2014 &27 &12 &1 &45 &13 &17 &12 &10 &4 &40 &6 &14 &10 &20 &14 &19 &19 &12 &8 &21 &11 &27 &15 &9 &18 &5 &10 &8  \\
2015 &16 &3 &. &27 &8 &4 &3 &4 &1 &22 &5 &8 &1 &9 &7 &10 &13 &8 &7 &11 &4 &19 &7 &4 &8 &3 &4 &3  \\
2016 &12 &2 &1 &27 &9 &5 &5 &4 &3 &23 &3 &4 &1 &7 &6 &4 &8 &4 &1 &10 &6 &17 &9 &8 &13 &6 &6 &7  \\
2017 &10 &7 &. &23 &7 &6 &4 &5 &2 &15 &3 &3 &5 &4 &5 &3 &4 &2 &1 &6 &3 &19 &7 &8 &15 &4 &5 &2  \\
2018 &7 &2 &. &19 &1 &1 &4 &7 &. &16 &3 &3 &1 &4 &5 &6 &3 &2 &2 &4 &2 &12 &4 &2 &7 &. &2 &4  \\  \hline
\end{tabular}
\end{table}

\section{Choices of models and methods for parameter estimation in
  BRNs}\label{sc:select}

We now evaluate what BRN models are preferred for different parameter
estimation strategies, and also to explore what parameter estimation methods
are assumed in different parameter estimation tasks. Dots in tables represent
zero counts to improve readability. The models, estimation tasks and estimation
methods considered are the same as in \tref{tb:6} and \tref{tb:7},
respectively.

\tref{tb:8} shows the number of papers concerning given BRN model and given
estimation strategy where we excluded papers which were deemed to only
marginally consider given combination of model and task or method. We can
observe that the parameter inference task has been considered for all models of
BRNs, however, some models have been investigated much more than others. The
most popular models for parameter inference and other related tasks are
generally models involving differential equations, Markov processes, and state
space representations. The second most popular group of models for parameter
estimation include S-system and polynomial models, and moment closure and
linear noise approximations.

Sensitivity analysis, using information theoretic measures and evaluation of
confidence and credible intervals have been considered for most BRN models.
Moreover, sensitivity analysis has somewhat similar distribution of models as
parameter inference, except the latter shows about ten times larger levels of
interest. In some cases, sensitivity analysis is combined with bifurcation
analysis, so the latter is not referred to explicitly in papers. Optimum
experiment design has been assumed for several models, but there seems to be no
clear model preference. Sum of squares measure is likely quit underestimated in
\tref{tb:8}, since it is often assumed without explicit reference.

Probabilistic MAP and ML measures have been assumed for all models. In many
cases, the corresponding inference tasks involve prior and posterior
distributions and probabilities, and parameter likelihoods. Variational
Bayesian and ABC methods are mostly used with Markov processes, since this is
where they were originally developed whereas Markov processes are derived from
differential equations. The EM method is mostly used with differential
equations. The MC based sampling methods including particle filters are
important for practical implementation of Bayesian inference strategies.
However, these methods were rarely used with less popular BRN models. Similar
comments can be made about Kalman filtering, LS regression, and most data
fitting methods considered. The PSO method has been mainly considered for
differential equation models, and to some extent also for several other models.
There are several BRN models which are not assumed with inference strategies
within other algorithms such as neural networks.

Statistical learning methods including MLR, DLR and TLR are still used
sporadically, compared to other methods discussed so far. Consequently, it is
still difficult to identify preferred models of BRNs in literature for
statistical learning algorithms. Statistical learning requires enough training
data as well as some level of time invariance in order to find generalized
descriptions of systems to make predictions from data. However, as interest in
applications of MLR techniques is growing and the efficiency of learning from
data improves, it will also affect suitability of MLR techniques for different
models of BRNs.

\begin{table}[!t]
\centering\footnotesize
\caption{Adjusted number of papers concerning given methods with different
models of BRNs.}
\label{tb:8}
\setlength{\tabcolsep}{2pt} \renewcommand{\arraystretch}{1.1}
\begin{tabular}{p{0.7cm}l*{28}{c}}
\headerBa
\multirow{3}{0.5cm}{\xrot{\xcell{Physical\\laws}}}
& kinetic rate laws &3 &2 &. &\bf{10} &3 &1 &. &2 &. &\bf{6} &1 &. &1 &1 &. &2 &2 &2 &2 &3 &2 &4 &1 &3 &3 &1 &2 &. \\
& mass action kinetics &\bf{7} &2 &. &\bf{19} &2 &1 &. &2 &1 &\bf{11} &. &3 &2 &4 &1 &. &. &. &. &4 &2 &4 &2 &3 &2 &1 &2 &. \\
& mechanistic models &\bf{5} &. &. &\bf{15} &2 &\bf{5} &1 &1 &1 &\bf{12} &3 &3 &. &3 &1 &3 &1 &1 &. &2 &1 &4 &2 &2 &2 &2 &2 &1 \\
\multirow{4}{0.5cm}{\xrot{\xcell{Random\\processes}}}
& Markov process &\bf{22} &3 &1 &\bf{99} &\bf{10} &\bf{11} &\bf{8} &\bf{9} &1 &\bf{85} &\bf{15} &\bf{7} &\bf{7} &\bf{49} &\bf{30} &\bf{27} &\bf{11} &4 &3 &\bf{16} &3 &\bf{12} &\bf{5} &3 &\bf{7} &. &3 &4 \\
& Poisson process &\bf{6} &2 &1 &\bf{22} &4 &2 &3 &\bf{6} &1 &\bf{17} &2 &3 &4 &\bf{12} &\bf{9} &4 &2 &. &. &4 &2 &4 &. &1 &3 &. &1 &. \\
& birth-death process &3 &1 &. &\bf{9} &1 &1 &. &1 &. &\bf{8} &. &1 &2 &3 &2 &1 &. &. &. &. &. &. &. &. &1 &. &. &. \\
& telegraph process &. &. &. &2 &. &. &. &. &. &1 &. &. &1 &. &. &. &. &. &. &. &. &. &. &. &1 &. &. &. \\
\multirow{8}{0.5cm}{\xrot{\xcell{Mathematical\\models}}}
& state space representation &\bf{16} &\bf{5} &1 &\bf{53} &\bf{7} &4 &\bf{5} &\bf{6} &1 &\bf{45} &\bf{5} &\bf{5} &3 &\bf{15} &\bf{13} &\bf{14} &\bf{15} &\bf{7} &\bf{7} &\bf{7} &2 &\bf{8} &3 &1 &\bf{5} &. &2 &2 \\
& ODEs, PDEs, SDEs, DDEs &\bf{43} &\bf{8} &1 &\bf{132} &\bf{16} &\bf{15} &\bf{12} &\bf{8} &3 &\bf{90} &\bf{8} &\bf{14} &\bf{6} &\bf{32} &\bf{16} &\bf{20} &\bf{19} &\bf{12} &\bf{9} &\bf{19} &\bf{8} &\bf{25} &\bf{11} &\bf{15} &\bf{14} &\bf{9} &\bf{7} &3 \\
& rational model &4 &. &. &\bf{6} &2 &2 &1 &. &. &\bf{5} &2 &1 &. &2 &2 &4 &1 &1 &1 &2 &1 &3 &3 &1 &1 &. &1 &2 \\
& differential algebraic equations &4 &1 &. &\bf{6} &1 &1 &. &. &. &4 &1 &. &. &. &1 &. &1 &. &. &3 &3 &3 &2 &2 &. &1 &. &. \\
& tensor representation &3 &2 &. &4 &2 &1 &1 &1 &. &3 &. &. &. &. &. &. &1 &1 &1 &1 &. &. &1 &. &. &. &1 &. \\
& S-system model &4 &2 &1 &\bf{25} &3 &2 &1 &2 &. &\bf{9} &. &2 &2 &2 &1 &. &3 &3 &2 &\bf{11} &\bf{5} &\bf{6} &3 &\bf{7} &. &2 &2 &. \\
& polynomial model &\bf{10} &3 &1 &\bf{25} &\bf{5} &4 &2 &3 &1 &\bf{12} &1 &2 &2 &\bf{5} &4 &1 &2 &1 &2 &\bf{8} &2 &\bf{11} &. &2 &1 &. &2 &. \\
& manifold map &3 &. &. &\bf{7} &2 &1 &. &1 &. &\bf{6} &3 &2 &. &3 &2 &4 &3 &2 &2 &4 &. &. &1 &. &1 &. &2 &2 \\
\multirow{4}{0.5cm}{\xrot{\xcell{Interaction\\models}}}
& Petri nets &1 &. &1 &3 &2 &. &1 &2 &. &2 &. &1 &. &1 &1 &. &. &. &. &1 &1 &1 &. &. &. &. &1 &. \\
& Boolean networks &2 &. &1 &2 &1 &1 &2 &1 &. &2 &. &1 &. &1 &1 &1 &. &. &. &2 &1 &2 &. &1 &. &. &. &1 \\
& neural networks &3 &. &. &\bf{9} &3 &1 &1 &2 &. &\bf{5} &1 &1 &. &1 &1 &1 &3 &3 &3 &\bf{5} &2 &2 &2 &1 &1 &1 &\bf{9} &3 \\
& agent based models &2 &. &1 &\bf{9} &2 &1 &1 &2 &. &\bf{7} &1 &2 &. &3 &3 &\bf{5} &2 &. &. &1 &2 &3 &2 &. &2 &. &1 &. \\
\multirow{6}{0.5cm}{\xrot{\xcell{CME based\\models}}}
& Langevin equation &4 &1 &. &\bf{17} &4 &2 &2 &3 &1 &\bf{15} &. &. &3 &\bf{9} &\bf{8} &4 &3 &1 &. &. &. &2 &1 &. &3 &. &1 &. \\
& Fokker-Planck equation &\bf{5} &2 &. &\bf{11} &1 &2 &2 &2 &1 &\bf{9} &. &. &4 &4 &2 &1 &1 &. &. &. &. &1 &. &. &3 &. &. &. \\
& reaction rate equation &3 &1 &. &3 &1 &1 &1 &. &. &3 &. &. &. &2 &1 &. &1 &. &. &. &. &1 &. &. &2 &. &. &. \\
& moment closure &\bf{8} &. &. &\bf{24} &4 &1 &3 &\bf{5} &2 &\bf{17} &1 &1 &4 &\bf{5} &2 &3 &2 &1 &. &1 &. &2 &. &. &\bf{5} &. &1 &1 \\
& linear noise approximation &\bf{10} &1 &. &\bf{29} &\bf{6} &3 &4 &4 &2 &\bf{25} &2 &2 &1 &\bf{12} &\bf{5} &\bf{6} &2 &. &. &1 &. &. &. &1 &\bf{5} &. &1 &. \\
& system size expansion &2 &1 &. &4 &1 &1 &. &1 &2 &4 &. &1 &1 &2 &. &. &. &. &. &. &. &. &. &. &1 &. &. &. \\  \hline
\end{tabular}
\end{table}

Another interesting viewpoint is to assess what inference methods are used with
different inference tasks. The numbers of cited papers for given combinations
of inference tasks and inference methods are provided in \tref{tb:9}. With one
exception, there is at least one paper for each such combination, however, the
level of interest appears to vary considerably. In particular, the largest
number of papers for all inference tasks considered assume Bayesian analysis
and methods for model fitting to data. On the other hand, sum of squared
errors, unscented Kalman filter (UKF), and PSO method are generally least
assumed in the papers cited. As discussed, sum of squared errors is used often,
but rarely mentioned explicitly whereas UKF and PSO methods are usually rather
difficult to implement.

Assuming \tref{tb:9}, we can also compare the levels of interest for two or
more methods across different inference tasks. For example, EM and MCMC methods
are used equally often for sensitivity analysis whereas MCMC is preferred over
EM for identifiability task. Also, LS and regression methods are always
preferred over Kalman filtering due to implementation complexity.
Interestingly, machine learning methods appear to be considered more often than
ABC, variational Bayesian inference, UKF, and PSO methods, but comparably often
to EKF.

\begin{table}[!t]
\centering\footnotesize
\caption{Number of papers concerning given tasks and different estimation
  methods for BRNs.}
\label{tb:9}
\setlength{\tabcolsep}{2pt} \renewcommand{\arraystretch}{1.1}
\begin{tabular}{p{0.3cm}l*{23}{c}}
\headerBb
\multirow{5}{0.5cm}{\xrot{Tasks}}
& identifi., observab., reachability &44 &54 &28 &10 &131 &20 &25 &21 &45 &37 &43 &49 &27 &15 &71 &42 &96 &56 &45 &60 &19 &23 &26 \\
& bifurcation analysis &16 &15 &15 &9 &55 &10 &17 &11 &25 &24 &23 &16 &8 &4 &27 &21 &48 &30 &16 &25 &10 &11 &9 \\
& optimum experiment &4 &7 &3 &2 &9 &. &4 &2 &2 &3 &2 &3 &1 &1 &5 &4 &7 &4 &5 &4 &2 &5 &3 \\
& inference, identification &63 &68 &50 &21 &233 &30 &57 &33 &82 &78 &77 &87 &53 &30 &110 &77 &191 &97 &83 &113 &40 &56 &45 \\
& sensitivity analysis &25 &36 &19 &6 &71 &13 &24 &15 &22 &21 &27 &28 &17 &12 &44 &28 &63 &37 &29 &35 &13 &14 &16 \\  \hline
\end{tabular}
\end{table}

\subsection{Future research problems}\label{sc:select:1}

Tables \ref{tb:8} and \ref{tb:9} can be used as guidelines to define new
problems which have not been sufficiently investigated in literature. We can
use data in \tref{tb:6} and \tref{tb:7} to identify such cases in \tref{tb:8}
where we excluded papers not clearly investigating given model and task or
method. We can separate models, tasks and methods into groups having smaller
and larger levels of interests. There are certainly research opportunities
where the number of papers in all dimensions is small. However, it is more
convenient to enumerate problems which have already been well investigated in
literature. Such cases of paper counts being equal to or larger than 5 are
highlighted in \tref{tb:8} using boldface, and they include:
\begin{enumerate}
\item identification and inference tasks with Markov processes, state space
  representation, differential equations, polynomial function, S-system,
  Langevin and Fokker-Planck equations, and CME approximation models; and
\item most inference methods with Markov processes, state space representation,
  and differential equation models; and
\item some inference methods with Poisson process, S-system, polynomial
  function, Langevin equation, and CME approximation models; and
\item Bayesian methods with MAP and ML inferences with most models considered;
  and
\item LS regression and optimization programming mainly with Markov processes,
  state space representation, differential equation, S-system and polynomial
  models; and
\item search methods with Markov processes, state space representation,
  differential equations, and CME approximation models.
\end{enumerate}

Furthermore, bifurcation analysis appears to be the least considered task for
all models. In many papers, bifurcation analysis may not be referred to
explicitly, but performed as part of sensitivity analysis. Similar comments can
be made about a sum of squared errors method. From \tref{tb:8}, also machine
learning methods have been considered sporadically and only for some BRN models
to solve inference problems. Comparing machine learning methods with
conventional methods of statistical inference may be one of the most
interesting research avenues in near future. It is likely that machine learning
is more beneficial for some models, depending on availability of observations
and training data. In addition, we can observe from \tref{tb:9} that optimum
experiment design is underrepresented in comparison to other inference tasks.

We can also point out other research opportunities which are not immediately
apparent from the tables presented in previous sections. In particular, there
are other types of inference algorithms and strategies than those listed e.g.
in \tref{tb:8}. For instance, minimum mean square error (MMSE) estimator is
only discussed in reference \citep{koeppl2012}. Since estimation errors may
have different distributions depending on BRN model considered, generalized
linear regression (GLR) can be assumed as a simple, universal and yet powerful
statistical learning technique. The GLR method has not been investigated
comprehensively in literature to make inferences in BRNs. In addition, also
distributions can be inferred from observations \cite{atitey2018}. Knowledge of
distributions greatly affects available choices of estimators and their
performance. Another unexplored strategy is compressive sensing (CS) which
exploits sparsity of parameter space in some transform domain. Among machine
learning methods, transfer learning has not been used for inferences in BRNs in
order to exploit increasing availability of omics data \citep{weiss2016}.

Furthermore, vast majority of inference problems in literature assume
well-stirred models of BRNs with reactions dependent solely on species
concentrations, but not species spatial distributions. Assuming spatially
resolved models of BRNs with diffusion and other means of molecule transport
through complex fluids is a rather realistic assumption. Such models are
usually described by RDME \citep{lotstedt2018}. Moreover, in many BRNs, the
reaction rates can be time varying. Inference of time varying parameters in
models of BRNs have not been explicitly considered in literature.

Most inference problems in literature assume simple models of measurements such
as obtaining noisy concentrations at discrete time instances. In order to
increase the sensitivity of measurements, observations are often integrated in
time \citep{folia2018}. Such data transformations representing various
measurement techniques cannot be ignored when devising inference strategies as
well as optimum experiment design for BRNs. Since observations may affect
biological processes, the number and duration of observations should be
minimized in space and also in time. In addition, measurement noise is often
(but not always) assumed to be independent of species concentrations and
Gaussian distributed. In realistic experiments, measurement noises can be
correlated in time, among other measurements, and dependent on reaction rates
as well as concentrations of species. It would be very useful to report
statistical properties of different measurement techniques in different lab
experiments. Having such statistical models of measurement noises can
considerably improves efficiency and accuracy of different inference methods in
BRNs.

More generally, performance of various inference strategies is greatly
dependent on the structure, parameter values as well as initial state of BRN
considered. These aspects were mainly taken into consideration to optimize data
fitting methods, but much less for other statistical parameter inference
strategies. There is a trade-off to mechanistically employ universal inference
methods against specializing these methods to specific scenarios of BRNs. The
latter approach may improve the performance and efficiency of inference at the
cost of increased implementation complexity. A useful area of research would be
to combine and jointly consider model simplification strategies as in
\citep{eghtesadi2014} with parameter estimation strategies. However, it is
important to test and validate all devised inference algorithms. In some
papers, inference algorithms are tested on multiple data sets, but general
methodology to test and validate these algorithms for case of BRNs have not
been presented in literature. It is also useful to separate inference concepts
and strategies form their implementations, for example, Bayesian inference can
be implemented using stochastic sampling, ABC, variational inference, EM and
other methods. Many papers on inferences in BRNs are concerned with
implementation aspects rather than concepts.

Finally, let's not forget that the ultimate goal of statistical inferences in
models of BRNs is to elucidate understanding of in vivo and in vitro biological
systems. This is primarily dependent on having accurate models of these systems
including knowing values of their parameters. As experimental techniques
improve, new observation data from experiments will stimulate development of
new biological models, and thus, there will be also need for new inference
methods and strategies in future.

\section{Conclusions}\label{sc:concl}

The aim of this review paper was to explore how various inference tasks and
methods are used with different models of BRNs. Dependency between tasks,
methods and models were captured in tables containing counts of relevant papers
among almost 300 cited references. More detailed results can be found in
supplementary tables including a list of many cited references with links to
their citations in Google Scholar. Basic concepts of modeling and parameter
inference for BRNs were discussed. In order to facilitate understanding of
inference methods used for BRNs, a survey of parameter estimation strategies
for general systems was also included.

Common models of BRNs and inference tasks and methods were identified by text
mining all cited papers. The text mining was accomplished partly manually and
partly it was automated using text processing scripts. Such automation is
indispensable when dealing with large number of references as is the case in
our paper. For convenience, both models and methods were presented in several
groups. Most common models for BRNs in literature assume mass action kinetics,
Markov processes, state space representation, and differential equations. Less
common but still popular are kinetic rate law, mechanistic, Poisson process,
polynomial and rational function, S-system, Langevin equation, and CME based
approximation models.

Several previously published review papers concerning inferences in BRNs were
outlined. Relevant research theses from past decade were also listed, since
these works tend to contain comprehensive literature surveys and tutorial style
explanations. We observed that the most common inference tasks are concerned
with identifiability, parameter inference and sensitivity analysis. The most
common inference methods are Bayesian analysis using MAP and ML estimators, MC
sampling techniques, LS, and evolutionary algorithms for data fitting,
especially optimization programming, simulated annealing, and scatter and other
searches. Main references concerning evolutionary algorithms were summarized.

In the last part of the paper, the levels of interest in different inference
tasks and methods for given BRN models were assessed. This allowed us to
identify inference problems in BRNs which were most often considered in
literature. The references cited in this paper show that the interest in
inference problems in BRNs peaked in year 2014. However, it is likely that the
current interest in machine learning methods, progress in experimental
techniques, and availability of omics data will stimulate new developments in
modeling and parameter inference for BRNs.

\begin{table}[!h]
  {ABC: approximate Bayesian computation, artificial bee colony; ABM: agent
    based model; AKF: augmented Kalman filter; AR: alternating regression;
    BLUE: best linear unbiased; CCA: canonical correlation analysis; CDIS:
    conditional density importance sampling; CGA: continuous genetic algorithm;
    CLE: chemical Langevin equation; CME: chemical master equation; CRO:
    chemical reaction optimization; CS: compressive sensing; CTMC: continuous
    time Markov chain; CTMP: continuous time Markov process; DE: differential
    evolution; DLR: deep learning; EKF: extended Kalman filter; EM:
    expectation-maximization; EP: expectation propagation; FA: firefly
    algorithm; FDM: finite differences method; GLR: generalized linear
    regression; GLR: generalized linear regression; GP: genetic programming;
    HDL: hardware description language; KF: Kalman filter; LFM: linear
    fractional model; LNA: linear noise approximation; LS: least squares; MAP:
    maximum a posterior; MC: Monte Carlo; MCEM: MC expectation-maximization;
    MCMC: MC Markov Chain; MES: maximum entropy sampling; ML: maximum
    likelihood; MLR: machine learning; MM: method of moments; MMSE: minimum
    mean square error; MVUB: minimum variance unbiased; NLP: non-linear
    programming; NLSQ: non-linear least squares; ODE: ordinary differential
    equation; PDF: portable document format, probability density function; PMC:
    population Monte Carlo; PSO: particle swarm optimization; QE:
    quasi-equilibrium; QSS: quasi-steady state; RDME: reaction-diffusion master
    equation; RRE: reaction rate equation; SA: simulated annealing; SMC:
    sequential Monte Carlo; SMCMC: sequential Markov chain Monte Carlo; SS:
    scatter search; SSE: sum of squared errors, system size expansion; TLR:
    transfer learning; UKF: unscented Kalman filter}
\end{table}

\label{sc:ref}
\bibliographystyle{FrontiersStyle}
\bibliography{refer}

\begin{thebibliography}{304}
\providecommand{\natexlab}[1]{#1}
\expandafter\ifx\csname urlstyle\endcsname\relax
  \providecommand{\doi}[1]{doi:\discretionary{}{}{}#1}\else
  \providecommand{\doi}{doi:\discretionary{}{}{}\begingroup
  \urlstyle{rm}\Url}\fi
\providecommand{\selectlanguage}[1]{\relax}
\providecommand{\bibAnnoteFile}[1]{%
  \IfFileExists{#1}{\begin{quotation}\noindent\textsc{Key:} #1\\
  \textsc{Annotation:}\ \input{#1}\end{quotation}}{}}
\providecommand{\bibAnnote}[2]{%
  \begin{quotation}\noindent\textsc{Key:} #1\\
  \textsc{Annotation:}\ #2\end{quotation}}

\bibitem[{Abdullah et~al.(2013{\natexlab{a}})Abdullah, Deris, Anwar, and
  Arjunan}]{abdullah2013b}
Abdullah, A., Deris, S., Anwar, S., and Arjunan, S. N.~V. (2013{\natexlab{a}}).
\newblock An evolutionary firefly algorithm for the estimation of nonlinear
  biological model parameters.
\newblock \emph{PLOS One} 8, 3/e56310.
\newblock \doi{10.1371/journal.pone.0056310}
\bibAnnoteFile{abdullah2013b}

\bibitem[{Abdullah et~al.(2013{\natexlab{b}})Abdullah, Deris, Mohamad, and
  Anwar}]{abdullah2013a}
Abdullah, A., Deris, S., Mohamad, M.~S., and Anwar, S. (2013{\natexlab{b}}).
\newblock An improved swarm optimization for parametera estimation and
  biological model selection.
\newblock \emph{PLOS One} 8, 4/e61258.
\newblock \doi{10.1371/journal.pone.0061258}
\bibAnnoteFile{abdullah2013a}

\bibitem[{Abdullah et~al.(2013{\natexlab{c}})Abdullah, Deris, Mohamad, and
  Hashim}]{abdullah2013}
Abdullah, A., Deris, S., Mohamad, M.~S., and Hashim, S. Z.~M.
  (2013{\natexlab{c}}).
\newblock A new particle swarm evolutionary optimization for parameter
  estimation of biological models.
\newblock \emph{IJCISIM} 5, 571--580
\bibAnnoteFile{abdullah2013}

\bibitem[{Alberton et~al.(2013)Alberton, Alberton, Maggio, D\'{i}az, and
  Secchi}]{alberton2013}
Alberton, K. P.~F., Alberton, A.~L., Maggio, J. A.~D., D\'{i}az, M.~S., and
  Secchi, A.~R. (2013).
\newblock Accelerating the parameters identifiability procedure: {S}et by set
  selection.
\newblock \emph{Comput. Chem. Eng.} 55, 181--197.
\newblock \doi{10.1016/j.compchemeng.2013.04.014}
\bibAnnoteFile{alberton2013}

\bibitem[{Ale et~al.(2013)Ale, Kirk, and Stumpf}]{ale2013}
Ale, A., Kirk, P., and Stumpf, M. P.~H. (2013).
\newblock A general moment expansion method for stochastic kinetic models.
\newblock \emph{Chem. Phys.} 138, 1--12.
\newblock \doi{10.1063/1.4802475}
\bibAnnoteFile{ale2013}

\bibitem[{Ali et~al.(2015)Ali, Hoang, Hussain, and Dochain}]{ali2015}
Ali, J.~M., Hoang, N.~H., Hussain, M.~A., and Dochain, D. (2015).
\newblock Review and classification of recent observers applied in chemical
  process systems.
\newblock \emph{Comput. Chem. Eng.} 76, 27--41.
\newblock \doi{10.1016/j.compchemeng.2015.01.019}
\bibAnnoteFile{ali2015}

\bibitem[{Amrein and K\"unsch(2012)}]{amrein2012}
Amrein, M. and K\"unsch, H.~R. (2012).
\newblock Rate estimation in partially observed {Markov} jump processes with
  measurement errors.
\newblock \emph{Stat. Comput.} 22, 513--526.
\newblock \doi{10.1007/s11222-011-9244-1}
\bibAnnoteFile{amrein2012}

\bibitem[{Anai et~al.(2006)Anai, Orii, and Horimoto}]{anai2006}
Anai, H., Orii, S., and Horimoto, K. (2006).
\newblock Symbolic-numeric estimation of parameters in biochemical models by
  quantifier elimination.
\newblock \emph{Bioinform. Comput. Biol.} 4, 1097--1117.
\newblock \doi{10.1142/S0219720006002351}
\bibAnnoteFile{anai2006}

\bibitem[{Andreychenko(2014)}]{andreychenko2014}
Andreychenko, A. (2014).
\newblock Model reconstruction for moment-based stochastic chemical kinetics.
\newblock ArXiv:1410.3177v1 [math.NA]
\bibAnnoteFile{andreychenko2014}

\bibitem[{Andreychenko et~al.(2011)Andreychenko, Mikeev, Spieler, and
  Wolf}]{andreychenko2011}
Andreychenko, A., Mikeev, L., Spieler, D., and Wolf, V. (2011).
\newblock Parameter identification for {Markov} models of biochemical
  reactions.
\newblock ArXiv:1102.2819v1 [q-bio.QM]
\bibAnnoteFile{andreychenko2011}

\bibitem[{Andreychenko et~al.(2012)Andreychenko, Mikeev, Spieler, and
  Wolf}]{andreychenko2012}
Andreychenko, A., Mikeev, L., Spieler, D., and Wolf, V. (2012).
\newblock Approximate maximum likelihood estimation for stochastic chemical
  kinetics.
\newblock \emph{EURASIP J. Bioinf. Sys. Biol.} 2012.
\newblock \doi{10.1186/1687-4153-2012-9}
\bibAnnoteFile{andreychenko2012}

\bibitem[{Andrieu et~al.(2010)Andrieu, Doucet, and Holenstein}]{andrieu2010}
Andrieu, C., Doucet, A., and Holenstein, R. (2010).
\newblock Particle {Markov} chain {Monte Carlo} methods.
\newblock \emph{J. R. Statist. Soc. B} 72, 269--342.
\newblock \doi{10.1111/j.1467-9868.2009.00736.x}
\bibAnnoteFile{andrieu2010}

\bibitem[{Angius and Horv\'ath(2011)}]{angius2011}
Angius, A. and Horv\'ath, A. (2011).
\newblock The {Monte Carlo} {EM} method for the parameter estimation of
  biological models.
\newblock \emph{El. Notes Theor. Comp. Sci.} 275, 23--36.
\newblock \doi{10.1016/j.entcs.2011.09.003}
\bibAnnoteFile{angius2011}

\bibitem[{Arnold et~al.(2014)Arnold, Calvetti, and Somersalo}]{arnold2014}
Arnold, A., Calvetti, D., and Somersalo, E. (2014).
\newblock Parameter estimation for stiff deterministic dynamical systems via
  ensemble {Kalman} filter.
\newblock \emph{Inv. Prob.} 30.
\newblock \doi{10.1088/0266-5611/30/10/105008}
\bibAnnoteFile{arnold2014}

\bibitem[{Ashyraliyev et~al.(2009)Ashyraliyev, Fomekong, Nanfack, Kaandorp, and
  Blom}]{ashyraliyev2009}
Ashyraliyev, M., Fomekong, Y., Nanfack, Kaandorp, J.~A., and Blom, J.~G.
  (2009).
\newblock Systems biology: {P}arameter estimation for biochemical models.
\newblock \emph{FEBS J.} 276, 886--902.
\newblock \doi{10.1111/j.1742-4658.2008.06844.x}
\bibAnnoteFile{ashyraliyev2009}

\bibitem[{Atitey et~al.(2018{\natexlab{a}})Atitey, Loskot, and
  Rees}]{atitey2018a}
Atitey, K., Loskot, P., and Rees, P. (2018{\natexlab{a}}).
\newblock Determining transcription rates yielding steady state production of
  {mRNA} in the lac genetic switch of {Escherichia} coli.
\newblock \emph{Comput. Biol.} 25, 1023--1039.
\newblock \doi{10.1089/cmb.2018.0055}
\bibAnnoteFile{atitey2018a}

\bibitem[{Atitey et~al.(2018{\natexlab{b}})Atitey, Loskot, and
  Rees}]{atitey2018}
Atitey, K., Loskot, P., and Rees, P. (2018{\natexlab{b}}).
\newblock Inferring distributions from observed {mRNA} and protein copy counts
  in genetic circuits.
\newblock \emph{Biomed. Phy. Eng. Express} 5, 015022.
\newblock \doi{10.1088/2057-1976/aaef5c}
\bibAnnoteFile{atitey2018}

\bibitem[{Atitey et~al.(2019)Atitey, Loskot, and Rees}]{atitey2019}
Atitey, K., Loskot, P., and Rees, P. (2019).
\newblock Elucidating effects of reaction rates on dynamics of the lac circuit
  in {Escherichia} coli.
\newblock \emph{Biosyst.} 175, 1--10.
\newblock \doi{10.1016/j.biosystems.2018.11.003}
\bibAnnoteFile{atitey2019}

\bibitem[{Azab et~al.(2018)Azab, Toth, Mihaylova, and Arvaneh}]{azab2018}
Azab, A.~M., Toth, J., Mihaylova, L.~S., and Arvaneh, M. (2018).
\newblock A review on transfer learning approaches in brain-computer interface.
\newblock In \emph{Signal Processing and Machine Learning for Brain-Machine
  Interfaces}, eds. T.~Tanaka and M.~Arvaneh (IET). 81--98
\bibAnnoteFile{azab2018}

\bibitem[{Babtie and Stumpf(2017)}]{babtie2017}
Babtie, A.~C. and Stumpf, M. P.~H. (2017).
\newblock How to deal with parameters for whole-cell modelling.
\newblock \emph{J. R. Soc. Interface} 14.
\newblock \doi{10.1098/rsif.2017.0237}
\bibAnnoteFile{babtie2017}

\bibitem[{Backenk\"ohler et~al.(2016)Backenk\"ohler, Bortolussi, and
  Wolf}]{backenkohler2016}
Backenk\"ohler, M., Bortolussi, L., and Wolf, V. (2016).
\newblock \emph{Generalized Method of Moments for Stochastic Reaction Networks
  in Equilibrium} (LNBI 9859), chap. CMSB 2016.
\newblock 15--29.
\newblock \doi{10.1007/978-3-319-45177-0 2}
\bibAnnoteFile{backenkohler2016}

\bibitem[{Backenk\"ohler et~al.(2018)Backenk\"ohler, Bortolussi, and
  Wolf}]{backenkohler2018}
Backenk\"ohler, M., Bortolussi, L., and Wolf, V. (2018).
\newblock Moment-based parameter estimation for stochastic reaction networks in
  equilibrium.
\newblock \emph{IEEE Trans. Comp. Biol. Bioinf.} 15, 1180--1192.
\newblock \doi{10.1109/TCBB.2017.2775219}
\bibAnnoteFile{backenkohler2018}

\bibitem[{Baker et~al.(2011)Baker, Poskar, and Junker}]{baker2011}
Baker, S.~M., Poskar, C.~H., and Junker, B.~H. (2011).
\newblock Unscented {Kalman} filter with parameter identifiability analysis for
  the estimation of multiple parameters in kinetic models.
\newblock \emph{EURASIP Bioinf. Sys. Biol.} 2011.
\newblock \doi{10.1186/1687-4153-2011-7}
\bibAnnoteFile{baker2011}

\bibitem[{Baker et~al.(2013)Baker, Poskar, Schreiber, and Junker}]{baker2013}
Baker, S.~M., Poskar, C.~H., Schreiber, F., and Junker, B.~H. (2013).
\newblock An improved constraint filtering technique for inferring hidden
  states and parameters of a biological model.
\newblock \emph{Bioinf.} 29, 1052--1059.
\newblock \doi{10.1093/bioinformatics/btt097}
\bibAnnoteFile{baker2013}

\bibitem[{Baker et~al.(2015)Baker, Poskar, Schreiber, and Junker}]{baker2015}
Baker, S.~M., Poskar, C.~H., Schreiber, F., and Junker, B.~H. (2015).
\newblock A unified framework for estimating parameters of kinetic biological
  models.
\newblock \emph{BMC Bioinf.} 16.
\newblock \doi{10.1186/s12859-015-0500-9}
\bibAnnoteFile{baker2015}

\bibitem[{Baker et~al.(133, 2010)Baker, Schallau, and Junker}]{baker2010}
Baker, S.~M., Schallau, K., and Junker, B.~H. (133, 2010).
\newblock Comparison of different algorithms for simultaneous estimation of
  multiple parameters in kinetic metabolic models.
\newblock \emph{Integr. Bioinf.} 7.
\newblock \doi{10.2390/biecoll-jib-2010-133}
\bibAnnoteFile{baker2010}

\bibitem[{Balsa-Canto et~al.(2008)Balsa-Canto, Peifer, Banga, Timmer, and
  Fleck}]{canto2008}
Balsa-Canto, E., Peifer, M., Banga, J.~R., Timmer, J., and Fleck, C. (2008).
\newblock Hybrid optimization method with general switching strategy for
  parameter estimation.
\newblock \emph{BMC Sys. Biol.} 2.
\newblock \doi{10.1186/1752-0509-2-26}
\bibAnnoteFile{canto2008}

\bibitem[{Banga and Canto(2008)}]{banga2008}
Banga, J.~R. and Canto, E.~B. (2008).
\newblock Parameter estimation and optimal experimental design.
\newblock \emph{Essays In Biochemistry} 45, 195--210.
\newblock \doi{10.1042/bse0450195}
\bibAnnoteFile{banga2008}

\bibitem[{Barnes et~al.(2011)Barnes, Silk†, and Stumpf}]{barnes2018}
Barnes, C.~P., Silk†, D., and Stumpf, M. P.~H. (2011).
\newblock Bayesian design strategies for synthetic biology.
\newblock \emph{Interface Focus} 1, 895--908.
\newblock \doi{10.1098/rsfs.2011.0056}
\bibAnnoteFile{barnes2018}

\bibitem[{Bayer et~al.(2015)Bayer, Moraes, Tempone, and Vilanova}]{bayer2015}
Bayer, C., Moraes, A., Tempone, R., and Vilanova, P. (2015).
\newblock An efficient forward-reverse expectation-maximization algorithm for
  statistical inference in stochastic reaction networks.
\newblock ArXiv:1504.04155v1 [math.NA]
\bibAnnoteFile{bayer2015}

\bibitem[{Berrones et~al.(2016)Berrones, Jim\'enez, Alcorta-Garc\'ia, Almaguer,
  and na}]{berrones2016}
Berrones, A., Jim\'enez, E., Alcorta-Garc\'ia, M.~A., Almaguer, F.-J., and na,
  B.~P. (2016).
\newblock Parameter inference of general nonlinear dynamical models of gene
  regulatory networks from small and noisy time series.
\newblock \emph{Neurocomputing} 175, 555--563.
\newblock \doi{10.1016/j.neucom.2015.10.095}
\bibAnnoteFile{berrones2016}

\bibitem[{Besozzi et~al.(2009)Besozzi, Cazzaniga, Pescini, and
  Vanneschi}]{besozzi2009}
Besozzi, D., Cazzaniga, P., Pescini, G. M.~D., and Vanneschi, L. (2009).
\newblock \emph{A Comparison of Genetic Algorithms And Particle Swarm
  Optimization for Parameter Estimation in Stochastic Biochemical Systems}
  (Springer, Berlin), chap. EvoBIO.
\newblock 116--127.
\newblock \doi{10.1007/978-3-642-01184-9_11}
\bibAnnoteFile{besozzi2009}

\bibitem[{Bhaskar et~al.(2010)Bhaskar, Mihaylova, and Achim}]{bhaskar2010}
Bhaskar, H., Mihaylova, L., and Achim, A. (2010).
\newblock Video foreground detection based on symmetric alpha-stable mixture
  models.
\newblock \emph{IEEE Trans. Circ. Systems Video Technol.} 20, 1133--1138.
\newblock \doi{10.1109/TCSVT.2010.2051282}
\bibAnnoteFile{bhaskar2010}

\bibitem[{Bogomolov et~al.(2015)Bogomolov, Henzinger, Podelski, Ruess, and
  Schilling}]{bogomolov2015}
Bogomolov, S., Henzinger, T.~A., Podelski, A., Ruess, J., and Schilling, C.
  (2015).
\newblock \emph{Adaptive Moment Closure for Parameter Inference of Biochemical
  Reaction Networks} (Springer, Cham), chap. CMSB 2015.
\newblock 77--89.
\newblock \doi{10.1007/978-3-319-23401-4_8}
\bibAnnoteFile{bogomolov2015}

\bibitem[{Bouraoui et~al.(2015)Bouraoui, Farza, M\'enard, Abdennour, M'Saad,
  and Mosrati}]{bouraoui2015}
Bouraoui, I., Farza, M., M\'enard, T., Abdennour, R.~B., M'Saad, M., and
  Mosrati, H. (2015).
\newblock Observer design for a class of uncertain nonlinear systems with
  sampled outputs application to the estimation of kinetic rates in
  bioreactors.
\newblock \emph{Automatica} 55, 78--87.
\newblock \doi{10.1016/j.automatica.2015.02.036}
\bibAnnoteFile{bouraoui2015}

\bibitem[{Boys et~al.(2008)Boys, Wilkinson, and Kirkwood}]{boys2008}
Boys, R.~J., Wilkinson, D.~J., and Kirkwood, T. B.~L. (2008).
\newblock Bayesian inference for a discretely observed stochastic kinetic
  model.
\newblock \emph{Stat. Comput.} 18, 125--135.
\newblock \doi{10.1007/s11222-007-9043-x}
\bibAnnoteFile{boys2008}

\bibitem[{Brim et~al.(2013)Brim, \v{C}eska, Dra\v{z}an, and
  \v{S}afr\'anek}]{brim2013}
Brim, L., \v{C}eska, M., Dra\v{z}an, S., and \v{S}afr\'anek, D. (2013).
\newblock On robustness analysis of stochastic biochemical systems by
  probabilistic model checking.
\newblock ArXiv:1310.4734v1 [cs.NA]
\bibAnnoteFile{brim2013}

\bibitem[{Bronstein and Koeppl(2017)}]{bronstein2017}
Bronstein, L. and Koeppl, H. (2017).
\newblock A variational approach to moment-closure approximations for the
  kinetics of biomolecular reaction networks.
\newblock ArXiv:1709.02963v2 [q-bio.QM]
\bibAnnoteFile{bronstein2017}

\bibitem[{Bronstein et~al.(2015)Bronstein, Zechner, and Koeppl}]{bronstein2015}
Bronstein, L., Zechner, C., and Koeppl, H. (2015).
\newblock Bayesian inference of reaction kinetics from single-cell recordings
  across a heterogeneous cell population.
\newblock \emph{Methods} 85, 22--35.
\newblock \doi{10.1016/j.ymeth.2015.05.012}
\bibAnnoteFile{bronstein2015}

\bibitem[{Brunel et~al.(2014)Brunel, Clairon, and d'Alch\'e Buc}]{brunel2014}
Brunel, N. J.-B., Clairon, Q., and d'Alch\'e Buc, F. (2014).
\newblock Parametric estimation of ordinary differential equations with
  orthogonality conditions.
\newblock \emph{Theory and Met.} 109, 173--185.
\newblock \doi{10.1080/01621459.2013.841583}
\bibAnnoteFile{brunel2014}

\bibitem[{Busetto and Buhmann(2009)}]{busetto2009}
Busetto, A.~G. and Buhmann, J.~M. (2009).
\newblock Stable {Bayesian} parameter estimation for biological dynamical
  systems.
\newblock In \emph{ICCSE}. 148--157.
\newblock \doi{10.1109/CSE.2009.134}
\bibAnnoteFile{busetto2009}

\bibitem[{Camacho et~al.(2018)Camacho, Collins, Powers, Costello, and
  Collins}]{camacho2018}
Camacho, D.~M., Collins, K.~M., Powers, R.~K., Costello, J.~C., and Collins,
  J.~J. (2018).
\newblock Next-generation machine learning for biological networks.
\newblock \emph{Cell} 173, 1--12.
\newblock \doi{10.1016/j.cell.2018.05.015}
\bibAnnoteFile{camacho2018}

\bibitem[{Carmi et~al.(2013)Carmi, Mihaylova, Gning, Gurfil, and
  Godsill}]{carmi2013}
Carmi, A., Mihaylova, L., Gning, A., Gurfil, P., and Godsill, S. (2013).
\newblock \emph{{Markov Chain Monte Carlo} Based Autonomous Tracking and
  Causality Reasoning} (Springer), chap. Advances in Intelligent Signal
  Processing and Data Mining.
\newblock 7--53.
\newblock \doi{10.1007/978-3-642-28696-4_2}
\bibAnnoteFile{carmi2013}

\bibitem[{Cazzaniga et~al.(2015)Cazzaniga, Nobile, and Besozzi}]{cazzaniga2015}
Cazzaniga, P., Nobile, M.~S., and Besozzi, D. (2015).
\newblock The impact of particles initialization in {PSO}: {P}arameter
  estimation as a case in point.
\newblock In \emph{CIBCB}. 1--8.
\newblock \doi{10.1109/CIBCB.2015.7300288}
\bibAnnoteFile{cazzaniga2015}

\bibitem[{Cedersund et~al.(2016)Cedersund, Samuelsson, Ball, Tegn\'er, and
  Gomez-Cabrero}]{cedersund2016}
Cedersund, G., Samuelsson, O., Ball, G., Tegn\'er, J., and Gomez-Cabrero, D.
  (2016).
\newblock \emph{Optimization in Biology Parameter Estimation And the Associated
  Optimization Problem} (Springer, Cham), chap. Uncertainty in Biology.
\newblock 177--197.
\newblock \doi{10.1007/978-3-319-21296-8_7}
\bibAnnoteFile{cedersund2016}

\bibitem[{Chen et~al.(2017)Chen, Cao, and Watson}]{chen2017}
Chen, M., Cao, Y., and Watson, L.~T. (2017).
\newblock Parameter estimation of stochastic models based on limited data.
\newblock \emph{ACM SIG Bioinf.} 7, 3:1--3:3.
\newblock \doi{10.1145/3183624.3183627}
\bibAnnoteFile{chen2017}

\bibitem[{Chen et~al.(2015)Chen, S{\"{a}}rkk{\"{a}}, and Godsill}]{chen2015}
Chen, X., S{\"{a}}rkk{\"{a}}, S., and Godsill, S.~J. (2015).
\newblock A {Bayesian} particle filtering method for brain source localisation.
\newblock \emph{Dig. Signal Proces.} 47, 192--204.
\newblock \doi{10.1016/j.dsp.2015.06.007}
\bibAnnoteFile{chen2015}

\bibitem[{Chevaliera and Samadb(2011)}]{chevalier2011}
Chevaliera, M.~W. and Samadb, H.~E. (2011).
\newblock A data-integrated method for analyzing stochastic biochemical
  networks.
\newblock \emph{Chem. Phy.} 135.
\newblock \doi{10.1063/1.3664126}
\bibAnnoteFile{chevalier2011}

\bibitem[{Chong et~al.(2014)Chong, Mohamad, Deris, Shamsir, Chai, and
  Choon}]{chong2014}
Chong, C.~K., Mohamad, M.~S., Deris, S., Shamsir, M.~S., Chai, L.~E., and
  Choon, Y.~W. (2014).
\newblock Parameter estimation by using an improved bee memory differential
  evolution algorithm {(IBMDE)} to simulate biochemical pathways.
\newblock \emph{Current Bioinf.} 9, 65--75.
\newblock \doi{10.2174/15748936113080990007}
\bibAnnoteFile{chong2014}

\bibitem[{Chong et~al.(2012)Chong, Mohamad, Deris, Shamsir, Choon, and
  Chai}]{chong2012}
Chong, C.~K., Mohamad, M.~S., Deris, S., Shamsir, M.~S., Choon, Y.~W., and
  Chai, L.~E. (2012).
\newblock Improved differential evolution algorithm for parameter estimation to
  improve the production of biochemical pathway.
\newblock \emph{Int. J. AI and Interac. Multimed.} 1, 22--29.
\newblock \doi{10.9781/ijimai.2012.153}
\bibAnnoteFile{chong2012}

\bibitem[{Chou et~al.(2006)Chou, Martens, and Voit}]{chou2006}
Chou, I.-C., Martens, H., and Voit, E.~O. (2006).
\newblock Parameter estimation in biochemical systems models with alternating
  regression.
\newblock \emph{Theor. Biol. Med. Model.} 3.
\newblock \doi{10.1186/1742-4682-3-25}
\bibAnnoteFile{chou2006}

\bibitem[{Chou and Voit(2009)}]{chun2009}
Chou, I.-C. and Voit, E.~O. (2009).
\newblock Recent developments in parameter estimation and structure
  identification of biochemical and genomic systems.
\newblock \emph{Math. Biosci.} 219, 57--83.
\newblock \doi{10.1016/j.mbs.2009.03.002}
\bibAnnoteFile{chun2009}

\bibitem[{Cseke et~al.(2016)Cseke, Schnoerr, Opper, and
  Sanguinetti}]{cseke2016}
Cseke, B., Schnoerr, D., Opper, M., and Sanguinetti, G. (2016).
\newblock Expectation propagation for continuous time stochastic processes.
\newblock ArXiv:1512.06098v2 [stat.ML]
\bibAnnoteFile{cseke2016}

\bibitem[{Dai and Lai(2010)}]{dai2010}
Dai, Z. and Lai, L. (2010).
\newblock Differential simulated annealing: {A} robust and efficient global
  optimization algorithm for parameter estimation of biological networks.
\newblock \emph{Mol. BioSyst.} 10, 1385--1392.
\newblock \doi{10.1039/c4mb00100a}
\bibAnnoteFile{dai2010}

\bibitem[{Daigle et~al.(2012)Daigle, Roh, Petzold, and Niemi}]{daigle2012}
Daigle, B.~J., Roh, M.~K., Petzold, L.~R., and Niemi, J. (2012).
\newblock Accelerated maximum likelihood parameter estimation for stochastic
  biochemical systems.
\newblock \emph{BMC Bioinf.} 13.
\newblock \doi{10.1186/1471-2105-13-68}
\bibAnnoteFile{daigle2012}

\bibitem[{Dargatz(2010)}]{dargatz2010th}
Dargatz, C. (2010).
\newblock \emph{Bayesian Inference for Diffusion Processes with Applications in
  Life Sciences}.
\newblock Ph.D. thesis, Ludwig Maximilians Universit\"at M\"unchen
\bibAnnoteFile{dargatz2010th}

\bibitem[{Dattner(2015)}]{dattner2015}
Dattner, I. (2015).
\newblock Optimal rate of direct estimators in systems of ordinary differential
  equations linear in functions of the parameters.
\newblock \emph{Electron. J. Statist.} 9, 1939--1973.
\newblock \doi{10.1214/15-EJS1053}
\bibAnnoteFile{dattner2015}

\bibitem[{Deng and Tian(2014)}]{deng2014}
Deng, Z. and Tian, T. (2014).
\newblock A continuous optimization approach for inferring parameters in
  mathematical models of regulatory networks.
\newblock \emph{BMC Bioinf.} 15.
\newblock \doi{10.1186/1471-2105-15-256}
\bibAnnoteFile{deng2014}

\bibitem[{Dey et~al.(2018)Dey, Chakrabarti, Gola, and Sen}]{dey2018}
Dey, A., Chakrabarti, K., Gola, K.~K., and Sen, S. (2018).
\newblock A {Kalman} filter approach for biomolecular systems with noise
  covariance updating.
\newblock ArXiv:1712.02150v2 [q-bio.QM]
\bibAnnoteFile{dey2018}

\bibitem[{Dinh and Sidje(2017)}]{dinh2017}
Dinh, K.~N. and Sidje, R.~B. (2017).
\newblock An application of the krylov-fsp-ssa method to parameter fitting with
  maximum likelihood.
\newblock \emph{Phys. Biol.} 14, 065001.
\newblock \doi{10.1088/1478-3975/aa868a}
\bibAnnoteFile{dinh2017}

\bibitem[{Dochain(2003)}]{dochain2003}
Dochain, D. (2003).
\newblock State observers for processes with uncertain kinetics.
\newblock \emph{Int. J. Control} 76, 1483--1492.
\newblock \doi{10.1080/00207170310001604936}
\bibAnnoteFile{dochain2003}

\bibitem[{Doucet et~al.(2001)Doucet, Freitas, and Gordon}]{doucet2001}
Doucet, A., Freitas, N., and Gordon, N. (2001).
\newblock \emph{Sequential {Monte Carlo} Methods in Practice} (New York:
  Springer-Verlag)
\bibAnnoteFile{doucet2001}

\bibitem[{Drovandi et~al.(2016)Drovandi, Pettitt, and Mccutchan}]{drovandi2016}
Drovandi, C.~C., Pettitt, A.~N., and Mccutchan, R.~A. (2016).
\newblock Exact and approximate {Bayesian} inference for low integer-valued
  time series models with intractable likelihoods.
\newblock \emph{Bayesian Anal.} 11, 325--352.
\newblock \doi{10.1214/15-BA950}
\bibAnnoteFile{drovandi2016}

\bibitem[{Eghtesadi and Mcauley(2014)}]{eghtesadi2014}
Eghtesadi, Z. and Mcauley, K.~B. (2014).
\newblock Mean square error based method for parameter ranking and selection to
  obtain accurate predictions at specified operating conditions.
\newblock \emph{Ind. Eng. Chem. Res.} 53, 6033--6046.
\newblock \doi{10.1021/ie5002444}
\bibAnnoteFile{eghtesadi2014}

\bibitem[{Eisenberg and Hayashi(2014)}]{eisenberg2014}
Eisenberg, M.~C. and Hayashi, M. A.~L. (2014).
\newblock Determining identifiable parameter combinations using subset
  profiling.
\newblock \emph{Math. Biosci.} 256, 116--126.
\newblock \doi{10.1016/j.mbs.2014.08.008}
\bibAnnoteFile{eisenberg2014}

\bibitem[{Emmert-Streib et~al.(2012)Emmert-Streib, Glazko, Altay, and
  Simoes}]{streib2012}
Emmert-Streib, F., Glazko, G.~V., Altay, G., and Simoes, R. D.~M. (2012).
\newblock Statistical inference and reverse engineering of gene regulatory
  networks from observational expression data.
\newblock \emph{Front. Genet.} 3.
\newblock \doi{10.3389/fgene.2012.00008}
\bibAnnoteFile{streib2012}

\bibitem[{Engl et~al.(2009)Engl, Flamm, K\"ugler, Lu, M\"uller, and
  Schuster}]{eng2009}
Engl, H.~W., Flamm, C., K\"ugler, P., Lu, J., M\"uller, S., and Schuster, P.
  (2009).
\newblock Inverse problems in systems biology.
\newblock \emph{Inverse Problems} 25, 1--51.
\newblock \doi{10.1088/0266-5611/25/12/123014}
\bibAnnoteFile{eng2009}

\bibitem[{Erguler and Stumpf(2011)}]{erguler2011}
Erguler, K. and Stumpf, M. P.~H. (2011).
\newblock Practical limits for reverse engineering of dynamical systems: {A}
  statistical analysis of sensitivity and parameter inferability in systems
  biology models.
\newblock \emph{Mol. BioSyst.} 7, 1593--1602.
\newblock \doi{10.1039/c0mb00107d}
\bibAnnoteFile{erguler2011}

\bibitem[{Fages et~al.(2015)Fages, Gay, and Soliman}]{fages2015}
Fages, F., Gay, S., and Soliman, S. (2015).
\newblock Inferring reaction systems from ordinary differential equations.
\newblock \emph{Theor. Comput. Sci.} 599, 64--78.
\newblock \doi{10.1016/j.tcs.2014.07.032}
\bibAnnoteFile{fages2015}

\bibitem[{Famili et~al.(2005)Famili, Mahadevan, and Palsson}]{famili2005}
Famili, I., Mahadevan, R., and Palsson, B.~O. (2005).
\newblock K-cone analysis: {D}etermining all candidate values for kinetic
  parameters on a network scale.
\newblock \emph{Biophys.} 88, 1616--1625.
\newblock \doi{10.1529/biophysj.104.050385}
\bibAnnoteFile{famili2005}

\bibitem[{Farina et~al.(2006)Farina, Findeisen, Bullinger, Bittanti,
  Allg\"ower, and Wellstead}]{farina2006}
Farina, M., Findeisen, R., Bullinger, E., Bittanti, S., Allg\"ower, F., and
  Wellstead, P. (2006).
\newblock Results towards identifiability properties of biochemical reaction
  networks.
\newblock In \emph{ICDC}. 2104--2109.
\newblock \doi{10.1109/CDC.2006.376925}
\bibAnnoteFile{farina2006}

\bibitem[{Farza et~al.(2016)Farza, Menard, Abdennour, and
  M'Saad}]{bouraoui2016}
Farza, I. B.~M., Menard, T., Abdennour, R.~B., and M'Saad, M. (2016).
\newblock On-line estimation of the reaction rates from sampled measurements in
  bioreactors.
\newblock In \emph{IFAC Symp.} vol.~49, 1205--1210.
\newblock \doi{10.1016/j.ifacol.2016.07.375}
\bibAnnoteFile{bouraoui2016}

\bibitem[{Fearnhead et~al.(2014)Fearnhead, Giagos, and
  Sherlock}]{fearnhead2014}
Fearnhead, P., Giagos, V., and Sherlock, C. (2014).
\newblock Inference for reaction networks using the linear noise approximation.
\newblock \emph{Biomet.} 70, 457--466.
\newblock \doi{10.1111/biom.12152}
\bibAnnoteFile{fearnhead2014}

\bibitem[{Fearnhead and Prangle(2012)}]{fearnhead2012}
Fearnhead, P. and Prangle, D. (2012).
\newblock Constructing summary statistics for approximate {Bayesian}
  computation: {S}emi-automatic approximate {Bayesian} computation.
\newblock \emph{J. R. Statist. Soc. B} 74, 419--474.
\newblock \doi{10.1111/j.1467-9868.2011.01010.x}
\bibAnnoteFile{fearnhead2012}

\bibitem[{Fey and Bullinger(2010)}]{fey2010}
Fey, D. and Bullinger, E. (2010).
\newblock Limiting the parameter search space for dynamic models with rational
  kinetics using semi-definite programming.
\newblock In \emph{CAB Symp.} 1--6.
\newblock \doi{10.3182/20100707-3-BE-2012.0088}
\bibAnnoteFile{fey2010}

\bibitem[{Fey et~al.(2008)Fey, Findeisen, and Bullinger}]{fey2008}
Fey, D., Findeisen, R., and Bullinger, E. (2008).
\newblock Parameter estimation in kinetic reaction models using nonlinear
  observers is facilitated by model extensions.
\newblock In \emph{IFAC}. vol.~41, 313--318.
\newblock \doi{10.3182/20080706-5-KR-1001.3521}
\bibAnnoteFile{fey2008}

\bibitem[{Flassig(2014)}]{flassig2014th}
Flassig, R.~J. (2014).
\newblock \emph{Statistical Model Identification: {D}ynamical Processes and
  Large-Scale Networks in Systems Biology}.
\newblock Ph.D. thesis, Otto-von-Guericke-Universit\"at Magdeburg
\bibAnnoteFile{flassig2014th}

\bibitem[{Folia and Rattray(2018)}]{folia2018}
Folia, M.~M. and Rattray, M. (2018).
\newblock Trajectory inference and parameter estimation in stochastic models
  with temporally aggregated data.
\newblock \emph{Stat. Comput.} 28, 1053--1072.
\newblock \doi{10.1007/s11222-017-9779-x}
\bibAnnoteFile{folia2018}

\bibitem[{Fr\"ohlich et~al.(2014)Fr\"ohlich, Hross, Theis, and
  Hasenauer}]{frohlich2014a}
Fr\"ohlich, F., Hross, S., Theis, F.~J., and Hasenauer, J. (2014).
\newblock \emph{Radial Basis Function Approximations of {Bayesian} Parameter
  Posterior Densities for Uncertainty Analysis} (Springer, Cham), chap. CMSB.
\newblock 73--85.
\newblock \doi{10.1007/978-3-319-12982-2_6}
\bibAnnoteFile{frohlich2014a}

\bibitem[{Fr\"ohlich et~al.(2017)Fr\"ohlich, Kaltenbacher, Theis, and
  Hasenauer}]{frohlich2017}
Fr\"ohlich, F., Kaltenbacher, B., Theis, F.~J., and Hasenauer, J. (2017).
\newblock Scalable parameter estimation for genomescale biochemical reaction
  networks.
\newblock \emph{PLOS Comput. Biol.} 13, e1005331.
\newblock \doi{10.1371/journal.pcbi.1005331}
\bibAnnoteFile{frohlich2017}

\bibitem[{Fr\"ohlich et~al.(2016)Fr\"ohlich, Thomas, Kazeroonian, Theis, Grima,
  and Hasenauer}]{frohlich2016}
Fr\"ohlich, F., Thomas, P., Kazeroonian, A., Theis, F.~J., Grima, R., and
  Hasenauer, J. (2016).
\newblock Inference for stochastic chemical kinetics using moment equations and
  system size expansion.
\newblock \emph{PLOS Comput. Biol.} 7.
\newblock \doi{10.1371/journal.pcbi.1005030}
\bibAnnoteFile{frohlich2016}

\bibitem[{G\'abor and Banga(2014)}]{gabor2014}
G\'abor, A. and Banga, J.~R. (2014).
\newblock \emph{Improved Parameter Estimation in Kinetic Models: {S}election
  And Tuning of Regularization Methods} (Springer, Cham), chap. CMSB.
\newblock 45--60.
\newblock \doi{10.1007/978-3-319-12982-2_4}
\bibAnnoteFile{gabor2014}

\bibitem[{G\'abor et~al.(2017)G\'abor, Villaverde, and Banga}]{gabor2017}
G\'abor, A., Villaverde, A.~F., and Banga, J.~R. (2017).
\newblock Parameter identifiability analysis and visualization in large-scale
  kinetic models of biosystems.
\newblock \emph{BMC Sys. Biol.} 11.
\newblock \doi{10.1186/s12918-017-0428-y}
\bibAnnoteFile{gabor2017}

\bibitem[{Galagali(2016)}]{galagali2016th}
Galagali, N. (2016).
\newblock \emph{Bayesian Inference of Chemical Reaction Networks}.
\newblock Ph.D. thesis, Massachusetts Inst. Techn.
\bibAnnoteFile{galagali2016th}

\bibitem[{Geffen et~al.(2008)Geffen, Findeisen, Schliemann, Allg\"ower, and
  Guay}]{geffen2008}
Geffen, D., Findeisen, R., Schliemann, M., Allg\"ower, F., and Guay, M. (2008).
\newblock Observability based parameter identifiability for biochemical
  reaction networks.
\newblock In \emph{ACC}. 2130--2135.
\newblock \doi{10.1109/ACC.2008.4586807}
\bibAnnoteFile{geffen2008}

\bibitem[{Gennemark and Wedelin(2007)}]{gennemark2007}
Gennemark, P. and Wedelin, D. (2007).
\newblock Efficient algorithms for ordinary differential equation model
  identification of biological systems.
\newblock \emph{IET Syst. Biol.} 1, 120--129.
\newblock \doi{10.1049/iet-syb:20050098}
\bibAnnoteFile{gennemark2007}

\bibitem[{Georgieva et~al.(2016)Georgieva, Bouaynaya, Silva, Mihaylova, and
  Jain}]{georgieva2016}
Georgieva, P., Bouaynaya, N., Silva, F., Mihaylova, L., and Jain, L.~C. (2016).
\newblock A beamformer-particle filter framework for localization of correlated
  eeg sources.
\newblock \emph{IEEE J. Biomed. Health Inf.} 20, 880--892.
\newblock \doi{10.1109/JBHI.2015.2413752}
\bibAnnoteFile{georgieva2016}

\bibitem[{Ghusinga et~al.(2017)Ghusinga, Vargas-Garcia, Lamperski, and
  Singh}]{ghusinga2017}
Ghusinga, K.~R., Vargas-Garcia, C.~A., Lamperski, A., and Singh, A. (2017).
\newblock Exact lower and upper bounds on stationary moments in stochastic
  biochemical systems.
\newblock \emph{Phys. Biol.} 14.
\newblock \doi{10.1088/1478-3975/aa75c6}
\bibAnnoteFile{ghusinga2017}

\bibitem[{Gillespie and Golightly(2012)}]{gillespie2014}
Gillespie, C.~S. and Golightly, A. (2012).
\newblock Bayesian inference for the chemical master equation using approximate
  models.
\newblock In \emph{WCSB}. 23--26
\bibAnnoteFile{gillespie2014}

\bibitem[{Golightly et~al.(2018)Golightly, Bradley, Lowe, and
  Gillespie}]{golightly2018}
Golightly, A., Bradley, E., Lowe, T., and Gillespie, C.~S. (2018).
\newblock Correlated pseudo-marginal schemes for time-discretised stochastic
  kinetic models.
\newblock ArXiv:1802.07148v2 [stat.CO]
\bibAnnoteFile{golightly2018}

\bibitem[{Golightly et~al.(2012)Golightly, Henderson, and
  Sherlock}]{golightly2012}
Golightly, A., Henderson, D.~A., and Sherlock, C. (2012).
\newblock Efficient particle {MCMC} for exact inference in stochastic
  biochemical network models through approximation of expensive likelihoods.
\newblock Newcastle Univ.
\bibAnnoteFile{golightly2012}

\bibitem[{Golightly et~al.(2015)Golightly, Henderson, and
  Sherlock}]{golightly2015}
Golightly, A., Henderson, D.~A., and Sherlock, C. (2015).
\newblock Delayed acceptance particle {MCMC} for exact inference in stochastic
  kinetic models.
\newblock \emph{Stat. Comput.} 25, 1039--1055.
\newblock \doi{10.1007/s11222-014-9469-x}
\bibAnnoteFile{golightly2015}

\bibitem[{Golightly and Kypraios(2017)}]{golightly2017}
Golightly, A. and Kypraios, T. (2017).
\newblock Efficient {SMC2} schemes for stochastic kinetic models.
\newblock \emph{Stat. Comput.} 28, 1215--1230.
\newblock \doi{10.1007/s11222-017-9789-8}
\bibAnnoteFile{golightly2017}

\bibitem[{Golightly and Wilkinson(2005)}]{golightly2005a}
Golightly, A. and Wilkinson, D.~J. (2005).
\newblock Bayesian inference for stochastic kinetic models using a diffusion
  approximation.
\newblock \emph{Biomet.} 61, 781--788.
\newblock \doi{10.1111/j.1541-0420.2005.00345.x}
\bibAnnoteFile{golightly2005a}

\bibitem[{Golightly and Wilkinson(2006)}]{golightly2005}
Golightly, A. and Wilkinson, D.~J. (2006).
\newblock Bayesian sequential inference for stochastic kinetic biochemical
  network models.
\newblock \emph{Comput. Biol.} 13, 838--851.
\newblock \doi{10.1089/cmb.2006.13.838}
\bibAnnoteFile{golightly2005}

\bibitem[{Golightly and Wilkinson(2011)}]{golightly2011}
Golightly, A. and Wilkinson, D.~J. (2011).
\newblock Bayesian parameter inference for stochastic biochemical network
  models using particle {Markov Chain Monte Carlo}.
\newblock \emph{Interf. Focus} 1, 807--820.
\newblock \doi{10.1098/rsfs.2011.0047}
\bibAnnoteFile{golightly2011}

\bibitem[{Golightly and Wilkinson(2014)}]{golightly2014}
Golightly, A. and Wilkinson, D.~J. (2014).
\newblock Bayesian inference for {Markov} jump processes with informative
  observations.
\newblock ArXiv:1409.4362v1 [stat.CO]
\bibAnnoteFile{golightly2014}

\bibitem[{Gonz\'alez et~al.(2013)Gonz\'alez, Vujac\v{c}i\'c, and
  Wit}]{gonzlez2013}
Gonz\'alez, J., Vujac\v{c}i\'c, I., and Wit, E. (2013).
\newblock Inferring latent gene regulatory network kinetics.
\newblock \emph{Stat. Appl. Gen. Mol. Biol.} 12, 109--127.
\newblock \doi{10.1515/sagmb-2012-0006}
\bibAnnoteFile{gonzlez2013}

\bibitem[{Gordon et~al.(1993)Gordon, Salmond, and Smith}]{gordon1993}
Gordon, N., Salmond, D., and Smith, A. (1993).
\newblock A novel approach to nonlinear/non-{G}aussian {B}ayesian state
  estimation.
\newblock \emph{IEE Pro.-{F}} 140, 107--113.
\newblock \doi{10.1049/ip-f-2.1993.0015}
\bibAnnoteFile{gordon1993}

\bibitem[{Goutsias and Jenkinson(2012)}]{goutsias2012}
Goutsias, J. and Jenkinson, G. (2012).
\newblock {Markov}ian dynamics on complex reaction networks.
\newblock ArXiv:1205.5524v3 [math-ph]
\bibAnnoteFile{goutsias2012}

\bibitem[{Gratie et~al.(2013)Gratie, Iancu, and Petre}]{gratie2013}
Gratie, D.-E., Iancu, B., and Petre, I. (2013).
\newblock \emph{{ODE} Analysis of Biological Systems} (Springer, Berlin,
  Heidelberg), chap. Formal Methods for Dynamical Systems.
\newblock 29--62.
\newblock \doi{10.1007/978-3-642-38874-3_2}
\bibAnnoteFile{gratie2013}

\bibitem[{Guill\'en-Gos\'albez et~al.(2013)Guill\'en-Gos\'albez, Mir\'o, Alves,
  Sorribas, and Jim\'enez}]{gosalbez2013}
Guill\'en-Gos\'albez, G., Mir\'o, A., Alves, R., Sorribas, A., and Jim\'enez,
  L. (2013).
\newblock Identification of regulatory structure and kinetic parameters of
  biochemical networks via mixed-integer dynamic optimization.
\newblock \emph{BMC Syst. Biol.} 7.
\newblock \doi{10.1186/1752-0509-7-113}
\bibAnnoteFile{gosalbez2013}

\bibitem[{Gupta(2013)}]{gupta2013th}
Gupta, A. (2013).
\newblock \emph{Parameter Estimation in Deterministic And Stochastic Models of
  Biological Systems}.
\newblock Ph.D. thesis, Univ. Wisconsin-Madison
\bibAnnoteFile{gupta2013th}

\bibitem[{Gupta and Rawlings(2014)}]{gupta2014}
Gupta, A. and Rawlings, J.~B. (2014).
\newblock Comparison of parameter estimation methods in stochastic chemical
  kinetic models: {E}xamples in systems biology.
\newblock \emph{AIChE} 60, 1253--1268.
\newblock \doi{10.1002/aic.14409}
\bibAnnoteFile{gupta2014}

\bibitem[{Hagen et~al.(2013)Hagen, White, and Tidor}]{hagen2018}
Hagen, D.~R., White, J.~K., and Tidor, B. (2013).
\newblock Convergence in parameters and predictions using computational
  experimental design.
\newblock \emph{Interf. Focus} 3.
\newblock \doi{10.1098/rsfs.2013.0008}
\bibAnnoteFile{hagen2018}

\bibitem[{Hasenauer(2013)}]{hasenauer2013th}
Hasenauer, J. (2013).
\newblock \emph{Modeling And Parameter Estimation for Heterogeneous Cell
  Populations}.
\newblock Ph.D. thesis, Univ. Stuttgart
\bibAnnoteFile{hasenauer2013th}

\bibitem[{Hasenauer et~al.(2010)Hasenauer, Waldherr, Wagner, and
  Allg\"ower}]{hasenauer2010}
Hasenauer, J., Waldherr, S., Wagner, K., and Allg\"ower, F. (2010).
\newblock Parameter identiﬁcation, experimental design and model
  falsiﬁcation for biological network models using semideﬁnite programming.
\newblock \emph{IET Syst. Biol.} 4, 119--130.
\newblock \doi{10.1049/iet-syb.2009.0030}
\bibAnnoteFile{hasenauer2010}

\bibitem[{He et~al.(2004)He, Sosonkina, Shaffer, Tyson, Watson, and
  Zwolak}]{sosonkina2004}
He, J., Sosonkina, M., Shaffer, C.~A., Tyson, J.~J., Watson, L.~T., and Zwolak,
  J.~W. (2004).
\newblock A hierarchical parallel scheme for global parameter estimation in
  systems biology.
\newblock In \emph{Paral. Distr. Comput. Symp.} 42--50.
\newblock \doi{10.1109/IPDPS.2004.1302958}
\bibAnnoteFile{sosonkina2004}

\bibitem[{Hussain(2016)}]{hussain2016th}
Hussain, F. (2016).
\newblock \emph{Techniques for Automated Parameter Estimation in Computational
  Models of Probabilistic Systems}.
\newblock Ph.D. thesis, Univ. Central Florida
\bibAnnoteFile{hussain2016th}

\bibitem[{Hussain et~al.(2015)Hussain, Langmead, Mi, Dutta-Moscato, Vodovotz,
  and Jha}]{hussain2015}
Hussain, F., Langmead, C.~J., Mi, Q., Dutta-Moscato, J., Vodovotz, Y., and Jha,
  S.~K. (2015).
\newblock Automated parameter estimation for biological models using {Bayesian}
  statistical model checking.
\newblock \emph{BMC Bioinf.} 16, PMC4674867.
\newblock \doi{10.1186/1471-2105-16-S17-S8}
\bibAnnoteFile{hussain2015}

\bibitem[{Iwata et~al.(2014)Iwata, Sriyudthsak, Hirai, and
  Shiraishi}]{iwata2014}
Iwata, M., Sriyudthsak, K., Hirai, M.~Y., and Shiraishi, F. (2014).
\newblock Estimation of kinetic parameters in an {S}-system equation model for
  a metabolic reaction system using the {N}ewton-{R}aphson method.
\newblock \emph{Math. Biosci.} 248, 11--21.
\newblock \doi{10.1016/j.mbs.2013.11.002}
\bibAnnoteFile{iwata2014}

\bibitem[{J.~O.~Ramsay and Cao(2007)}]{ramsay2007}
J.~O.~Ramsay, D.~C., G.~Hooker and Cao, J. (2007).
\newblock Parameter estimation for differential equations: {A} generalized
  smoothing approach.
\newblock \emph{J. R. Statist. Soc. B} 69, 741--796.
\newblock \doi{10.1111/j.1467-9868.2007.00610.x}
\bibAnnoteFile{ramsay2007}

\bibitem[{Jagiella et~al.(2017)Jagiella, Rickert, Theis, and
  Hasenauer}]{jagiella2017}
Jagiella, N., Rickert, D., Theis, F.~J., and Hasenauer, J. (2017).
\newblock Parallelization and high-performance computing enables automated
  statistical inference of multiscale models.
\newblock \emph{Cell Systems} 4, 194--206.
\newblock \doi{10.1016/j.cels.2016.12.002}
\bibAnnoteFile{jagiella2017}

\bibitem[{Jaime and Denis(2015)}]{moreno2005}
Jaime, M. and Denis, D. (2015).
\newblock Global observability and detectability analysis of uncertain reaction
  systems.
\newblock In \emph{IFAC World Cong.} 1062--1070.
\newblock \doi{10.1080/00207170701636534}
\bibAnnoteFile{moreno2005}

\bibitem[{Jang et~al.(2016)Jang, Kim, Braatz, Gopaluni, and Lee}]{jang2016}
Jang, H., Kim, K.-K.~K., Braatz, R.~D., Gopaluni, R.~B., and Lee, J.~H. (2016).
\newblock Regularized maximum likelihood estimation of sparse stochastic
  monomolecular biochemical reaction networks.
\newblock \emph{Comput. Chem. Eng.} 90, 111--120.
\newblock \doi{10.1016/j.compchemeng.2016.03.018}
\bibAnnoteFile{jang2016}

\bibitem[{Jaqaman and Danuser(2006)}]{jaqaman2006}
Jaqaman, K. and Danuser, G. (2006).
\newblock Linking data to models: {D}ata regression.
\newblock \emph{Nat. Rev. Mol. Cell Biol.} 7, 813--819.
\newblock \doi{10.1038/nrm2030}
\bibAnnoteFile{jaqaman2006}

\bibitem[{Ji and Brown(2009)}]{ji2009}
Ji, Z. and Brown, M. (2009).
\newblock Joint state and parameter estimation for biochemical dynamic pathways
  with iterative extended {Kalman} filter: {C}omparison with dual state and
  parameter estimation.
\newblock \emph{Open Aut. Contr. Syst.} 2, 69--77.
\newblock \doi{10.2174/1874444300902010069}
\bibAnnoteFile{ji2009}

\bibitem[{Jia et~al.(2011)Jia, Stephanopoulos, and Gunawan}]{jia2011}
Jia, G., Stephanopoulos, G.~N., and Gunawan, R. (2011).
\newblock Parameter estimation of kinetic models from metabolic profiles:
  {T}wo-phase dynamic decoupling method.
\newblock \emph{Syst. Biol.} 27, 1964--1970.
\newblock \doi{10.1093/bioinformatics/btr293}
\bibAnnoteFile{jia2011}

\bibitem[{Joshia et~al.(2006)Joshia, Morgensterna, and Kremlingb}]{joshi2006}
Joshia, M., Morgensterna, A.~S., and Kremlingb, A. (2006).
\newblock Exploiting the bootstrap method for quantifying parameter confidence
  intervals in dynamical systems.
\newblock \emph{Metab. Eng.} 8, 447--455.
\newblock \doi{10.1016/j.ymben.2006.04.003}
\bibAnnoteFile{joshi2006}

\bibitem[{Jr. et~al.(2012)Jr., Roh, Wheeler, Petzold, and Niemi}]{daigle20xx}
Jr., B. J.~D., Roh, M.~K., Wheeler, M., Petzold, L.~R., and Niemi, J. (2012).
\newblock Accelerated maximum likelihood estimation for stochastic biochemical
  systems.
\newblock \emph{BMC Bioinf.} 13.
\newblock \doi{10.1186/1471-2105-13-68}
\bibAnnoteFile{daigle20xx}

\bibitem[{Karimi and Mcauley(2013)}]{karimi2013}
Karimi, H. and Mcauley, K.~B. (2013).
\newblock An approximate expectation maximization algorithm for estimating
  parameters, noise variances, and stochastic disturbance intensities in
  nonlinear dynamic models.
\newblock \emph{Ind. Eng. Chem. Res.} 52, 18303--18323.
\newblock \doi{10.1021/ie4023989}
\bibAnnoteFile{karimi2013}

\bibitem[{Karimi and Mcauley(2014{\natexlab{a}})}]{karimi2014a}
Karimi, H. and Mcauley, K.~B. (2014{\natexlab{a}}).
\newblock An approximate expectation maximisation algorithm for estimating
  parameters in nonlinear dynamic models with process disturbances.
\newblock \emph{Can. J. Chem. Eng.} 92, 835--850.
\newblock \doi{10.1002/cjce.21932}
\bibAnnoteFile{karimi2014a}

\bibitem[{Karimi and Mcauley(2014{\natexlab{b}})}]{karimi2014}
Karimi, H. and Mcauley, K.~B. (2014{\natexlab{b}}).
\newblock A maximum-likelihood method for estimating parameters, stochastic
  disturbance intensities and measurement noise variances in nonlinear dynamic
  models with process disturbances.
\newblock \emph{Comp. Chem. Eng.} 67, 178--198.
\newblock \doi{10.1016/j.compchemeng.2014.04.007}
\bibAnnoteFile{karimi2014}

\bibitem[{Karnaukhov et~al.(2007)Karnaukhov, Karnaukhova, and
  Williamson}]{kamaukhov2007}
Karnaukhov, A.~V., Karnaukhova, E.~V., and Williamson, J.~R. (2007).
\newblock Numerical matrices method for nonlinear system identification and
  description of dynamics of biochemical reaction networks.
\newblock \emph{Biophys.} 92, 3459--3473.
\newblock \doi{10.1529/biophysj.106.093344}
\bibAnnoteFile{kamaukhov2007}

\bibitem[{Kay(1993)}]{kay93}
Kay, S.~M. (1993).
\newblock \emph{Fundamentals of Statistical Signal Processing: {E}stimation
  Theory}, vol.~I (Prentice Hall)
\bibAnnoteFile{kay93}

\bibitem[{Kay(1998)}]{kay98}
Kay, S.~M. (1998).
\newblock \emph{Fundamentals of Statistical Signal Processing, Volume II:
  {D}etection Theory}, vol.~II (Prentice Hall)
\bibAnnoteFile{kay98}

\bibitem[{Kimura et~al.(2015)Kimura, Celani, Nagao, Stasevich, and
  Nakamura}]{kimura2015}
Kimura, A., Celani, A., Nagao, H., Stasevich, T., and Nakamura, K. (2015).
\newblock Estimating cellular parameters through optimization procedures:
  {E}lementary principles and applications.
\newblock \emph{Front. Physiol.} 6.
\newblock \doi{10.3389/fphys.2015.00060}
\bibAnnoteFile{kimura2015}

\bibitem[{Klein et~al.(2011)Klein, Mihaylova, and Nour-Eddin}]{klein2011}
Klein, L., Mihaylova, L., and Nour-Eddin, E.-F. (2011).
\newblock Sensor and data fusion: {T}axonomy, challenges and applications.
\newblock In \emph{Handbook on Soft Computing for Video Surveillance}, eds.
  S.~Pal, A.~Petrosino, and L.~Maddalena (Taylor and Francis). 139--183.
\newblock \doi{10.1201/b11631-7}
\bibAnnoteFile{klein2011}

\bibitem[{Kleinstein et~al.(2006)Kleinstein, Bottino, Georgieva, and
  Sarangapani}]{kleinstein2006}
Kleinstein, S.~H., Bottino, D., Georgieva, A., and Sarangapani, R. (2006).
\newblock Nonuniform sampling for global optimization of kinetic rate constants
  in biological pathways.
\newblock In \emph{Winter Simul. Conf.} 1611--1616.
\newblock \doi{10.1109/WSC.2006.322934}
\bibAnnoteFile{kleinstein2006}

\bibitem[{Ko et~al.(2009)Ko, Voit, and Wang}]{ko2009}
Ko, C.~L., Voit, E.~O., and Wang, F.-S. (2009).
\newblock Estimating parameters for generalized mass action models with
  connectivity information.
\newblock \emph{BMC Bioinf.} 10.
\newblock \doi{10.1186/1471-2105-10-140}
\bibAnnoteFile{ko2009}

\bibitem[{Koblents and M\'iguez(2011)}]{koblents2011}
Koblents, E. and M\'iguez, J. (2011).
\newblock A population {Monte Carlo} method for {Bayesian} inference and its
  application to stochastic kinetic models.
\newblock In \emph{EUSIPCO}. 679--683
\bibAnnoteFile{koblents2011}

\bibitem[{Koblents and M\'iguez(2014)}]{koblents2014}
Koblents, E. and M\'iguez, J. (2014).
\newblock A comparison of nonlinear population {Monte Carlo} and particle
  {Markov Chain Monte Carlo} algorithms for {Bayesian} inference in stochastic
  kinetic models.
\newblock ArXiv:1404.5218v1 [stat.ME]
\bibAnnoteFile{koblents2014}

\bibitem[{Koeppl et~al.(2010)Koeppl, Setti, Pelet, Mangia, Petrov, and
  Peter}]{koeppl2010}
Koeppl, H., Setti, G., Pelet, S., Mangia, M., Petrov, T., and Peter, M. (2010).
\newblock Probability metrics to calibrate stochastic chemical kinetics.
\newblock In \emph{ISCS}. 541--544.
\newblock \doi{10.1109/ISCAS.2010.5537549}
\bibAnnoteFile{koeppl2010}

\bibitem[{Koeppl et~al.(2012)Koeppl, Zechner, Ganguly, Pelet, and
  Peter}]{koeppl2012}
Koeppl, H., Zechner, C., Ganguly, A., Pelet, S., and Peter, M. (2012).
\newblock Accounting for extrinsic variability in the estimation of stochastic
  rate constants.
\newblock \emph{Int. J. Robust. Nonlinear Contr.} 22, 1103--1119.
\newblock \doi{10.1002/rnc.2804}
\bibAnnoteFile{koeppl2012}

\bibitem[{Komorowski et~al.(2011)Komorowski, Costa, Rand, and
  Stumpf}]{komorowski2011}
Komorowski, M., Costa, M.~J., Rand, D.~A., and Stumpf, M. P.~H. (2011).
\newblock Sensitivity, robustness, and identifiability in stochastic chemical
  kinetics models.
\newblock \emph{PNAS} 108, 8645--8650.
\newblock \doi{10.1073/pnas.1015814108}
\bibAnnoteFile{komorowski2011}

\bibitem[{Komorowski et~al.(2009)Komorowski, Finkenst\"adt, Harper, and
  Rand}]{komorowski2009}
Komorowski, M., Finkenst\"adt, B., Harper, C.~V., and Rand, D.~A. (2009).
\newblock Bayesian inference of biochemical kinetic parameters using the linear
  noise approximation.
\newblock \emph{BMC Bioinf.} 10.
\newblock \doi{10.1186/1471-2105-10-343}
\bibAnnoteFile{komorowski2009}

\bibitem[{Kravaris et~al.(2013)Kravaris, Hahn, and Chu}]{kravaris2013}
Kravaris, C., Hahn, J., and Chu, Y. (2013).
\newblock Advances and selected recent developments in state and parameter
  estimation.
\newblock \emph{Comput. Chem. Eng.} 51, 111--123.
\newblock \doi{10.1016/j.compchemeng.2012.06.001}
\bibAnnoteFile{kravaris2013}

\bibitem[{Kuepfer et~al.(2007)Kuepfer, Sauer, and Parrilo}]{kuepfer2007a}
Kuepfer, L., Sauer, U., and Parrilo, P.~A. (2007).
\newblock Methodology article efficient classification of complete parameter
  regions based on semidefinite programming.
\newblock \emph{BMC Bioinf.} 8.
\newblock \doi{10.1186/1471-2105-8-12}
\bibAnnoteFile{kuepfer2007a}

\bibitem[{K\"ugler(2012)}]{kugler2012}
K\"ugler, P. (2012).
\newblock Moment fitting for parameter inference in repeatedly and partially
  observed stochastic biological models.
\newblock \emph{PLOS One} 7, e43001.
\newblock \doi{10.1371/journal.pone.0043001}
\bibAnnoteFile{kugler2012}

\bibitem[{Kulikov and Kulikova(2017)}]{kulikov2017}
Kulikov, G. and Kulikova, M. (2017).
\newblock Accurate state estimation of stiff continuous-time stochastic models
  in chemical and other engineering.
\newblock \emph{Math. Comput. Simul.} 142, 62--81.
\newblock \doi{10.1016/j.matcom.2017.04.006}
\bibAnnoteFile{kulikov2017}

\bibitem[{Kulikov and Kulikova(2015{\natexlab{a}})}]{kulikov2015}
Kulikov, G.~Y. and Kulikova, M.~V. (2015{\natexlab{a}}).
\newblock High-order accurate continuous-discrete extended {Kalman} ﬁlter for
  chemical engineering.
\newblock \emph{Europ. J. Control.} 21, 14--26.
\newblock \doi{10.1016/j.ejcon.2014.11.003}
\bibAnnoteFile{kulikov2015}

\bibitem[{Kulikov and Kulikova(2015{\natexlab{b}})}]{kulikov2015a}
Kulikov, G.~Y. and Kulikova, M.~V. (2015{\natexlab{b}}).
\newblock State estimation in chemical systems with infrequent measurements.
\newblock In \emph{ECC}. 2688--2693.
\newblock \doi{10.1109/ECC.2015.7330944}
\bibAnnoteFile{kulikov2015a}

\bibitem[{Kuntz et~al.(2017)Kuntz, Thomas, Stan, and Barahona}]{kuntz2017}
Kuntz, J., Thomas, P., Stan, G.-B., and Barahona, M. (2017).
\newblock Rigorous bounds on the stationary distributions of the chemical
  master equation via mathematical programming.
\newblock ArXiv:1702.05468v1 [math.PR]
\bibAnnoteFile{kuntz2017}

\bibitem[{Kutalik et~al.(2007)Kutalik, Tucker, and Moulton}]{kutalik2007}
Kutalik, Z., Tucker, W., and Moulton, V. (2007).
\newblock {S}-system parameter estimation for noisy metabolic profiles using
  {N}ewton-flow analysis.
\newblock \emph{IET Syst. Biol.} 1, 174--180.
\newblock \doi{10.1049/iet-syb:20060064}
\bibAnnoteFile{kutalik2007}

\bibitem[{Kuwahara et~al.(2013)Kuwahara, Fan, Wang, and Gao}]{kuwahara2013}
Kuwahara, H., Fan, M., Wang, S., and Gao, X. (2013).
\newblock A framework for scalable parameter estimation of gene circuit models
  using structural information.
\newblock In \emph{ISMB/ECCB}. vol.~29, i98--i107.
\newblock \doi{10.1093/bioinformatics/btt232}
\bibAnnoteFile{kuwahara2013}

\bibitem[{Kyriakopoulos and Wolf(2015)}]{kyriakopoulos2014}
Kyriakopoulos, C. and Wolf, V. (2015).
\newblock \emph{Optimal Observation Time Points in Stochastic Chemical
  Kinetics} (Spriner, Cham), chap. Hybrid Syst. Biol.
\newblock 83--96.
\newblock \doi{10.1007/978-3-319-27656-4_5}
\bibAnnoteFile{kyriakopoulos2014}

\bibitem[{Lakatos(2017)}]{lakatos2017th}
Lakatos, E. (2017).
\newblock \emph{Stochastic analysis and control methods for molecular cell
  biology}.
\newblock Ph.D. thesis, Imperial College London
\bibAnnoteFile{lakatos2017th}

\bibitem[{Lakatos et~al.(2015)Lakatos, Ale, Kirk, and Stumpf}]{lakatos2015}
Lakatos, E., Ale, A., Kirk, P. D.~W., and Stumpf, M. P.~H. (2015).
\newblock Multivariate moment closure techniques for stochastic kinetic models.
\newblock \emph{Chem. Phys.} 143, 1--13.
\newblock \doi{10.1063/1.4929837}
\bibAnnoteFile{lakatos2015}

\bibitem[{Lang and Stelling(2016)}]{lang2016}
Lang, M. and Stelling, J. (2016).
\newblock Modular parameter identification of biomolecular networks.
\newblock \emph{SIAM J. Sci. Comput.} 38, B988--B1008.
\newblock \doi{10.1137/15M103306X}
\bibAnnoteFile{lang2016}

\bibitem[{Lecca et~al.(2009)Lecca, Palmisano, Priami, and
  Sanguinetti}]{lecca2009}
Lecca, P., Palmisano, A., Priami, C., and Sanguinetti, G. (2009).
\newblock A new probabilistic generative model of parameter inference in
  biochemical networks.
\newblock In \emph{SAC}. 758--765.
\newblock \doi{10.1145/1529282.1529442}
\bibAnnoteFile{lecca2009}

\bibitem[{Li and Vu(2013)}]{li2013}
Li, P. and Vu, Q.~D. (2013).
\newblock Identification of parameter correlations for parameter estimation in
  dynamic biological models.
\newblock \emph{BMC Syst. Biol.} 7.
\newblock \doi{10.1186/1752-0509-7-91}
\bibAnnoteFile{li2013}

\bibitem[{Li and Vu(2015)}]{li2015}
Li, P. and Vu, Q.~D. (2015).
\newblock A simple method for identifying parameter correlations in partially
  observed linear dynamic models.
\newblock \emph{BMC Syst. Biol.} 9.
\newblock \doi{10.1186/s12918-015-0234-3}
\bibAnnoteFile{li2015}

\bibitem[{Liao(2017)}]{liao2017th}
Liao, S. (2017).
\newblock \emph{High-Dimensional Problems in Stochastic Modelling of Biological
  Processes}.
\newblock Ph.D. thesis, University of Oxford
\bibAnnoteFile{liao2017th}

\bibitem[{Liao et~al.(2015{\natexlab{a}})Liao, Vejchodsk\'y, and
  Erban}]{liao2015}
Liao, S., Vejchodsk\'y, T., and Erban, R. (2015{\natexlab{a}}).
\newblock Tensor methods for parameter estimation and bifurcation analysis of
  stochastic reaction networks.
\newblock \emph{J. R. Soc. Interface} 12.
\newblock \doi{10.1098/rsif.2015.0233}
\bibAnnoteFile{liao2015}

\bibitem[{Liao et~al.(2015{\natexlab{b}})Liao, Vejchodsky, and
  Erban}]{liao2018}
Liao, S., Vejchodsky, T., and Erban, R. (2015{\natexlab{b}}).
\newblock Tensor methods for parameter estimation and bifurcation analysis of
  stochastic reaction networks.
\newblock ArXiv:1406.7825v2 [q-bio.MN]
\bibAnnoteFile{liao2018}

\bibitem[{Liepe et~al.(2014)Liepe, Kirk, Filippi, Toni, Barnes, and
  Stumpf}]{liepe2014}
Liepe, J., Kirk, P., Filippi, S., Toni, T., Barnes, C.~P., and Stumpf, M.~P.
  (2014).
\newblock A framework for parameter estimation and model selection from
  experimental data in systems biology using approximate {Bayesian}
  computation.
\newblock \emph{Nat. Protoc.} 9, 439--456.
\newblock \doi{10.1038/nprot.2014.025}
\bibAnnoteFile{liepe2014}

\bibitem[{Lillacci and Khammash(2010{\natexlab{a}})}]{lilacci2010}
Lillacci, G. and Khammash, M. (2010{\natexlab{a}}).
\newblock Parameter estimation and model selection in computational biology.
\newblock \emph{PLOS Comp. Biol.} 6, e1000696.
\newblock \doi{10.1371/journal.pcbi.1000696}
\bibAnnoteFile{lilacci2010}

\bibitem[{Lillacci and Khammash(2010{\natexlab{b}})}]{lillacci2010}
Lillacci, G. and Khammash, M. (2010{\natexlab{b}}).
\newblock Parameter identification of biological networks using extended
  {Kalman} filtering and $\chi^2$ criteria.
\newblock In \emph{CDC}. 3367--3372.
\newblock \doi{10.1109/CDC.2010.5717460}
\bibAnnoteFile{lillacci2010}

\bibitem[{Lillacci and Khammash(2012)}]{lillacci2012}
Lillacci, G. and Khammash, M. (2012).
\newblock A distribution-matching method for parameter estimation and model
  selection in computational biology.
\newblock \emph{Int. J. Robust Nonlinear Contr.} 22, 1065--1081.
\newblock \doi{10.1002/rnc.2794}
\bibAnnoteFile{lillacci2012}

\bibitem[{Linder(2013)}]{linder2013th}
Linder, D.~F. (2013).
\newblock \emph{Penalized Least Squares And The Algebraic Statistical Model For
  Biochemical Reaction Networks}.
\newblock Ph.D. thesis, Georgia Regents University
\bibAnnoteFile{linder2013th}

\bibitem[{Lindera and Rempala(2015)}]{linder2014}
Lindera, D.~F. and Rempala, G.~A. (2015).
\newblock Bootstrapping least-squares estimates in biochemical reaction
  networks.
\newblock \emph{Biol. Dynamics} 9, 125--146.
\newblock \doi{10.1080/17513758.2015.1033022}
\bibAnnoteFile{linder2014}

\bibitem[{Liu et~al.(2006)Liu, Farza, and Saad}]{liu2006}
Liu, F., Farza, M., and Saad, M.~M. (2006).
\newblock Unknown input observers design for a class of nonlinear systems -
  application to biochemical processes.
\newblock In \emph{ROCOND}. vol.~39, 131--136
\bibAnnoteFile{liu2006}

\bibitem[{Liu et~al.(2012)Liu, Wu, and Zhang}]{liu2012}
Liu, L.-Z., Wu, F.-X., and Zhang, W.~J. (2012).
\newblock Inference of biological {S}-system using the separable estimation
  method and the genetic algorithm.
\newblock \emph{IEEE/ACM Trans. Comput. Biol. Bioinf.} 9, 955--965.
\newblock \doi{10.1109/TCBB.2011.126}
\bibAnnoteFile{liu2012}

\bibitem[{Liu and Wang(2008{\natexlab{a}})}]{liu2008}
Liu, P.-K. and Wang, F.-S. (2008{\natexlab{a}}).
\newblock Inference of biochemical network models in {S}-system using
  multiobjective optimization approach.
\newblock \emph{Systems Biol.} 24, 1085--1092.
\newblock \doi{10.1093/bioinformatics/btn075}
\bibAnnoteFile{liu2008}

\bibitem[{Liu and Wang(2008{\natexlab{b}})}]{liu2008a}
Liu, P.~K. and Wang, F.-S. (2008{\natexlab{b}}).
\newblock Inverse problems of biological systems using multi-objective
  optimization.
\newblock \emph{J. Chinese Inst. Chem. Eng.} 39, 399--406.
\newblock \doi{10.1016/j.jcice.2008.05.001}
\bibAnnoteFile{liu2008a}

\bibitem[{Liu and Wang(2009)}]{liu2009}
Liu, P.~K. and Wang, F.~S. (2009).
\newblock Hybrid differential evolution with geometric mean mutation in
  parameter estimation of bioreaction systems with large parameter search
  space.
\newblock \emph{Comput. Chem. Eng.} 33, 1851--1860.
\newblock \doi{10.1016/j.compchemeng.2009.05.008}
\bibAnnoteFile{liu2009}

\bibitem[{Liu(2014)}]{liu2014th}
Liu, X. (2014).
\newblock \emph{Probabilistic Inference in Models of Systems Biology}.
\newblock Ph.D. thesis, Univ. of Southampton
\bibAnnoteFile{liu2014th}

\bibitem[{Liu and Gunawan(2014)}]{liu2014}
Liu, Y. and Gunawan, R. (2014).
\newblock Parameter estimation of dynamic biological network models using
  integrated fluxes.
\newblock \emph{BMC Syst. Biol.} 8.
\newblock \doi{10.1186/s12918-014-0127-x}
\bibAnnoteFile{liu2014}

\bibitem[{Loos et~al.(2016)Loos, Fiedler, and Hasenauer}]{loos2016}
Loos, C., Fiedler, A., and Hasenauer, J. (2016).
\newblock \emph{Parameter Estimation for Reaction Rate Equation Constrained
  Mixture Models} (Springer, Cham), chap. CMSB.
\newblock 186--200.
\newblock \doi{10.1007/978-3-319-45177-0_12}
\bibAnnoteFile{loos2016}

\bibitem[{L\"otstedt(2018)}]{lotstedt2018}
L\"otstedt, P. (2018).
\newblock The linear noise approximation for spatially dependent biochemical
  networks.
\newblock \emph{Bull. Math. Biol.} 81, 1--29.
\newblock \doi{10.1007/s11538-018-0428-0}
\bibAnnoteFile{lotstedt2018}

\bibitem[{L\"uck and Wolf(2016)}]{luck2016}
L\"uck, A. and Wolf, V. (2016).
\newblock Generalized method of moments for estimating parameters of stochastic
  reaction networks.
\newblock \emph{BMC Syst. Biol.} 10.
\newblock \doi{10.1186/s12918-016-0342-8}
\bibAnnoteFile{luck2016}

\bibitem[{Mancini et~al.(2015)Mancini, Tronci, Salvo, Mari, Massini, and
  Melatti}]{mancini2015}
Mancini, T., Tronci, E., Salvo, I., Mari, F., Massini, A., and Melatti, I.
  (2015).
\newblock Computing biological model parameters by parallel statistical model
  checking.
\newblock In \emph{IWBBIO}. 542--554.
\newblock \doi{10.1007/978-3-319-16480-9_52}
\bibAnnoteFile{mancini2015}

\bibitem[{Mannakee et~al.(2016)Mannakee, Ragsdale, Transtrum, and
  Gutenkunst}]{mannakee2016}
Mannakee, B.~K., Ragsdale, A.~P., Transtrum, M.~K., and Gutenkunst, R.~N.
  (2016).
\newblock \emph{Sloppiness and the Geometry of Parameter Space} (Springer,
  Cham), chap. Uncertainty in Biology. Studies in Mechanobiology, Tissue
  Engineering and Biomaterials.
\newblock 271--299.
\newblock \doi{10.1007/978-3-319-21296-8_11}
\bibAnnoteFile{mannakee2016}

\bibitem[{Mansouri et~al.(2015)Mansouri, Avci, Nounou, and
  Nounou}]{mansouri2015}
Mansouri, M., Avci, O., Nounou, H., and Nounou, M. (2015).
\newblock Parameter identification for nonlinear biological phenomena modeled
  by {S}-systems.
\newblock In \emph{SSD}. 1--6.
\newblock \doi{10.1109/SSD.2015.7348187}
\bibAnnoteFile{mansouri2015}

\bibitem[{Mansouri et~al.(2014)Mansouri, Nounou, Nounou, and
  Datta}]{mansouri2014}
Mansouri, M.~M., Nounou, H.~N., Nounou, M.~N., and Datta, A.~A. (2014).
\newblock Modeling of nonlinear biological phenomena modeled by {S}-systems.
\newblock \emph{Math. Biosci.} 249, 75--91.
\newblock \doi{10.1016/j.mbs.2014.01.011}
\bibAnnoteFile{mansouri2014}

\bibitem[{Matsubara et~al.(2006)Matsubara, Kikuchi, Sugimoto, and
  Tomita}]{matsubara2006}
Matsubara, Y., Kikuchi, S., Sugimoto, M., and Tomita, M. (2006).
\newblock Parameter estimation for stiff equations of biosystems using radial
  basis function networks.
\newblock \emph{BMC Bioinf.} 7.
\newblock \doi{10.1186/1471-2105-7-230}
\bibAnnoteFile{matsubara2006}

\bibitem[{Mazur(2012)}]{mazur2012th}
Mazur, J. (2012).
\newblock \emph{Bayesian Inference of Gene Regulatory Networks: {F}rom
  Parameter Estimation to Experimental Design}.
\newblock Ph.D. thesis, Ruprecht-Karls Univ.
\bibAnnoteFile{mazur2012th}

\bibitem[{Mazur and Kaderali(2013)}]{mazur2013}
Mazur, J. and Kaderali, L. (2013).
\newblock \emph{the Importance And Challenges of {Bayesian} Parameter Learning
  in Systems Biology} (Springer-Verlag, Berlin), chap. Model Based Param.
  Estim.
\newblock 145--156.
\newblock \doi{10.1007/978-3-642-30367-8 6}
\bibAnnoteFile{mazur2013}

\bibitem[{McGoff et~al.(2015)McGoff, Mukherjee, and Pillai}]{mcgoff2015}
McGoff, K., Mukherjee, S., and Pillai, N. (2015).
\newblock Statistical inference for dynamical systems: {A} review.
\newblock \emph{Stat. Surv.} 9, 209--252.
\newblock \doi{10.1214/15-SS111}
\bibAnnoteFile{mcgoff2015}

\bibitem[{Meskin et~al.(2011)Meskin, , Nounou, Nounou, Datta, and
  Dougherty}]{meskin2011}
Meskin, N., , Nounou, H., Nounou, M., Datta, A., and Dougherty, E.~R. (2011).
\newblock Parameter estimation of biological phenomena modeled by {S}-systems:
  {A}n extended {Kalman} filter approach.
\newblock In \emph{CDC-ECC}. 4424--4429.
\newblock \doi{10.1109/CDC.2011.6160690}
\bibAnnoteFile{meskin2011}

\bibitem[{Meskin et~al.(2013)Meskin, Nounou, Nounou, and Datta}]{meskin2013}
Meskin, N., Nounou, H., Nounou, M., and Datta, A. (2013).
\newblock Parameter estimation of biological phenomena: {A}n unscented {Kalman}
  filter approach.
\newblock \emph{Trans. Comput. Biol. Bioinf.} 10, 537--543.
\newblock \doi{10.1109/TCBB.2013.19}
\bibAnnoteFile{meskin2013}

\bibitem[{Meyer et~al.(2014)Meyer, Cokelaer, Chandran, Kim, Loh, Tucker
  et~al.}]{meyer2014}
Meyer, P., Cokelaer, T., Chandran, D., Kim, K.~H., Loh, P.-R., Tucker, G.,
  et~al. (2014).
\newblock Network topology and parameter estimation: {F}rom experimental design
  methods to gene regulatory network kinetics using a community based approach.
\newblock \emph{BMC Syst. Biol.} 8.
\newblock \doi{10.1186/1752-0509-8-13}
\bibAnnoteFile{meyer2014}

\bibitem[{Michailidis and d’Alch\'e Buc(2013)}]{michailidis2013}
Michailidis, G. and d’Alch\'e Buc, F. (2013).
\newblock Autoregressive models for gene regulatory network inference:
  {S}parsity, stability and causality issues.
\newblock \emph{Math. Biosci.} 246, 326--334.
\newblock \doi{10.1016/j.mbs.2013.10.003}
\bibAnnoteFile{michailidis2013}

\bibitem[{Michalik et~al.(2009)Michalik, Chachuat, and
  Marquardt}]{michalik2009}
Michalik, C., Chachuat, B., and Marquardt, W. (2009).
\newblock Incremental global parameter estimation in dynamical systems.
\newblock \emph{Ind. Eng. Chem. Res.} 48, 5489--5497.
\newblock \doi{10.1021/ie8015472}
\bibAnnoteFile{michalik2009}

\bibitem[{Mihaylova et~al.(2011)Mihaylova, Angelova, Bull, and
  Canagarajah}]{mihaylova2011}
Mihaylova, L., Angelova, D., Bull, D.~R., and Canagarajah, N. (2011).
\newblock Localization of mobile nodes in wireless networks with correlated in
  time measurement noise.
\newblock \emph{IEEE Trans. Mobile Comput.} 10, 44--53.
\newblock \doi{10.1109/TMC.2010.132}
\bibAnnoteFile{mihaylova2011}

\bibitem[{Mihaylova et~al.(2014)Mihaylova, Carmi, Septier, Gning, Pang, and
  Godsill}]{mihaylova2014}
Mihaylova, L., Carmi, A.~Y., Septier, F., Gning, A., Pang, S.~K., and Godsill,
  S. (2014).
\newblock Overview of {Bayesian} sequential {Monte Carlo} methods for group and
  extended object tracking.
\newblock \emph{Dig. Signal Proces.} 25, 1--16.
\newblock \doi{10.1016/j.dsp.2013.11.006}
\bibAnnoteFile{mihaylova2014}

\bibitem[{Mihaylova et~al.(2012)Mihaylova, Hegyi, Gning, and
  Boel}]{mihaylova2012}
Mihaylova, L., Hegyi, A., Gning, A., and Boel, R. (2012).
\newblock Parallelized particle and {Gaussian} sum particle filters for large
  scale freeway traffic systems.
\newblock \emph{IEEE Trans. Intel. Transport. Systems} 13, 36--48.
\newblock \doi{10.1109/TITS.2011.2178833}
\bibAnnoteFile{mihaylova2012}

\bibitem[{Mikeev and Wolf(2012)}]{mikeev2012}
Mikeev, L. and Wolf, V. (2012).
\newblock Parameter estimation for stochastic hybrid models of biochemical
  reaction networks.
\newblock In \emph{HSCC}. 155--166.
\newblock \doi{10.1145/2185632.2185657}
\bibAnnoteFile{mikeev2012}

\bibitem[{Mikelson and Khammash(2016)}]{mikelson2016}
Mikelson, J. and Khammash, M. (2016).
\newblock A parallelizable sampling method for parameter inference of large
  biochemical reaction models.
\newblock ArXiv:1606.08281v1 [q-bio.QM]
\bibAnnoteFile{mikelson2016}

\bibitem[{Milios et~al.(2018)Milios, Sanguinetti, and Schnoerr}]{milios2018}
Milios, D., Sanguinetti, G., and Schnoerr, D. (2018).
\newblock Probabilistic model checking for continuous-time {Markov} chains via
  sequential {Bayesian} inference.
\newblock ArXiv:1711.01863v2 [cs.LO]
\bibAnnoteFile{milios2018}

\bibitem[{Milner et~al.(2013)Milner, Gillespie, and Wilkinson}]{milner2013}
Milner, P., Gillespie, C.~S., and Wilkinson, D.~J. (2013).
\newblock Moment closure based parameter inference of stochastic kinetic
  models.
\newblock \emph{Stat. Comput.} 23.
\newblock \doi{10.1007/s11222-011-9310-8}
\bibAnnoteFile{milner2013}

\bibitem[{Mizera et~al.(2014)Mizera, Pang, and Yuan}]{mizera2014}
Mizera, A., Pang, J., and Yuan, Q. (2014).
\newblock Model-checking based approaches to parameter estimation of gene
  regulatory networks.
\newblock In \emph{Int. Conf. Eng. Compl. Comput. Syst.} 206--209.
\newblock \doi{10.1109/ICECCS.2014.38}
\bibAnnoteFile{mizera2014}

\bibitem[{Moles et~al.(2003)Moles, Mendes, and Banga}]{moles2003}
Moles, C.~G., Mendes, P., and Banga, J.~R. (2003).
\newblock Parameter estimation in biochemical pathways: {A} comparison of
  global optimization methods.
\newblock \emph{Genome Res.} 13, 2467--2474.
\newblock \doi{10.1101/gr.1262503}
\bibAnnoteFile{moles2003}

\bibitem[{Moritz(2014)}]{moritz2014th}
Moritz, L. (2014).
\newblock \emph{Modular Identification and Analysis of Biomolecular Networks}.
\newblock Ph.D. thesis, ETH Zurich
\bibAnnoteFile{moritz2014th}

\bibitem[{Mozgunov et~al.(2018)Mozgunov, Beccuti, Horvath, Jaki, Sirovich, and
  Bibbona}]{mozgunov2018}
Mozgunov, P., Beccuti, M., Horvath, A., Jaki, T., Sirovich, R., and Bibbona, E.
  (2018).
\newblock A review of the deterministic and diffusion approximations for
  stochastic chemical reaction networks.
\newblock \emph{Reac. Kinet. Mech. Cat.} 123, 289--312.
\newblock \doi{10.1007/s11144-018-1351-y}
\bibAnnoteFile{mozgunov2018}

\bibitem[{Mu(2010)}]{mu2010th}
Mu, L. (2010).
\newblock \emph{Parameter Estimation Methods for Biological Systems}.
\newblock Ph.D. thesis, Univ. of Saskatchewan
\bibAnnoteFile{mu2010th}

\bibitem[{M\"uller et~al.(2011)M\"uller, Ramaswamy, and
  Sbalzarini}]{muller2011}
M\"uller, C.~L., Ramaswamy, R., and Sbalzarini, I.~F. (2011).
\newblock Global parameter identification of stochastic reaction networks from
  single trajectories.
\newblock ArXiv:1111.4785v1 [q-bio.MN]
\bibAnnoteFile{muller2011}

\bibitem[{Murakami(2014)}]{murakami2014}
Murakami, Y. (2014).
\newblock Bayesian parameter inference and model selection by population
  annealing in systems biology.
\newblock \emph{PLOS One} 9.
\newblock \doi{10.1371/journal.pone.0104057}
\bibAnnoteFile{murakami2014}

\bibitem[{Mustafa et~al.(2013)Mustafa, Khammash, and Hara}]{hori2013}
Mustafa, Y.~H., Khammash, H., and Hara, S. (2013).
\newblock Efficient parameter identification for stochastic biochemical
  networks using a reduced-order realization.
\newblock In \emph{ECC}. 4154--4159.
\newblock \doi{10.23919/ECC.2013.6669455}
\bibAnnoteFile{hori2013}

\bibitem[{Nemeth et~al.(2014)Nemeth, Fearnhead, and Mihaylova}]{nemeth2014}
Nemeth, C., Fearnhead, P., and Mihaylova, L. (2014).
\newblock {Sequential Monte Carlo} methods for state and parameter estimation
  in abruptly changing environments.
\newblock \emph{IEEE Trans. Signal Proces.} 62, 1245--1255.
\newblock \doi{10.1109/TSP.2013.2296278}
\bibAnnoteFile{nemeth2014}

\bibitem[{Nienaltowski et~al.(2015)Nienaltowski, Wlodarczyk, Lipniacki, and
  Komorowski}]{nienaltowski2015}
Nienaltowski, K., Wlodarczyk, M., Lipniacki, T., and Komorowski, M. (2015).
\newblock Clustering reveals limits of parameter identifiability in
  multi-parameter models of biochemical dynamics.
\newblock \emph{BMC Syst. Biol.} 9.
\newblock \doi{10.1186/s12918-015-0205-8}
\bibAnnoteFile{nienaltowski2015}

\bibitem[{Nim et~al.(2013)Nim, Luo, Cl\'ement, White, and Kellogg}]{nim2013}
Nim, T.~H., Luo, L., Cl\'ement, M.-V., White, J.~K., and Kellogg, L.~T. (2013).
\newblock Systematic parameter estimation in data-rich environments for cell
  signalling dynamics.
\newblock \emph{Syst. Biol.} 29, 1044--1051.
\newblock \doi{10.1093/bioinformatics/btt083}
\bibAnnoteFile{nim2013}

\bibitem[{Nobile et~al.(2013)Nobile, Besozzi, Cazzaniga, Pescini, and
  Mauri}]{nobile2013}
Nobile, M.~S., Besozzi, D., Cazzaniga, P., Pescini, D., and Mauri, G. (2013).
\newblock Reverse engineering of kinetic reaction networks by means of
  cartesian genetic programming and particle swarm optimization.
\newblock In \emph{Congress Evol. Comput.} 1594--1601.
\newblock \doi{10.1109/CEC.2013.6557752}
\bibAnnoteFile{nobile2013}

\bibitem[{Nobile et~al.(2016)Nobile, Tangherloni, Besozzi, and
  Cazzaniga}]{nobile2016}
Nobile, M.~S., Tangherloni, A., Besozzi, D., and Cazzaniga, P. (2016).
\newblock {GPU}-powered and settings-free parameter estimation of biochemical
  systems.
\newblock In \emph{CEC}. 32--39.
\newblock \doi{10.1109/CEC.2016.7743775}
\bibAnnoteFile{nobile2016}

\bibitem[{Pahle et~al.(2012)Pahle, Challenger, Mendes, and McKane}]{pahle2012}
Pahle, J., Challenger, J.~D., Mendes, P., and McKane, A.~J. (2012).
\newblock Biochemical fluctuations, optimisation and the linear noise
  approximation.
\newblock \emph{BMC Syst. Biol.} 6, 1--13.
\newblock \doi{10.1186/1752-0509-6-86}
\bibAnnoteFile{pahle2012}

\bibitem[{Palmisano(2010)}]{palmisano2010th}
Palmisano, A. (2010).
\newblock \emph{Modelling and Inference strategies for Biological Systems}.
\newblock Ph.D. thesis, Universit\`a degli Studi di Trento
\bibAnnoteFile{palmisano2010th}

\bibitem[{Pan and Yang(2010)}]{pan2010}
Pan, S.~J. and Yang, Q. (2010).
\newblock A survey on transfer learning.
\newblock \emph{IEEE Trans. Knowl. Data Eng.} 22, 1345--1359.
\newblock \doi{10.1109/TKDE.2009.191}
\bibAnnoteFile{pan2010}

\bibitem[{Pantazis et~al.(2013)Pantazis, Katsoulakis, and
  Vlachos}]{pantazis2013}
Pantazis, Y., Katsoulakis, M.~A., and Vlachos, D.~G. (2013).
\newblock Parametric sensitivity analysis for biochemical reaction networks
  based on pathwise information theory.
\newblock \emph{BMC Bioinf.} 14, 1--19.
\newblock \doi{10.1186/1471-2105-14-311}
\bibAnnoteFile{pantazis2013}

\bibitem[{Paul(2014)}]{paul2014th}
Paul, D. (2014).
\newblock \emph{Efficient Parameter Inference for Stochastic Chemical
  Kinetics}.
\newblock Ph.D. thesis, Uppsala University
\bibAnnoteFile{paul2014th}

\bibitem[{Penas et~al.(2017)Penas, Gonz\'alez, Egea, Doallo, and
  Banga}]{penas2017}
Penas, D.~R., Gonz\'alez, P., Egea, J.~A., Doallo, R., and Banga, J.~R. (2017).
\newblock Parameter estimation in large-scale systems biology models: {A}
  parallel and self-adaptive cooperative strategy.
\newblock \emph{BMC Bioinf.} 18.
\newblock \doi{10.1186/s12859-016-1452-4}
\bibAnnoteFile{penas2017}

\bibitem[{Plesa et~al.(2017)Plesa, Vejchodsk\'y, and Erban}]{plesa2017}
Plesa, T., Vejchodsk\'y, T., and Erban, R. (2017).
\newblock \emph{Test Models for Statistical Inference: {T}wo-Dimensional
  Reaction Systems Displaying Limit Cycle Bifurcations and Bistability}
  (Springer, Cham), chap. Stochastic Processes, Multiscale Modeling, and
  Numerical Methods for Computational Cellular Biology.
\newblock 3--27.
\newblock \doi{10.1007/978-3-319-62627-7_1}
\bibAnnoteFile{plesa2017}

\bibitem[{Poovathingal and Gunawan(2010)}]{poovathingal2010}
Poovathingal, S.~K. and Gunawan, R. (2010).
\newblock Global parameter estimation methods for stochastic biochemical
  systems.
\newblock \emph{BMC Bioinf.} 11.
\newblock \doi{10.1186/1471-2105-11-414}
\bibAnnoteFile{poovathingal2010}

\bibitem[{Pullen and Morris(2014)}]{pullen2014}
Pullen, N. and Morris, R.~J. (2014).
\newblock Bayesian model comparison and parameter inference in systems biology
  using nested sampling.
\newblock \emph{PLOS One} 9, e88419.
\newblock \doi{10.1371/journal.pone.0088419}
\bibAnnoteFile{pullen2014}

\bibitem[{Quach et~al.(2007)Quach, Brunel, and D'Alch\'e-Buc}]{quach2007}
Quach, M., Brunel, N., and D'Alch\'e-Buc, F. (2007).
\newblock Estimating parameters and hidden variables in non-linear state-space
  models based on odes for biological networks inference.
\newblock \emph{Syst. Biol.} 23, 3209--3216.
\newblock \doi{10.1093/bioinformatics/btm510}
\bibAnnoteFile{quach2007}

\bibitem[{Radulescu et~al.(2012)Radulescu, Gorban, Zinovyev, and
  Noel}]{radulescu2012}
Radulescu, Gorban, A.~N., Zinovyev, A., and Noel, V. (2012).
\newblock Reduction of dynamical biochemical reactions networks in
  computational biology.
\newblock \emph{Front. Genet.} 3, 1--17.
\newblock \doi{10.3389/fgene.2012.00131}
\bibAnnoteFile{radulescu2012}

\bibitem[{Rakhshania et~al.(2016)Rakhshania, Dehghanianb, and
  Rahatia}]{rakhshani2016}
Rakhshania, H., Dehghanianb, E., and Rahatia, A. (2016).
\newblock Hierarchy cuckoo search algorithm for parameter estimation in
  biological systems.
\newblock \emph{Chemomet. Intel. Lab. Syst.} 159, 97--107.
\newblock \doi{10.1016/j.chemolab.2016.10.011}
\bibAnnoteFile{rakhshani2016}

\bibitem[{Rapaport and Dochain(2005)}]{rapaport2005}
Rapaport, A. and Dochain, D. (2005).
\newblock Interval observers for biochemical processes with uncertain kinetics
  and inputs.
\newblock \emph{Math. Biosci.} 193, 235--253.
\newblock \doi{10.1016/j.mbs.2004.07.004}
\bibAnnoteFile{rapaport2005}

\bibitem[{Reinker et~al.(2006)Reinker, Altman, and Timmer}]{reinker2006}
Reinker, S., Altman, R., and Timmer, J. (2006).
\newblock Parameter estimation in stochastic biochemical reactions.
\newblock \emph{IEE Proc. Syst. Biol.} 153, 168--178.
\newblock \doi{10.1049/ip-syb:20050105}
\bibAnnoteFile{reinker2006}

\bibitem[{Reis et~al.(2018)Reis, Kromer, and Klipp}]{reis2018}
Reis, M., Kromer, J.~A., and Klipp, E. (2018).
\newblock General solution of the chemical master equation and modality of
  marginal distributions for hierarchic first-order reaction networks.
\newblock \emph{J. Math. Biol.} 77, 377--419.
\newblock \doi{10.1007/s00285-018-1205-2}
\bibAnnoteFile{reis2018}

\bibitem[{Remlia et~al.(2017)Remlia, Derisb, Mohamada, Omatuc, and
  Corchadod}]{remli2017}
Remlia, M.~A., Derisb, S., Mohamada, M.~S., Omatuc, S., and Corchadod, J.~M.
  (2017).
\newblock An enhanced scatter search with combined opposition-based learning
  for parameter estimation in large-scale kinetic models of biochemical
  systems.
\newblock \emph{Eng. Appl. AI} 62, 164--180.
\newblock \doi{10.1016/j.engappai.2017.04.004}
\bibAnnoteFile{remli2017}

\bibitem[{Rempala(2012)}]{rempala2012}
Rempala, G.~A. (2012).
\newblock Least squares estimation in stochastic biochemical networks.
\newblock \emph{Bull. Math. Biol.} 74, 1938--1955.
\newblock \doi{10.1007/s11538-012-9744-y}
\bibAnnoteFile{rempala2012}

\bibitem[{Revell and Zuliani(2018)}]{revell2018}
Revell, J. and Zuliani, P. (2018).
\newblock \emph{Stochastic Rate Parameter Inference Using the Cross-Entropy
  Method} (Springer, Cham), chap. CMSB.
\newblock 146--164.
\newblock \doi{10.1007/978-3-319-99429-1_9}
\bibAnnoteFile{revell2018}

\bibitem[{Rodriguez-Fernandez et~al.(2006{\natexlab{a}})Rodriguez-Fernandez,
  Egea, and Banga}]{fernandez2006}
Rodriguez-Fernandez, M., Egea, J.~A., and Banga, J.~R. (2006{\natexlab{a}}).
\newblock Novel metaheuristic for parameter estimation in nonlinear dynamic
  biological systems.
\newblock \emph{BMC Bioinf.} 7.
\newblock \doi{10.1186/1471-2105-7-483}
\bibAnnoteFile{fernandez2006}

\bibitem[{Rodriguez-Fernandez et~al.(2006{\natexlab{b}})Rodriguez-Fernandez,
  Mendes, and Banga}]{fernandez2005}
Rodriguez-Fernandez, M., Mendes, P., and Banga, J.~R. (2006{\natexlab{b}}).
\newblock A hybrid approach for efficient and robust parameter estimation in
  biochemical pathways.
\newblock \emph{BioSyst.} 83, 248--265.
\newblock \doi{10.1016/j.biosystems.2005.06.016}
\bibAnnoteFile{fernandez2005}

\bibitem[{Rodriguez-Fernandez et~al.(2013)Rodriguez-Fernandez, Rehberg,
  Kremling, and Banga}]{fernandez2013}
Rodriguez-Fernandez, M., Rehberg, M., Kremling, A., and Banga, J.~R. (2013).
\newblock Simultaneous model discrimination and parameter estimation in dynamic
  models of cellular systems.
\newblock \emph{BMC Syst. Biol.} 7.
\newblock \doi{10.1186/1752-0509-7-76}
\bibAnnoteFile{fernandez2013}

\bibitem[{Rosati et~al.(2018)Rosati, Madec, Kammerer, H\'ébrard, Lallement,
  and Haiech}]{rosati2018}
Rosati, E., Madec, M., Kammerer, J.-B., H\'ébrard, L., Lallement, C., and
  Haiech, J. (2018).
\newblock Efficient modeling and simulation of space-dependent biological
  systems.
\newblock \emph{Comput. Biol.} 25, 917--933.
\newblock \doi{10.1089/cmb.2018.0012}
\bibAnnoteFile{rosati2018}

\bibitem[{Ruess(2014)}]{ruess2014th}
Ruess, J. (2014).
\newblock \emph{Moment-Based Methods for the Analysis and Identification of
  Stochastic Models Of Biochemical Reaction Networks}.
\newblock Ph.D. thesis, ETH
\bibAnnoteFile{ruess2014th}

\bibitem[{Ruess and Lygeros(2015)}]{ruess2015}
Ruess, J. and Lygeros, J. (2015).
\newblock Moment-based methods for parameter inference and experiment design
  for stochastic biochemical reaction networks.
\newblock \emph{ACM Trans. Model. Comput. Simul.} 25, Article 8.
\newblock \doi{10.1145/2688906}
\bibAnnoteFile{ruess2015}

\bibitem[{Ruess et~al.(2011)Ruess, Milias-Argeitis, Summers, and
  Lygeros}]{ruess2011}
Ruess, J., Milias-Argeitis, A., Summers, S., and Lygeros, J. (2011).
\newblock Moment estimation for chemically reacting systems by extended
  {Kalman} filtering.
\newblock \emph{Chem. Phys.} 135.
\newblock \doi{10.1063/1.3654135}
\bibAnnoteFile{ruess2011}

\bibitem[{Rumschinski et~al.(2010)Rumschinski, Borchers, Bosio, Weismantel, and
  Findeisen}]{rumschinski2010}
Rumschinski, P., Borchers, S., Bosio, S., Weismantel, R., and Findeisen, R.
  (2010).
\newblock Set-base dynamical parameter estimation and model invalidation for
  biochemical reaction networks.
\newblock \emph{BMC Syst. Biol.} 4.
\newblock \doi{10.1186/1752-0509-4-69}
\bibAnnoteFile{rumschinski2010}

\bibitem[{Ruttor and Opper(2010)}]{ruttor2010}
Ruttor, A. and Opper, M. (2010).
\newblock Efficient statistical inference for stochastic reaction processes.
\newblock ArXiv:0906.5321v2 [physics.data-an]
\bibAnnoteFile{ruttor2010}

\bibitem[{Sadamoto et~al.(2017)Sadamoto, Ishizaki, and ichi
  Imura}]{sadamoto2017}
Sadamoto, T., Ishizaki, T., and ichi Imura, J. (2017).
\newblock Average state observers for large-scale network systems.
\newblock \emph{IEEE Trans. Contr. Net. Syst.} 4, 761--769.
\newblock \doi{10.1109/TCNS.2016.2550866}
\bibAnnoteFile{sadamoto2017}

\bibitem[{Sagar et~al.(2017)Sagar, LeCover, Shoemaker, and Varner}]{sagar2017}
Sagar, A., LeCover, R., Shoemaker, C., and Varner, J. (2017).
\newblock Dynamic optimization with particle swarms {(DOPS)}: {A}
  meta-heuristic for parameter estimation in biochemical models.
\newblock \emph{BMC Syst. Biol.} 12, 1--15.
\newblock \doi{10.1186/s12918-018-0610-x}
\bibAnnoteFile{sagar2017}

\bibitem[{Saltelli et~al.(2005)Saltelli, Ratto, Tarantola, and
  Campolongo}]{saltelli2005}
Saltelli, A., Ratto, M., Tarantola, S., and Campolongo, F. (2005).
\newblock Sensitivity analysis for chemical models.
\newblock \emph{Chem. Rev.} 105, 2811--2828.
\newblock \doi{10.1021/cr040659d}
\bibAnnoteFile{saltelli2005}

\bibitem[{Saltelli et~al.(2004)Saltelli, Tarantola, Campolongo, and
  Ratto}]{saltelli2004}
Saltelli, A., Tarantola, S., Campolongo, F., and Ratto, M. (2004).
\newblock \emph{Sensitivity Analysis in Practice} (Wiley)
\bibAnnoteFile{saltelli2004}

\bibitem[{Schenkendorf(2014)}]{schenkendorf2014th}
Schenkendorf, R. (2014).
\newblock \emph{Optimal Experimental Design for Parameter Identification and
  Model Selection}.
\newblock Ph.D. thesis, Otto-von-Guericke-Universit\"at Magdeburg
\bibAnnoteFile{schenkendorf2014th}

\bibitem[{Schilling et~al.(2016)Schilling, Bogomolov, Henzinger, Podelski, and
  Ruess}]{schilling2016}
Schilling, C., Bogomolov, S., Henzinger, T.~A., Podelski, A., and Ruess, J.
  (2016).
\newblock Adaptive moment closure for parameter inference of biochemical
  reaction networks.
\newblock \emph{BioSyst.} \doi{10.1016/j.biosystems.2016.07.005}
\bibAnnoteFile{schilling2016}

\bibitem[{Schnoerr(2016)}]{schnoerr2016th}
Schnoerr, D. (2016).
\newblock \emph{Approximation methods and inference for stochastic biochemical
  kinetics}.
\newblock Ph.D. thesis, University of Edinburgh
\bibAnnoteFile{schnoerr2016th}

\bibitem[{Schnoerr et~al.(2017)Schnoerr, Sanguinetti, and Grima}]{schnoerr2017}
Schnoerr, D., Sanguinetti, G., and Grima, R. (2017).
\newblock Approximation and inference methods for stochastic biochemical
  kinetics - {A} tutorial review.
\newblock \emph{J. Physics A: Math. Theor.} \doi{10.1088/1751-8121/aa54d9}
\bibAnnoteFile{schnoerr2017}

\bibitem[{Septier and Peters(2016)}]{septier2016}
Septier, F. and Peters, G.~W. (2016).
\newblock Langevin and {Hamiltonian} based sequential {MCMC} for efficient
  {Bayesian} filtering in high-dimensional spaces.
\newblock \emph{IEEE J. Sel. Topics Sig. Proces.} 10, 312--327.
\newblock \doi{10.1109/JSTSP.2015.2497211}
\bibAnnoteFile{septier2016}

\bibitem[{Shacham and Brauner(2014)}]{shacham2014}
Shacham, M. and Brauner, N. (2014).
\newblock Application of stepwise regression for dynamic parameter estimation.
\newblock \emph{Comput. Chem. Eng.} 69, 26--38.
\newblock \doi{10.1016/j.compchemeng.2014.06.013}
\bibAnnoteFile{shacham2014}

\bibitem[{Sherlock et~al.(2014)Sherlock, Golightly, and
  Gillespie}]{sherlock2014}
Sherlock, C., Golightly, A., and Gillespie, C.~S. (2014).
\newblock Bayesian inference for hybrid discrete-continuous stochastic kinetic
  models.
\newblock ArXiv:1402.6602v1 [stat.CO]
\bibAnnoteFile{sherlock2014}

\bibitem[{Shiang(2009)}]{shiang2009}
Shiang, K.~D. (2009).
\newblock A perturbation-based estimate algorithm for parameters of coupled
  ordinary differential equations, applications from chemical reactions to
  metabolic dynamics.
\newblock \emph{Comput. Method Prog. Biomed.} 94, 118--142.
\newblock \doi{10.1016/j.cmpb.2008.12.001}
\bibAnnoteFile{shiang2009}

\bibitem[{Siegal-Gaskins et~al.(2015)Siegal-Gaskins, Kim, and
  Szederk\'enyi}]{tuza2015}
Siegal-Gaskins, Z. A. T.~D., Kim, J., and Szederk\'enyi, G. (2015).
\newblock Analysis-based parameter estimation of an in vitro
  transcription-translation system.
\newblock In \emph{ECC}. 1560--1566.
\newblock \doi{10.1109/ECC.2015.7330760}
\bibAnnoteFile{tuza2015}

\bibitem[{Singh and Hahn(2005)}]{singh2005}
Singh, A.~K. and Hahn, J. (2005).
\newblock State estimation for high-dimensional chemical processes.
\newblock \emph{Comput. Chem. Eng.} 29, 2326--2334.
\newblock \doi{10.1016/j.compchemeng.2005.05.009}
\bibAnnoteFile{singh2005}

\bibitem[{Slezak et~al.(2010)Slezak, Su\'arez, Cecchi, Marshall, and
  Stolovitzky}]{slezak2010}
Slezak, D.~F., Su\'arez, C., Cecchi, G.~A., Marshall, G., and Stolovitzky, G.
  (2010).
\newblock When the optimal is not the best: {P}arameter estimation in complex
  biological models.
\newblock \emph{PLOS One} 5, e13283.
\newblock \doi{10.1371/journal.pone.0013283}
\bibAnnoteFile{slezak2010}

\bibitem[{Smadbeck(2014)}]{smadbeck2014th}
Smadbeck, P. (2014).
\newblock \emph{Chemical Master Equations for Non-linear Stochastic Reaction
  Networks: {C}losure Schemes and Implications for Discovery in the Biological
  Sciences}.
\newblock Ph.D. thesis, University Of Minnesota
\bibAnnoteFile{smadbeck2014th}

\bibitem[{Smet and Marchal(2010)}]{smet2010}
Smet, R.~D. and Marchal, K. (2010).
\newblock Advantages and limitations of current network inference methods.
\newblock \emph{Nature Rev. Microbiol.} 8, 717--729.
\newblock \doi{10.1038/nrmicro2419}
\bibAnnoteFile{smet2010}

\bibitem[{Smith and Grima(2018)}]{smith2018}
Smith, S. and Grima, R. (2018).
\newblock Spatial stochastic intracellular kinetics: {A} review of modelling
  approaches.
\newblock \emph{Bull. Math. Biol.} , 1--50\doi{10.1007/s11538-018-0443-1}
\bibAnnoteFile{smith2018}

\bibitem[{Srinath and Gunawan(2010)}]{srinath2010}
Srinath, S. and Gunawan, R. (2010).
\newblock Parameter identifiability of power-law biochemical system models.
\newblock \emph{Biotech.} 149, 132--140.
\newblock \doi{10.1016/j.jbiotec.2010.02.019}
\bibAnnoteFile{srinath2010}

\bibitem[{Srinivas and Rangaiah(2007)}]{srinivas2007}
Srinivas, M. and Rangaiah, G.~P. (2007).
\newblock Differential evolution with tabu list for global optimization and its
  application to phase equilibrium and parameter estimation problems.
\newblock \emph{Ind. Eng. Chem. Res.} 46, 3410--3421.
\newblock \doi{10.1021/ie0612459}
\bibAnnoteFile{srinivas2007}

\bibitem[{Srivastava(2012)}]{srivastava2012th}
Srivastava, R. (2012).
\newblock \emph{Parameter Estimation in Stochastic Chemical Kinetic Models}.
\newblock Ph.D. thesis, Univ. Wisconsin-Madison
\bibAnnoteFile{srivastava2012th}

\bibitem[{Srivastavaa and Rawlingsb(2014)}]{srivastava2014}
Srivastavaa, R. and Rawlingsb, J.~B. (2014).
\newblock Parameter estimation in stochastic chemical kinetic models using
  derivative free optimization and bootstrapping.
\newblock \emph{Comput. Chem. Eng.} 63, 152--158.
\newblock \doi{10.1016/j.compchemeng.2014.01.006}
\bibAnnoteFile{srivastava2014}

\bibitem[{Sun et~al.(2012)Sun, Garibaldi, and Hodgman}]{sun2012}
Sun, J., Garibaldi, J.~M., and Hodgman, C. (2012).
\newblock Parameter estimation using metaheuristics in systems biology: {A}
  comprehensive review.
\newblock \emph{ACM Trans. Comput. Biol. Bioinf.} 9, 185--202.
\newblock \doi{10.1109/TCBB.2011.63}
\bibAnnoteFile{sun2012}

\bibitem[{Sun et~al.(2014)Sun, Palade, Cai, Fang, and Wu}]{sun2014}
Sun, J., Palade, V., Cai, Y., Fang, W., and Wu, X. (2014).
\newblock Biochemical systems identification by a random drift particle swarm
  optimization approach.
\newblock \emph{BMC Bioinf.} 15, S1.
\newblock \doi{10.1186/1471-2105-15-S6-S1}
\bibAnnoteFile{sun2014}

\bibitem[{Sun et~al.(2008)Sun, Jin, and Xiong}]{sun2008}
Sun, X., Jin, L., and Xiong, M. (2008).
\newblock Extended {Kalman} filter for estimation of parameters in nonlinear
  state-space models of biochemical networks.
\newblock \emph{PLOS One} 3, e3758.
\newblock \doi{10.1371/journal.pone.0003758}
\bibAnnoteFile{sun2008}

\bibitem[{Swaminathan and Murray(2014)}]{swaminathan2014}
Swaminathan, A. and Murray, R.~M. (2014).
\newblock Identification of {Markov} chains from distributional measurements
  and applications to systems biology.
\newblock In \emph{IFAC World Cong.} vol.~47, 4400--4405.
\newblock \doi{10.3182/20140824-6-ZA-1003.02771}
\bibAnnoteFile{swaminathan2014}

\bibitem[{Tanevski et~al.(2010)Tanevski, D\v{z}eroski, and
  Kocarev}]{tanevski2010}
Tanevski, J., D\v{z}eroski, S., and Kocarev, L. (2010).
\newblock Approximate {Bayesian} parameter inference for dynamical systems in
  systems biology.
\newblock \emph{Sec. Math. Tech. Sci.} XXXI, 73--98.
\newblock \doi{10.20903/csnmbs.masa.2010.31.1-2.24}
\bibAnnoteFile{tanevski2010}

\bibitem[{Tangherloni et~al.(2016)Tangherloni, Nobile, and
  Cazzanigah}]{tangherloni2016}
Tangherloni, A., Nobile, M.~S., and Cazzanigah, P. (2016).
\newblock {GPU}-powered bat algorithm for the parameter estimation of
  biochemical kinetic values.
\newblock In \emph{CIBCB}. 1--6.
\newblock \doi{10.1109/CIBCB.2016.7758103}
\bibAnnoteFile{tangherloni2016}

\bibitem[{Teijeiro et~al.(2017)Teijeiro, Pardo, Penas, Gonz\'alez, Banga, and
  Doallo}]{teijeiro2017}
Teijeiro, D., Pardo, X.~C., Penas, D.~R., Gonz\'alez, P., Banga, J.~R., and
  Doallo, R. (2017).
\newblock A cloud-based enhanced differential evolution algorithm for parameter
  estimation problems in computational systems biology.
\newblock \emph{Cluster Comput.} 20, 1937--1950.
\newblock \doi{10.1007/s10586-017-0860-1}
\bibAnnoteFile{teijeiro2017}

\bibitem[{Tenazinha and Vinga(2011)}]{tenazinha2011}
Tenazinha, N. and Vinga, S. (2011).
\newblock A survey on methods for modeling and analyzing integrated biological
  networks.
\newblock \emph{Trans. Comput. Biol. Bioinf.} 8, 943--958.
\newblock \doi{10.1109/TCBB.2010.117}
\bibAnnoteFile{tenazinha2011}

\bibitem[{Th and Manini(2008)}]{horvath2008}
Th, A. S.~H. and Manini, D. (2008).
\newblock Parameter estimation of kinetic rates in stochastic reaction networks
  by the em method.
\newblock In \emph{CBEI}. 713--717.
\newblock \doi{10.1109/BMEI.2008.237}
\bibAnnoteFile{horvath2008}

\bibitem[{Thomas et~al.(2012)Thomas, Straube, and Grima}]{thomas2012}
Thomas, P., Straube, A.~V., and Grima, R. (2012).
\newblock The slow-scale linear noise approximation: {A}n accurate, reduced
  stochastic description of biochemical networks under timescale separation
  conditions.
\newblock \emph{BMC Syst. Biol.} 6, 1--23.
\newblock \doi{10.1186/1752-0509-6-39}
\bibAnnoteFile{thomas2012}

\bibitem[{Tian et~al.(2010)Tian, Mu, and Wu}]{tian2010}
Tian, L.-P., Mu, L., and Wu, F.~X. (2010).
\newblock Iterative linear least squares method of parameter estimation for
  linear-fractional models of molecular biological systems.
\newblock In \emph{ICBBE}. 1--4.
\newblock \doi{10.1109/ICBBE.2010.5516994}
\bibAnnoteFile{tian2010}

\bibitem[{Tian et~al.(2007)Tian, Xu, Gao, and Burrage}]{tian2007}
Tian, T., Xu, S., Gao, J., and Burrage, K. (2007).
\newblock Simulated maximum likelihood method for estimating kinetic rates in
  gene expression.
\newblock \emph{Syst. Biol.} 23, 84--91.
\newblock \doi{10.1093/bioinformatics/btl552}
\bibAnnoteFile{tian2007}

\bibitem[{Toni and Stumpf(2009)}]{toni2009}
Toni, T. and Stumpf, M. P.~H. (2009).
\newblock Parameter inference and model selection in signaling pathway models.
\newblock ArXiv:0905.4468v1 [q-bio.QM]
\bibAnnoteFile{toni2009}

\bibitem[{Transtrum and Qiu(2012)}]{transtrum2012}
Transtrum, M.~K. and Qiu, P. (2012).
\newblock Optimal experiment selection for parameter estimation in biological
  differential equation models.
\newblock \emph{BMC Bioinf.} 13.
\newblock \doi{10.1186/1471-2105-13-181}
\bibAnnoteFile{transtrum2012}

\bibitem[{Vanlier et~al.(2013)Vanlier, Tiemann, Hilbers, and van
  Riel}]{vanlier2013}
Vanlier, J., Tiemann, C.~A., Hilbers, P. A.~J., and van Riel, N. A.~W. (2013).
\newblock Parameter uncertainty in biochemical models described by ordinary
  differential equations.
\newblock \emph{Math. Biosci.} 246, 305--314.
\newblock \doi{10.1016/j.mbs.2013.03.006}
\bibAnnoteFile{vanlier2013}

\bibitem[{Vargas et~al.(2014)Vargas, Moreno, and Wouwer}]{vargas2014}
Vargas, A., Moreno, J., and Wouwer, A.~V. (2014).
\newblock A weighted variable gain super-twisting observer for the estimation
  of kinetic rates in biological systems.
\newblock \emph{Process Control} \doi{10.1016/j.jprocont.2014.04.018}
\bibAnnoteFile{vargas2014}

\bibitem[{\v{C}eska et~al.(2014)\v{C}eska, \v{S}afr\'anek, Dra\v{z}an, and
  Brim}]{ceska2014}
\v{C}eska, M., \v{S}afr\'anek, D., Dra\v{z}an, S., and Brim, L. (2014).
\newblock Robustness analysis of stochastic biochemical systems.
\newblock \emph{PLOS One} 9, e94553.
\newblock \doi{10.1371/journal.pone.0094553}
\bibAnnoteFile{ceska2014}

\bibitem[{\v{C}e\v{s}ka et~al.(2017)\v{C}e\v{s}ka, Dannenberg, Paoletti,
  Kwiatkowska, and Brim}]{ceska2017}
\v{C}e\v{s}ka, M., Dannenberg, F., Paoletti, N., Kwiatkowska, M., and Brim, L.
  (2017).
\newblock Precise parameter synthesis for stochastic biochemical systems.
\newblock \emph{Acta Informatica} 54, 589--623.
\newblock \doi{10.1007/s00236-016-0265-2}
\bibAnnoteFile{ceska2017}

\bibitem[{Veerman et~al.(2018)Veerman, Popovic, and Marr}]{veerman2018}
Veerman, F., Popovic, N., and Marr, C. (2018).
\newblock Parameter inference with analytical propagators for stochastic models
  of autoregulated gene expression.
\newblock \doi{10.1101/349431}.
\newblock BioRxiv
\bibAnnoteFile{veerman2018}

\bibitem[{Venayak et~al.(2018)Venayak, von Kamp, Klamt, and
  Mahadevan}]{venayak2018}
Venayak, N., von Kamp, A., Klamt, S., and Mahadevan, R. (2018).
\newblock Move identifies metabolic valves to switch between phenotypic states.
\newblock \emph{Nature Commun.} 9, 5332.
\newblock \doi{10.1038/s41467-018-07719-4}
\bibAnnoteFile{venayak2018}

\bibitem[{Villaverde and Barreiro(2016)}]{villaverde2016a}
Villaverde, A.~F. and Barreiro, A. (2016).
\newblock Identifiability of large nonlinear biochemical networks.
\newblock \emph{MATCH Commun. Math. Comput. Chem.} 76, 259--296
\bibAnnoteFile{villaverde2016a}

\bibitem[{Villaverde et~al.(2016)Villaverde, Barreiro, and
  Papachristodoulou}]{villaverde2016}
Villaverde, A.~F., Barreiro, A., and Papachristodoulou, A. (2016).
\newblock Structural identifiability of dynamic systems biology models.
\newblock \emph{PLOS Comput. Biol.} 12, e1005153.
\newblock \doi{10.1371/journal.pcbi.1005153}
\bibAnnoteFile{villaverde2016}

\bibitem[{Villaverde et~al.(2012)Villaverde, Egea, and Banga}]{villaverde2012}
Villaverde, A.~F., Egea, J.~A., and Banga, J.~R. (2012).
\newblock A cooperative strategy for parameter estimation in large scale
  systems biology models.
\newblock \emph{BMC Syst. Biol.} 6.
\newblock \doi{10.1186/1752-0509-6-75}
\bibAnnoteFile{villaverde2012}

\bibitem[{Villaverde et~al.(2014)Villaverde, Ross, Mor\'an, and
  Banga}]{villaverde2014}
Villaverde, A.~F., Ross, J., Mor\'an, F., and Banga, J.~R. (2014).
\newblock {MIDER}: {N}etwork inference with mutual information distance and
  entropy reduction.
\newblock \emph{PLOS One} 9, e96732.
\newblock \doi{10.1371/journal.pone.0096732}
\bibAnnoteFile{villaverde2014}

\bibitem[{Voit(2013)}]{voit2013}
Voit, E.~O. (2013).
\newblock Biochemical systems theory: {A} review.
\newblock \emph{ISRN Biomath.} , 1--53\doi{10.1155/2013/897658}
\bibAnnoteFile{voit2013}

\bibitem[{von Stosch et~al.(2014)von Stosch, Oliveira, Peres, and
  de~Azevedo}]{stosch2014}
von Stosch, M., Oliveira, R., Peres, J., and de~Azevedo, S.~F. (2014).
\newblock Hybrid semi-parametric modeling in process systems engineering:
  {P}ast, present and future.
\newblock \emph{Comput. Chem. Eng.} 60, 86--101.
\newblock \doi{10.1016/j.compchemeng.2013.08.008}
\bibAnnoteFile{stosch2014}

\bibitem[{Vrettas et~al.(2011)Vrettas, Cornford, and Opper}]{vrettas2011}
Vrettas, M.~D., Cornford, D., and Opper, M. (2011).
\newblock Estimating parameters in stochastic systems: {A} variational
  {Bayesian} approach.
\newblock \emph{Physica D: Nonlin. Phenom.} 240, 1877--1900.
\newblock \doi{10.1016/j.physd.2011.08.013}
\bibAnnoteFile{vrettas2011}

\bibitem[{Wang et~al.(2010)Wang, Christley, Mjolsness, and Xie}]{wang2010}
Wang, Y., Christley, S., Mjolsness, E., and Xie, X. (2010).
\newblock Parameter inference for discretely observed stochastic kinetic models
  using stochastic gradient descent.
\newblock \emph{BMC Syst. Biol.} 4.
\newblock \doi{10.1186/1752-0509-4-99}
\bibAnnoteFile{wang2010}

\bibitem[{Weber and Frey(2017)}]{weber2017}
Weber, M.~F. and Frey, E. (2017).
\newblock Master equations and the theory of stochastic path integrals.
\newblock \emph{Rep. Prog. Phys.} 80, 046601.
\newblock \doi{10.1088/1361-6633/aa5ae2}
\bibAnnoteFile{weber2017}

\bibitem[{Weiss et~al.(2016)Weiss, Khoshgoftaar, and Wang}]{weiss2016}
Weiss, K., Khoshgoftaar, T.~M., and Wang, D.-D. (2016).
\newblock A survey of transfer learning.
\newblock \emph{J. Big Data} 3, 1--40.
\newblock \doi{10.1186/s40537-016-0043-6}
\bibAnnoteFile{weiss2016}

\bibitem[{Whitaker et~al.(2017)Whitaker, Golightly, Boys, and
  Sherlock}]{whitaker2017}
Whitaker, G.~A., Golightly, A., Boys, R.~J., and Sherlock, C. (2017).
\newblock Bayesian inference for diffusion-driven mixed-effects models.
\newblock \emph{Bayesian Analysis} 12, 435--463.
\newblock \doi{10.1214/16-BA1009}
\bibAnnoteFile{whitaker2017}

\bibitem[{White et~al.(2016)White, Tolman, Thames, Withers, Mason, and
  Transtrum}]{white2016}
White, A., Tolman, M., Thames, H.~D., Withers, H.~R., Mason, K.~A., and
  Transtrum, M.~K. (2016).
\newblock The limitations of model-based experimental design and parameter
  estimation in sloppy systems.
\newblock \emph{PLOS Comput. Biol.} 12, e1005227.
\newblock \doi{10.1371/journal.pcbi.1005227}
\bibAnnoteFile{white2016}

\bibitem[{White et~al.(2015)White, Kypraios, and Preston}]{white2015}
White, S., Kypraios, T., and Preston, S. (2015).
\newblock Piecewise approximate {Bayesian} computation: {F}ast inference for
  discretely observed {Markov} models using a factorised posterior
  distribution.
\newblock \emph{Stat. Comput.} 25, 289--301.
\newblock \doi{10.1007/s11222-013-9432-2}
\bibAnnoteFile{white2015}

\bibitem[{Wong et~al.(2015)Wong, Krycer, Burchﬁeld, James, and
  Kuncic}]{wong2015}
Wong, M. K.~L., Krycer, J.~R., Burchﬁeld, J.~G., James, D.~E., and Kuncic, Z.
  (2015).
\newblock A generalised enzyme kinetic model for predicting the behaviour of
  complex biochemical systems.
\newblock \emph{FEBS Open Bio.} 5, 226--239.
\newblock \doi{10.1016/j.fob.2015.03.002}
\bibAnnoteFile{wong2015}

\bibitem[{Woodcock et~al.(2011)Woodcock, Komorowski, Finkenstadt, Harper,
  Davis, White et~al.}]{woodcock2011}
Woodcock, D., Komorowski, M., Finkenstadt, B., Harper, C., Davis, J., White,
  M., et~al. (2011).
\newblock A {Bayesian} hierarchical diffusion model for estimating kinetic
  parameters and cell-to-cell variability.
\newblock Research Article
\bibAnnoteFile{woodcock2011}

\bibitem[{Xiong and Zhou(2013)}]{xiong2013}
Xiong, J. and Zhou, T. (2013).
\newblock Parameter identification for nonlinear state-space models of a
  biological network via linearization and robust state estimation.
\newblock In \emph{Chinese Control. Conf.} 8235--8240
\bibAnnoteFile{xiong2013}

\bibitem[{Yang et~al.(2012)Yang, Dent, and Nardini}]{yang2015}
Yang, X., Dent, J.~E., and Nardini, C. (2012).
\newblock An {S}-system parameter estimation method {(SPEM)} for biological
  networks.
\newblock \emph{Comput. Biol.} 19, 175--187.
\newblock \doi{10.1089/cmb.2011.0269}
\bibAnnoteFile{yang2015}

\bibitem[{Yang et~al.(2014)Yang, Guo, and Guo}]{yang2014}
Yang, X., Guo, Y., and Guo, L. (2014).
\newblock An iterative parameter estimation method for biological systems and
  its parallel implementation.
\newblock \emph{Concur. Comput.: Pract. Exper.} 26, 1249--1267.
\newblock \doi{10.1002/cpe.3071}
\bibAnnoteFile{yang2014}

\bibitem[{Yenkie et~al.(2016)Yenkie, Diwekar, and Linninger}]{yenkie2016}
Yenkie, K., Diwekar, U., and Linninger, A. (2016).
\newblock Simulation-free estimation of reaction propensities in cellular
  reactions and gene signaling networks.
\newblock \emph{Comput. Chem. Eng.} 87, 154--163.
\newblock \doi{10.1016/j.compchemeng.2016.01.010}
\bibAnnoteFile{yenkie2016}

\bibitem[{Zamora-Sillero et~al.(2011)Zamora-Sillero, Hafner, Ibig, Stelling,
  and Wagner}]{sillero2011}
Zamora-Sillero, E., Hafner, M., Ibig, A., Stelling, J., and Wagner, A. (2011).
\newblock Efficient characterization of high-dimensional parameter spaces for
  systems biology.
\newblock \emph{BMC Syt. Biol.} 5.
\newblock \doi{10.1186/1752-0509-5-142}
\bibAnnoteFile{sillero2011}

\bibitem[{Zechner(2014)}]{zechner2014th}
Zechner, C. (2014).
\newblock \emph{Stochastic Biochemical Networks in Random Environments:
  {P}robabilistic Modeling And Inference}.
\newblock Ph.D. thesis, ETH.
\newblock \doi{10.3929/ethz-a-010256930}
\bibAnnoteFile{zechner2014th}

\bibitem[{Zechner et~al.(2012)Zechner, Nandy, Unger, and Koeppl}]{zechner2012}
Zechner, C., Nandy, P., Unger, M., and Koeppl, H. (2012).
\newblock Optimal variational perturbations for the inference of stochastic
  reaction dynamics.
\newblock In \emph{CDC}. 5336--5341.
\newblock \doi{10.1109/CDC.2012.6426738}
\bibAnnoteFile{zechner2012}

\bibitem[{Zechner et~al.(2011)Zechner, Pelet, Peter, and Koeppl}]{zechner2011}
Zechner, C., Pelet, S., Peter, M., and Koeppl, H. (2011).
\newblock Recursive {Bayesian} estimation of stochastic rate constants from
  heterogeneous cell populations.
\newblock In \emph{CDC-ECC}. 5837--5843.
\newblock \doi{10.1109/CDC.2011.6161329}
\bibAnnoteFile{zechner2011}

\bibitem[{Zhan et~al.(2014)Zhan, Situ, Yeung, Tsang, and Yang}]{zhan2014}
Zhan, C., Situ, W., Yeung, L.~F., Tsang, P. W.-M., and Yang, G. (2014).
\newblock A parameter estimation method for biological systems modeled by
  {ODEs/DDEs} models using spline approximation and differential evolution
  algorithm.
\newblock \emph{ACM Trans. Comput. Biol. Bioinf.} 11, 1066--1076.
\newblock \doi{10.1109/TCBB.2014.2322360}
\bibAnnoteFile{zhan2014}

\bibitem[{Zhan and Yeung(2011)}]{zhan2011}
Zhan, C. and Yeung, L.~F. (2011).
\newblock Parameter estimation in systems biology models using spline
  approximation.
\newblock \emph{BMC Syst. Biol.} 5.
\newblock \doi{10.1186/1752-0509-5-14}
\bibAnnoteFile{zhan2011}

\bibitem[{Zimmer(2015)}]{zimmer2015}
Zimmer, C. (2015).
\newblock Reconstructing the hidden states in time course data of stochastic
  models.
\newblock \emph{Math. Biosci.} 269, 117--129.
\newblock \doi{10.1016/j.mbs.2015.08.015}
\bibAnnoteFile{zimmer2015}

\bibitem[{Zimmer(2016)}]{zimmer2016}
Zimmer, C. (2016).
\newblock Experimental design for stochastic models of nonlinear signaling
  pathways using an interval-wise linear noise approximation and state
  estimation.
\newblock \emph{PLOS One} 11, 1--37.
\newblock \doi{10.1371/journal.pone.0159902}
\bibAnnoteFile{zimmer2016}

\bibitem[{Zimmer et~al.(2016)Zimmer, Bergmann, and Sahle}]{zimmer2016a}
Zimmer, C., Bergmann, F.~T., and Sahle, S. (2016).
\newblock Reducing local minima in fitness landscapes of parameter estimation
  by using piecewise evaluation and state estimation.
\newblock ArXiv:1601.04458v1 [q-bio.QM]
\bibAnnoteFile{zimmer2016a}

\bibitem[{Zimmer and Sahle(2012)}]{zimmer2012}
Zimmer, C. and Sahle, S. (2012).
\newblock Parameter estimation for stochastic models of biochemical reactions.
\newblock \emph{Comput. Sci. Syst. Biol.} 6.
\newblock \doi{10.4172/jcsb.1000095}
\bibAnnoteFile{zimmer2012}

\bibitem[{Zimmer and Sahle(2015)}]{zimmer2014}
Zimmer, C. and Sahle, S. (2015).
\newblock Deterministic inference for stochastic systems using multiple
  shooting and a linear noise approximation for the transition probabilities.
\newblock \emph{IET Syst. Biol.} 9, 181--192.
\newblock \doi{10.1049/iet-syb.2014.0020}
\bibAnnoteFile{zimmer2014}

\bibitem[{Zimmer et~al.(2014)Zimmer, Sahle, and Pahle}]{zimmer2010}
Zimmer, C., Sahle, S., and Pahle, J. (2014).
\newblock Exploiting intrinsic fluctuations to identify model parameters.
\newblock \emph{IET Syst. Biol.} 9, 64--73.
\newblock \doi{10.1049/iet-syb.2014.0010}
\bibAnnoteFile{zimmer2010}

\end{thebibliography}

\clearpage

\renewcommand{\thetable}{S\arabic{table}}
\newcommand{\sscite}[1]{{\footnotesize\cite{#1}}}

\section*{Supplementary tables}\label{sc:suppl}

\toc{tb:S1}{Coverage of modeling strategies of BRNs.}
\toc{tb:S2}{Coverage of parameter estimation strategies for BRNs.}
\toc{tb:S3}{References with citation links in Google Scholar.}
\toc{tb:S4}{Selected authors on Google Scholar.}

\tref{tb:S1} lists all references cited in the main text indicating how many
times given model was mentioned in each reference.

\tref{tb:S2} lists all references cited in the main text indicating how many
times given task or method was mentioned in each reference.

\tref{tb:S3} contains links to citing references on Google Scholar for selected
papers.

\tref{tb:S4} contains links for selected authors having papers concerning
parameter estimation in BRNs or in dynamic systems.

\begin{table}[h!]
\centering\footnotesize
\caption{Coverage of modeling strategies for BRNs.}
\label{tb:S1}
\setlength{\tabcolsep}{2pt} \renewcommand{\arraystretch}{1.1}
\begin{tabular}{p{4cm}*{25}{c}}
\headerA{Reference}
\sscite{abdullah2013} &. &. &. &. &. &. &. &1 &1 &1 &1 &. &. &. &. &. &. &. &. &. &. &. &. &. &. \\
\sscite{abdullah2013a} &. &. &. &. &. &. &. &2 &4 &4 &. &. &. &1 &. &. &. &. &1 &. &. &. &. &. &. \\
\sscite{abdullah2013b} &. &. &. &. &. &. &. &. &7 &1 &. &. &. &. &. &. &. &. &3 &. &. &. &. &. &. \\
\sscite{alberton2013} &. &. &1 &. &. &. &. &. &. &. &. &. &. &. &. &. &. &1 &. &. &. &. &. &. &. \\
\sscite{ale2013} &1 &. &. &. &. &. &. &. &10 &. &. &. &. &3 &. &. &. &. &. &. &. &. &7 &15 &. \\
\sscite{ali2015} &. &. &2 &. &. &. &. &4 &1 &. &. &. &. &3 &. &. &. &21 &. &. &. &. &. &. &. \\
\sscite{amrein2012} &. &2 &. &33 &3 &. &. &9 &1 &. &. &. &. &. &. &. &. &. &. &. &. &. &. &. &. \\
\sscite{anai2006} &. &. &. &. &. &. &. &. &1 &. &. &. &. &2 &. &. &. &. &. &. &. &. &. &. &. \\
\hline \end{tabular}
\end{table}

\begin{table}[h!]\ContinuedFloat
\centering\footnotesize
\caption{Coverage of modeling strategies for BRNs. (cont.)}
\setlength{\tabcolsep}{2pt} \renewcommand{\arraystretch}{1.1}
\begin{tabular}{p{4cm}*{25}{c}}
\headerA{Reference}
\sscite{chong2012} &. &. &. &. &. &. &. &. &4 &. &. &. &1 &. &. &. &. &1 &. &. &. &. &. &. &. \\
\sscite{chong2014} &. &. &. &. &. &. &. &. &7 &. &. &. &. &. &. &. &. &. &. &. &. &. &. &. &. \\
\sscite{chou2006} &1 &. &. &. &. &. &. &. &. &. &. &. &23 &. &. &. &. &1 &. &. &. &. &. &. &. \\
\sscite{chun2009} &9 &6 &10 &. &. &. &. &. &5 &. &. &. &82 &. &2 &1 &2 &6 &1 &. &. &. &. &. &. \\
\sscite{cseke2016} &. &. &. &22 &. &. &. &4 &14 &. &. &. &. &2 &. &. &. &. &. &6 &5 &. &14 &. &. \\
\sscite{dai2010} &. &. &. &. &. &. &. &. &7 &1 &. &. &. &. &1 &. &. &. &. &. &. &. &. &. &. \\
\sscite{daigle2012} &. &2 &5 &2 &47 &17 &. &. &. &. &. &. &. &. &. &. &. &. &. &. &. &. &. &. &. \\
\sscite{dargatz2010th} &. &1 &. &195 &25 &. &. &83 &144 &. &. &. &. &9 &1 &. &. &. &1 &33 &13 &. &. &2 &. \\
\hline \end{tabular}
\end{table}

\begin{table}[h!]\ContinuedFloat
\centering\footnotesize
\caption{Coverage of modeling strategies for BRNs. (cont.)}
\setlength{\tabcolsep}{2pt} \renewcommand{\arraystretch}{1.1}
\begin{tabular}{p{4cm}*{25}{c}}
\headerA{Reference}
\sscite{gosalbez2013} &3 &. &1 &. &. &. &. &. &1 &. &. &. &2 &. &. &. &. &. &. &. &. &. &. &. &. \\
\sscite{goutsias2012} &1 &2 &. &139 &24 &. &. &12 &1 &. &. &. &. &2 &. &8 &. &20 &10 &15 &2 &. &13 &22 &. \\
\sscite{gratie2013} &. &5 &4 &8 &. &. &. &1 &55 &1 &. &. &. &1 &. &2 &. &. &1 &. &. &. &. &. &. \\
\sscite{gupta2013th} &. &2 &1 &72 &6 &. &. &17 &48 &. &. &. &. &1 &. &. &. &. &. &. &. &. &. &. &. \\
\sscite{gupta2014} &. &2 &1 &17 &. &. &. &11 &4 &1 &. &. &. &. &. &. &. &. &. &. &. &. &. &. &. \\
\sscite{hagen2018} &. &1 &4 &. &. &. &. &. &6 &4 &. &. &. &. &. &. &. &. &. &. &. &. &. &2 &. \\
\sscite{hasenauer2010} &. &2 &. &. &. &. &. &. &. &1 &. &. &. &9 &. &. &. &. &. &. &. &. &. &. &. \\
\sscite{hasenauer2013th} &3 &3 &12 &43 &1 &4 &. &2 &99 &3 &. &. &. &6 &2 &. &. &. &2 &21 &30 &10 &1 &. &. \\
\hline \end{tabular}
\end{table}

\begin{table}[h!]\ContinuedFloat
\centering\footnotesize
\caption{Coverage of modeling strategies for BRNs. (cont.)}
\setlength{\tabcolsep}{2pt} \renewcommand{\arraystretch}{1.1}
\begin{tabular}{p{4cm}*{25}{c}}
\headerA{Reference}
\sscite{liao2017th} &. &. &1 &10 &18 &19 &. &10 &35 &. &. &535 &. &46 &1 &. &. &. &. &13 &42 &13 &. &14 &19 \\
\sscite{liepe2014} &. &. &3 &5 &. &. &. &. &19 &. &. &. &. &. &. &. &. &. &. &. &. &. &. &2 &. \\
\sscite{lillacci2010} &. &1 &. &2 &. &. &. &2 &2 &2 &. &. &. &. &. &. &. &. &1 &. &. &. &. &. &. \\
\sscite{lillacci2012} &. &1 &. &2 &. &. &. &4 &4 &1 &. &. &. &. &. &. &. &. &. &. &. &. &. &. &. \\
\sscite{linder2013th} &. &23 &. &6 &16 &. &. &. &56 &. &. &. &. &. &. &. &. &. &1 &2 &. &4 &. &53 &. \\
\sscite{linder2014} &. &10 &. &9 &2 &. &. &. &13 &. &. &. &. &. &. &. &. &. &. &1 &. &2 &. &18 &. \\
\sscite{liu2006} &. &. &. &. &. &. &. &. &. &. &. &. &. &. &. &. &. &. &. &. &. &. &. &. &. \\
\sscite{liu2008} &. &. &. &. &. &. &. &. &1 &. &. &. &42 &2 &. &. &. &1 &. &. &. &. &. &. &. \\
\hline \end{tabular}
\end{table}

\begin{table}[h!]\ContinuedFloat
\centering\footnotesize
\caption{Coverage of modeling strategies for BRNs. (cont.)}
\setlength{\tabcolsep}{2pt} \renewcommand{\arraystretch}{1.1}
\begin{tabular}{p{4cm}*{25}{c}}
\headerA{Reference}
\sscite{pan2010} &. &. &. &6 &. &. &. &. &. &. &. &. &. &. &4 &. &. &. &. &. &. &. &. &. &. \\
\sscite{pantazis2013} &. &2 &. &15 &2 &3 &. &. &8 &. &. &. &. &1 &. &. &. &. &. &6 &1 &. &. &15 &. \\
\sscite{paul2014th} &. &1 &. &12 &16 &26 &. &2 &12 &. &. &. &. &. &. &. &. &. &. &1 &. &. &2 &. &. \\
\sscite{penas2017} &. &. &. &. &. &. &. &. &10 &1 &4 &. &. &. &. &. &. &. &. &. &. &. &. &. &. \\
\sscite{plesa2017} &. &. &. &. &. &. &. &5 &35 &. &. &1 &. &9 &2 &. &. &. &. &. &. &. &. &. &. \\
\sscite{poovathingal2010} &. &. &. &1 &. &. &. &. &1 &. &. &. &. &. &. &. &. &. &. &1 &. &. &. &. &. \\
\sscite{pullen2014} &. &. &7 &6 &. &. &. &1 &6 &. &. &. &. &. &. &. &. &1 &1 &. &. &. &. &. &. \\
\sscite{quach2007} &. &3 &. &3 &. &. &. &16 &23 &. &. &. &. &. &. &. &. &. &. &. &. &. &. &. &. \\
\hline \end{tabular}
\end{table}

\begin{table}[h!]\ContinuedFloat
\centering\footnotesize
\caption{Coverage of modeling strategies for BRNs. (cont.)}
\setlength{\tabcolsep}{2pt} \renewcommand{\arraystretch}{1.1}
\begin{tabular}{p{4cm}*{25}{c}}
\headerA{Reference}
\sscite{swaminathan2014} &. &. &. &21 &. &. &. &1 &. &. &. &. &. &. &. &. &. &. &. &. &. &. &. &. &. \\
\sscite{tanevski2010} &. &2 &. &6 &5 &. &. &1 &1 &. &. &. &. &. &. &. &. &. &. &. &. &1 &. &. &. \\
\sscite{tangherloni2016} &1 &. &1 &1 &. &. &. &. &6 &. &. &. &. &. &. &. &. &. &. &. &. &. &. &. &. \\
\sscite{teijeiro2017} &. &. &. &. &. &. &. &. &48 &2 &. &. &. &. &. &. &. &. &. &. &. &. &. &. &. \\
\sscite{tenazinha2011} &1 &1 &3 &. &. &. &. &1 &39 &. &. &. &. &. &. &. &. &. &. &1 &. &. &. &. &. \\
\sscite{thomas2012} &5 &. &. &. &4 &. &. &. &. &. &. &. &. &. &. &. &. &. &. &35 &4 &. &. &175 &. \\
\sscite{tian2007} &. &. &. &1 &7 &. &. &. &29 &. &. &. &. &. &. &. &. &. &. &1 &. &. &. &. &. \\
\sscite{tian2010} &. &. &. &. &. &. &. &. &. &. &. &. &. &. &. &. &. &. &. &. &. &. &. &. &. \\
\hline \end{tabular}
\end{table}

\begin{table}[h!]
\centering\footnotesize
\caption{Coverage of parameter estimation methods for BRNs.}
\label{tb:S2}
\setlength{\tabcolsep}{2pt} \renewcommand{\arraystretch}{1.1}
\begin{tabular}{p{4cm}*{28}{c}}
\headerB{Reference}
\sscite{abdullah2013} &. &. &. &37 &. &. &. &. &. &2 &. &. &. &. &. &. &1 &4 &. &. &5 &. &. &35 &2 &59 &. &. \\
\sscite{abdullah2013a} &11 &. &. &60 &. &. &2 &. &. &2 &. &. &. &. &. &. &4 &4 &. &. &4 &. &1 &27 &2 &6 &9 &1 \\
\sscite{abdullah2013b} &15 &. &. &57 &. &. &1 &. &. &4 &. &. &. &. &. &. &. &. &. &1 &3 &. &2 &12 &2 &27 &12 &1 \\
\sscite{alberton2013} &60 &. &1 &66 &5 &. &2 &. &. &1 &. &. &. &. &. &. &. &. &. &. &2 &2 &. &. &. &2 &1 &1 \\
\sscite{ale2013} &1 &1 &. &20 &6 &. &. &2 &. &4 &. &. &. &. &. &. &. &. &. &. &. &. &. &. &2 &. &. &. \\
\sscite{ali2015} &45 &. &. &119 &. &. &. &. &. &1 &. &1 &. &. &. &1 &25 &23 &7 &. &4 &2 &1 &. &1 &. &22 &. \\
\sscite{amrein2012} &. &1 &. &18 &. &. &. &. &. &16 &. &. &. &9 &4 &1 &. &. &. &. &. &. &. &. &. &. &. &. \\
\sscite{anai2006} &. &. &. &33 &. &3 &. &. &. &. &. &. &. &. &. &. &. &. &. &1 &. &2 &. &1 &. &. &. &. \\
\hline \end{tabular}
\end{table}

\begin{table}[h!]\ContinuedFloat
\centering\footnotesize
\caption{Coverage of parameter estimation methods for BRNs. (cont.)}
\setlength{\tabcolsep}{2pt} \renewcommand{\arraystretch}{1.1}
\begin{tabular}{p{4cm}*{28}{c}}
\headerB{Reference}
\sscite{chong2012} &. &1 &. &38 &. &. &. &. &. &. &. &. &. &. &. &. &10 &. &1 &. &5 &3 &9 &38 &. &1 &1 &. \\
\sscite{chong2014} &. &. &. &65 &. &. &. &. &. &. &. &. &. &. &. &1 &6 &. &1 &. &6 &. &15 &61 &3 &7 &14 &. \\
\sscite{chou2006} &. &. &. &43 &. &. &. &. &3 &1 &. &. &. &. &. &. &. &. &. &42 &2 &. &. &. &. &. &1 &. \\
\sscite{chun2009} &3 &1 &. &232 &. &. &. &3 &2 &6 &. &. &. &. &. &. &1 &. &. &25 &15 &26 &10 &8 &. &10 &7 &. \\
\sscite{cseke2016} &. &. &. &46 &. &. &. &. &. &49 &. &. &23 &1 &. &. &1 &. &. &1 &. &. &. &1 &2 &. &. &1 \\
\sscite{dai2010} &1 &. &. &25 &. &. &1 &. &1 &1 &. &. &. &. &3 &. &. &. &. &. &1 &4 &18 &10 &2 &. &. &. \\
\sscite{daigle2012} &. &. &. &32 &1 &6 &. &15 &. &89 &4 &16 &. &. &3 &1 &. &. &. &. &3 &1 &. &. &2 &. &. &. \\
\sscite{dargatz2010th} &1 &2 &. &527 &1 &17 &3 &. &. &251 &. &3 &1 &100 &22 &11 &1 &. &. &1 &. &1 &. &. &. &. &. &. \\
\hline \end{tabular}
\end{table}

\begin{table}[h!]\ContinuedFloat
\centering\footnotesize
\caption{Coverage of parameter estimation methods for BRNs. (cont.)}
\setlength{\tabcolsep}{2pt} \renewcommand{\arraystretch}{1.1}
\begin{tabular}{p{4cm}*{28}{c}}
\headerB{Reference}
\sscite{gosalbez2013} &4 &. &. &43 &. &. &. &. &. &1 &. &. &. &. &. &. &. &. &. &1 &. &5 &. &. &. &. &. &. \\
\sscite{goutsias2012} &. &2 &. &34 &11 &. &. &42 &. &2 &. &. &. &. &1 &. &1 &. &. &. &. &1 &2 &. &. &. &20 &. \\
\sscite{gratie2013} &9 &. &. &23 &16 &. &. &. &. &. &. &. &. &. &. &. &. &. &. &. &. &. &1 &. &. &. &. &. \\
\sscite{gupta2013th} &. &8 &. &156 &. &. &. &1 &1 &280 &6 &. &. &174 &64 &. &. &. &. &10 &. &6 &. &. &1 &. &. &1 \\
\sscite{gupta2014} &. &2 &. &56 &. &. &. &. &. &93 &. &. &. &54 &10 &. &. &. &. &. &. &. &. &. &. &. &. &. \\
\sscite{hagen2018} &1 &. &. &7 &. &. &15 &1 &. &10 &. &. &. &. &. &. &. &. &. &. &. &2 &. &. &. &. &. &. \\
\sscite{hasenauer2010} &4 &. &. &38 &. &3 &4 &. &. &26 &. &. &. &. &. &. &. &. &. &. &. &10 &1 &2 &. &. &. &. \\
\sscite{hasenauer2013th} &16 &4 &. &210 &1 &49 &1 &2 &. &256 &. &. &1 &37 &10 &2 &2 &2 &. &1 &. &7 &. &. &10 &1 &. &. \\
\hline \end{tabular}
\end{table}

\begin{table}[h!]\ContinuedFloat
\centering\footnotesize
\caption{Coverage of parameter estimation methods for BRNs. (cont.)}
\setlength{\tabcolsep}{2pt} \renewcommand{\arraystretch}{1.1}
\begin{tabular}{p{4cm}*{28}{c}}
\headerB{Reference}
\sscite{liao2017th} &13 &40 &. &34 &7 &. &. &. &3 &8 &. &. &1 &. &. &. &. &. &. &1 &. &2 &. &. &. &. &. &. \\
\sscite{liepe2014} &2 &1 &. &67 &1 &. &. &1 &. &78 &129 &. &. &4 &1 &30 &. &. &. &1 &. &5 &1 &. &1 &. &. &. \\
\sscite{lillacci2010} &11 &. &. &93 &. &. &. &. &. &20 &. &. &. &. &. &2 &46 &41 &1 &. &8 &1 &2 &. &1 &. &. &. \\
\sscite{lillacci2012} &20 &. &. &56 &. &. &. &. &. &21 &. &. &. &. &1 &5 &9 &6 &. &. &2 &. &3 &. &. &. &. &. \\
\sscite{linder2013th} &7 &. &. &73 &. &. &. &. &. &40 &. &9 &. &6 &. &. &. &. &. &6 &. &. &. &5 &1 &. &. &. \\
\sscite{linder2014} &3 &. &. &12 &. &. &. &. &. &8 &. &4 &. &. &. &. &. &. &. &4 &. &. &. &. &. &. &. &. \\
\sscite{liu2006} &2 &. &. &14 &. &. &. &. &. &. &. &. &. &. &. &. &. &. &. &. &. &. &1 &1 &. &. &. &. \\
\sscite{liu2008} &. &. &. &81 &. &. &. &. &. &. &. &. &. &. &. &. &. &. &. &5 &3 &2 &2 &6 &1 &. &1 &. \\
\hline \end{tabular}
\end{table}

\begin{table}[h!]\ContinuedFloat
\centering\footnotesize
\caption{Coverage of parameter estimation methods for BRNs. (cont.)}
\setlength{\tabcolsep}{2pt} \renewcommand{\arraystretch}{1.1}
\begin{tabular}{p{4cm}*{28}{c}}
\headerB{Reference}
\sscite{pan2010} &. &. &. &14 &. &. &1 &. &. &20 &. &. &. &. &2 &. &. &. &. &12 &. &3 &. &. &. &. &. &145 \\
\sscite{pantazis2013} &13 &. &. &18 &49 &. &8 &47 &. &7 &. &. &1 &. &. &. &. &. &. &. &. &1 &. &3 &. &. &. &. \\
\sscite{paul2014th} &. &1 &. &25 &. &. &. &. &. &29 &. &. &. &42 &5 &. &. &. &. &. &. &1 &. &. &. &. &. &. \\
\sscite{penas2017} &. &. &. &39 &. &1 &. &. &. &1 &. &. &. &. &. &. &. &. &. &. &. &10 &1 &9 &39 &1 &. &. \\
\sscite{plesa2017} &. &87 &. &19 &. &. &. &. &. &. &. &. &. &. &. &. &. &. &. &. &. &1 &. &. &. &. &. &. \\
\sscite{poovathingal2010} &6 &1 &. &63 &2 &. &. &. &. &40 &. &1 &. &. &. &. &. &. &. &. &. &. &. &13 &. &. &. &. \\
\sscite{pullen2014} &1 &1 &. &46 &. &. &2 &1 &. &77 &. &. &. &20 &. &1 &1 &. &. &3 &1 &1 &6 &. &. &1 &2 &2 \\
\sscite{quach2007} &. &. &. &52 &. &1 &. &. &. &10 &. &. &. &1 &. &1 &13 &3 &19 &2 &. &. &1 &. &. &. &1 &. \\
\hline \end{tabular}
\end{table}

\begin{table}[h!]\ContinuedFloat
\centering\footnotesize
\caption{Coverage of parameter estimation methods for BRNs. (cont.)}
\setlength{\tabcolsep}{2pt} \renewcommand{\arraystretch}{1.1}
\begin{tabular}{p{4cm}*{28}{c}}
\headerB{Reference}
\sscite{swaminathan2014} &2 &. &. &23 &. &. &. &1 &. &15 &. &. &. &. &. &. &. &. &. &. &. &3 &. &. &. &. &. &2 \\
\sscite{tanevski2010} &. &1 &. &54 &. &. &. &. &. &57 &32 &. &. &3 &4 &8 &. &. &. &. &. &. &. &. &. &. &. &. \\
\sscite{tangherloni2016} &. &. &. &15 &1 &. &. &. &. &. &. &. &. &. &. &. &. &. &. &. &1 &1 &. &1 &. &43 &. &. \\
\sscite{teijeiro2017} &. &. &. &25 &. &. &. &. &. &. &. &. &. &. &. &. &. &. &. &. &. &14 &. &94 &14 &. &. &. \\
\sscite{tenazinha2011} &12 &3 &. &13 &1 &. &. &. &. &1 &. &11 &. &. &. &. &. &. &. &. &. &6 &1 &. &. &. &. &. \\
\sscite{thomas2012} &. &2 &. &1 &. &. &. &. &. &. &. &. &. &. &. &. &. &. &. &. &. &. &. &. &3 &. &. &. \\
\sscite{tian2007} &. &1 &. &66 &. &. &. &. &. &24 &. &. &. &. &. &. &. &. &. &. &14 &8 &. &. &. &. &. &2 \\
\sscite{tian2010} &. &. &. &24 &1 &. &. &. &. &. &. &. &. &. &. &. &. &. &. &. &. &1 &. &1 &. &. &. &. \\
\hline \end{tabular}
\end{table}

\begin{table}[h!]
\centering\footnotesize
\caption{References with citation links in Google Scholar.}
\label{tb:S3}
\setlength{\tabcolsep}{2pt} \renewcommand{\arraystretch}{1.1}
\medskip
\begin{tabular}[b]{p{4cm}c}
\cite{abdullah2013a} & \href{https://scholar.google.com/scholar?cites=10570837309197599144}{Cited by} \\
\cite{abdullah2013b} & \href{https://scholar.google.com/scholar?cites=7034188236591388967}{Cited by} \\
\cite{abdullah2013} & \href{https://scholar.google.com/scholar?cites=1765020018039400029}{Cited by} \\
\cite{alberton2013} & \href{https://scholar.google.com/scholar?cites=16204885638161406614}{Cited by} \\
\cite{ale2013} & \href{https://scholar.google.co.uk/scholar?cites=9654143282456951974}{Cited by} \\
\cite{ali2015} & \href{https://scholar.google.com/scholar?cites=1844180024481874978}{Cited by} \\
\cite{amrein2012} & \href{https://scholar.google.com/scholar?cites=115813755976429088}{Cited by} \\
\cite{anai2006} & \href{https://scholar.google.com/scholar?cites=6491868545884082479}{Cited by} \\
\end{tabular}\hspace{2cm}
\begin{tabular}[b]{p{4cm}c}
\cite{farina2006} & \href{https://scholar.google.com/scholar?cites=17993853061875092248}{Cited by} \\
\cite{fearnhead2012} & \href{https://scholar.google.co.uk/scholar?cites=9442788587044255939}{Cited by} \\
\cite{fearnhead2014} & \href{https://scholar.google.com/scholar?cites=1370170108587356242}{Cited by} \\
\cite{fernandez2005} & \href{https://scholar.google.com/scholar?cites=6578107654303571783}{Cited by} \\
\cite{fernandez2006} & \href{https://scholar.google.com/scholar?cites=14507744740078460589}{Cited by} \\
\cite{fernandez2013} & \href{https://scholar.google.com/scholar?cites=6583978075398386440}{Cited by} \\
\cite{fey2008} & \href{https://scholar.google.com/scholar?cites=10223995924379058215}{Cited by} \\
\cite{fey2010} & \href{https://scholar.google.com/scholar?cites=1273741637653706188}{Cited by} \\
\end{tabular}
\end{table}

\begin{table}[h!]\ContinuedFloat
\centering\footnotesize
\caption{References with citation links in Google Scholar. (cont.)}
\setlength{\tabcolsep}{2pt} \renewcommand{\arraystretch}{1.1}
\medskip
\begin{tabular}[b]{p{4cm}c}
\cite{kulikov2015} & \href{https://scholar.google.com/scholar?cites=1124157258227926081}{Cited by} \\
\cite{kulikov2017} & \href{https://scholar.google.com/scholar?cites=2855842962037700139}{Cited by} \\
\cite{kutalik2007} & \href{https://scholar.google.com/scholar?cites=1431488917076358365}{Cited by} \\
\cite{kuwahara2013} & \href{https://scholar.google.com/scholar?cites=15279087509228183902}{Cited by} \\
\cite{lakatos2015} & \href{https://scholar.google.co.uk/scholar?cites=11484023247295381795}{Cited by} \\
\cite{lang2016} & \href{https://scholar.google.com/scholar?cites=5252688847958826449}{Cited by} \\
\cite{li2013} & \href{https://scholar.google.com/scholar?cites=17516147301820104580}{Cited by} \\
\cite{li2015} & \href{https://scholar.google.com/scholar?cites=13688090178561832628}{Cited by} \\
\end{tabular}\hspace{2cm}
\begin{tabular}[b]{p{4cm}c}
\cite{radulescu2012} & \href{https://scholar.google.co.uk/scholar?cites=5123118791102915346}{Cited by} \\
\cite{rakhshani2016} & \href{https://scholar.google.com/scholar?cites=2123784919802140547}{Cited by} \\
\cite{ramsay2007} & \href{https://scholar.google.co.uk/scholar?cites=4794354911441576056}{Cited by} \\
\cite{rapaport2005} & \href{https://scholar.google.com/scholar?cites=668256977737481682}{Cited by} \\
\cite{reinker2006} & \href{https://scholar.google.com/scholar?cites=4195713876598631397}{Cited by} \\
\cite{reis2018} & \href{https://scholar.google.co.uk/scholar?cites=8378163191462724015}{Cited by} \\
\cite{remli2017} & \href{https://scholar.google.com/scholar?cites=15755057400145272690}{Cited by} \\
\cite{rempala2012} & \href{https://scholar.google.com/scholar?cites=12543233544086480158}{Cited by} \\
\end{tabular}
\end{table}

\begin{table}[h!]\ContinuedFloat
\centering\footnotesize
\caption{References with citation links in Google Scholar. (cont.)}
\setlength{\tabcolsep}{2pt} \renewcommand{\arraystretch}{1.1}
\medskip
\begin{tabular}[b]{p{4cm}c}
\cite{yang2014} & \href{https://scholar.google.com/scholar?cites=16274967709199694966}{Cited by} \\
\cite{yang2015} & \href{https://scholar.google.com/scholar?cites=17431728010711388579}{Cited by} \\
\cite{yenkie2016} & \href{https://scholar.google.com/scholar?cites=7188145514754128802}{Cited by} \\
\cite{zechner2011} & \href{https://scholar.google.com/scholar?cites=4385181190161354418}{Cited by} \\
\cite{zechner2012} & \href{https://scholar.google.com/scholar?cites=7001520248721553514}{Cited by} \\
\cite{zhan2011} & \href{https://scholar.google.com/scholar?cites=17390795037657585883}{Cited by} \\
\cite{zhan2014} & \href{https://scholar.google.com/scholar?cites=8272587224150367927}{Cited by} \\
\end{tabular}\hspace{2cm}
\begin{tabular}[b]{p{4cm}c}
\cite{zimmer2010} & \href{https://scholar.google.com/scholar?cites=902592463732622750}{Cited by} \\
\cite{zimmer2012} & \href{https://scholar.google.com/scholar?cites=11960062007382896208}{Cited by} \\
\cite{zimmer2014} & \href{https://scholar.google.com/scholar?cites=5299915778902021142}{Cited by} \\
\cite{zimmer2015} & \href{https://scholar.google.com/scholar?cites=11973287426086788711}{Cited by} \\
\cite{zimmer2016a} & \href{https://scholar.google.com/scholar?cites=9655919672435685759}{Cited by} \\
\cite{zimmer2016} & \href{https://scholar.google.co.uk/scholar?cites=9815856314965043654}{Cited by} \\
& \\
\end{tabular} 
\end{table}

\begin{table}[t!]
\centering\footnotesize
\caption{Selected authors on Google Scholar with papers concerning parameter
  estimation for BRNs and dynamic systems.}
\label{tb:S4}
\setlength{\tabcolsep}{2pt} \renewcommand{\arraystretch}{1.1}
\medskip
\begin{tabular}[b]{p{4cm}}
\href{https://scholar.google.com/citations?user=lSO0DLoAAAAJ}{Afnizanfaizal Abdullah} \\
\href{https://scholar.google.co.uk/citations?user=d27Ji6kAAAAJ}{Reka Albert} \\
\href{https://scholar.google.com/citations?user=mUQhSJUAAAAJ}{André Lu\'is Alberton} \\
\href{https://scholar.google.com/citations?user=NRDNuAQAAAAJ}{Jarinah Mohd Ali} \\
\href{https://scholar.google.com/citations?user=P2LqxTkAAAAJ}{Goekman Altay} \\
\href{https://scholar.google.com/citations?user=rIX6oiMAAAAJ}{James Anderson} \\
\href{https://scholar.google.de/citations?user=rbRdAM0AAAAJ}{Alexander Andreychenko} \\
\href{https://scholar.google.com/citations?user=kcsbLrAAAAAJ}{Christophe Andrieu} \\
\end{tabular}
\begin{tabular}[b]{p{4cm}}
\href{https://scholar.google.com/citations?user=_21NhC8AAAAJ}{Paul Fearnhead} \\
\href{https://scholar.google.com/citations?user=AIo0lxYAAAAJ}{Maria Rodriguez-Fernandez} \\
\href{https://scholar.google.com/citations?user=pGYETGQAAAAJ}{Fabian Fr\"ohlich} \\
  \href{https://scholar.google.com/citations?user=tdvvRoMAAAAJ}{Attila Gabor} \\
\href{https://scholar.google.com/citations?user=yVjqbPAAAAAJ}{Arnab Ganguly} \\
\href{https://scholar.google.com/citations?user=qyOs7OMAAAAJ}{Colin S Gillespie} \\
\href{https://scholar.google.com/citations?user=Pdt6-qMAAAAJ}{Andrew Golightly} \\
\href{https://scholar.google.com/citations?user=ynIWXnUAAAAJ}{Javier Gonz\'alez} \\
\end{tabular}
\begin{tabular}[b]{p{4cm}}
\href{https://scholar.google.com/citations?user=e0tI6ZcAAAAJ}{Hojjat Rakhshani} \\
\href{https://scholar.google.com/citations?user=3TVtgBAAAAAJ}{Rajesh Ramaswamy} \\
\href{https://scholar.google.com/citations?user=gTXxQAsAAAAJ}{Rangaiah} \\
\href{https://scholar.google.com/citations?user=JVzehi0AAAAJ}{Andreas Raue} \\
\href{https://scholar.google.com/citations?user=yjJoP8gAAAAJ}{Muhammad Akmal Remli} \\
\href{https://scholar.google.com/citations?user=G5jzkhkAAAAJ}{Michal Rewienski} \\
\href{https://scholar.google.com/citations?user=VDVhQ1EAAAAJ}{Jakob Ruess} \\
\href{https://scholar.google.com/citations?user=G-jg20sAAAAJ}{Philipp Rumschinski} \\
\end{tabular} 
\end{table}

\end{document}